\documentclass[runningheads]{svmult}

\usepackage{makeidx}   % allows index generation
\usepackage{graphicx}  % standard LaTeX graphics tool
\usepackage{subeqnar}  % subnumbers individual equations
\usepackage{multicol}  % used for the two-column index
\usepackage{physprbb}  % modified textarea for proceedings,
\makeindex             % used for the subject index
\begin{document}
\title*{MHD Turbulence: Scaling Laws and Astrophysical Implications}
\toctitle{MHD Turbulence:
\protect\newline Scaling Laws and Astrophysical Implications}

\titlerunning{MHD Turbulence}

\author{Jungyeon Cho\inst{1}
\and A. Lazarian\inst{1}
\and Ethan T.~Vishniac\inst{2}  }

\authorrunning{Cho, Lazarian, \& Vishniac}

\institute{Univ. of Wisconsin, Madison WI53706, USA
\and Johns Hopkins Univ., Baltimore MD21218, USA}

\maketitle              % typesets the title of the contribution

\begin{abstract}
%optional. 70-150 words.
Turbulence is the most common state of astrophysical flows.
In typical astrophysical fluids, turbulence is accompanied by
strong magnetic fields, which has a large
impact on the dynamics of the turbulent cascade.
Recently, there has been a significant
breakthrough on the theory of magnetohydrodynamic (MHD) turbulence.
For the first time we have
a scaling model that is supported by both observations and
numerical simulations.
We review recent progress in studies of both incompressible
and compressible turbulence.
We compare Iroshnikov-Kraichnan and
Goldreich-Sridhar models, and discuss scalings of
Alfv\'en, slow, and fast waves.
We also discuss the completely new regime of MHD turbulence that happens
below the scale at which hydrodynamic
turbulent motions are damped by viscosity.
In the case of the partially ionized diffuse interstellar gas
the viscosity is due to neutrals
and truncates the turbulent cascade at $\sim$parsec scales. We show
that below this scale magnetic fluctuations with a shallow spectrum
persist and discuss the possibility of a resumption of the
MHD cascade after ions and neutrals decouple.
We discuss the implications of this new insight into MHD turbulence
for cosmic ray transport, grain dynamics, etc., and how to test 
theoretical predictions against observations.
\end{abstract}

\section{Introduction} 

Most astrophysical systems, e.g. accretion disks, stellar winds, 
the interstellar medium (ISM) and intercluster medium 
are turbulent 
with an embedded magnetic field that influences almost all
of their properties. This turbulence which spans from km to many
kpc (see discussion in \cite{Arm95,Sca87,Lpe02})
holds the key to many astrophysical
processes (e.g., transport of mass and angular momentum,
star formation, fragmentation of molecular
clouds, heat and cosmic ray transport, magnetic reconnection).
Statistics of turbulence is also essential for the cosmic microwave
background (CMB) radiation foreground
studies \cite{Laz02}.

All turbulent systems have one thing in common: they have a large
``Reynolds number" ($Re\equiv LV/\nu$; L= the characteristic
scale or driving scale 
of the system, V=the velocity difference over this scale, and $\nu$=viscosity),
the ratio of
the viscous drag time on the largest scales ($L^2/\nu$)
to the eddy turnover time of a parcel of gas ($L/V$).
A similar parameter, the ``magnetic
Reynolds number", $Rm$ ($\equiv LV/\eta$; $\eta$=magnetic diffusion),
is the ratio of
the magnetic field decay time ($L^2/\eta$)
to the eddy turnover time ($L/V$).
The properties of the flows on all scales
depend on $Re$ and $Rm$. Flows with $Re<100$ are laminar; chaotic
structures develop gradually as $Re$ increases, and those with
$Re\sim10^3$ are appreciably less chaotic than those with
$Re\sim10^7$. Observed features such as star forming clouds 
and accretion disks are very
chaotic with $Re>10^8$ and $Rm>10^{16}$.

{}Let us start by considering incompressible hydrodynamic turbulence,
which can be described by the
Kolmogorov theory \cite{Kol41}.
Suppose that we excite fluid motions at a scale $L$.
We call this scale the {\it energy injection scale} or the
{\it largest energy containing eddy scale}. For instance, an obstacle
in a flow excites motions on scales of the order of its size.
Then the energy injected at the scale $L$ cascades
to progressively
smaller and
smaller scales
at the eddy turnover rate, i.e. $\tau_l^{-1}\approx v_l/l$,
with negligible energy losses along the cascade \footnote{This
is easy to see as the motions at the scales of large eddies
have $Re\gg 1$.}.
Ultimately, the energy reaches the molecular dissipation scale $l_d$,
i.e. the scale where the local $Re\sim 1$,
and is dissipated there.
The scales between $L$ and $l_d$ are called the {\it inertial range}
and it typically covers many decades. The motions over the inertial
range are {\it self-similar} and this provides tremendous advantages
for theoretical description.

The beauty of the Kolmogorov theory is that it does provide a simple
scaling for hydrodynamic motions. If
the velocity at a scale $l$ from the inertial range is $v_l$, the
Kolmogorov theory states that the kinetic energy ($\rho v_l^2\sim v_l^2$
as the density is constant) is
transferred to next scale within one eddy turnover
time ($l/v_l$). Thus within the Kolmogorov theory the energy
transfer rate ($v_l^2/(l/v_l)$) is scale-independent,
\begin{equation}
  \frac{ v_l^2}{t_{cas}}  \sim \frac{ v_l^2}{(l/v_l)} = \mbox{~constant}, \label{scale_indep}
\end{equation}
 and
we get the famous Kolmogorov scaling
\begin{equation}
v_l \propto l^{1/3}.
\end{equation} 

The one-dimensional\footnote{Dealing with observational data, e.g. in
LPE02 \cite{Lpe02},
we deal with three dimensional energy spectrum $P(k)$, which, 
for isotropic turbulence, is given by
$E(k)=4\pi k^2 P(k)$.}
energy spectrum $E(k)$ is
the amount of energy between the wavenumber $k$
and $k + dk$ divided by $dk$.
When $E(k)$ is a power law, $kE(k)$ is the energy {\it near} the 
wavenumber $k\sim 1/l$.
Since $v_l^2 \approx kE(k)$,
Kolmogorov scaling implies
\begin{equation}
E(k) \propto k^{-5/3}.
\end{equation}

Kolmogorov scalings were the first major
advance in the theory of incompressible turbulence.
They have led to numerous applications
in different branches of science (see \cite{Mon75}). 
However, astrophysical fluids are magnetized and the a 
dynamically important magnetic field should interfere with eddy
motions.    

Paradoxically, astrophysical measurements are consistent with
Kolmogorov spectra (see LPE02 \cite{Lpe02}).
For instance, interstellar scintillation observations
indicate an electron density spectrum  is very close to $-5/3$ for
$10^8 cm$ - $10^{15} cm$ (see \cite{Arm95}).   %%%(see Armstrong et al.~1995).
At larger scales LPE02 summarizes the evidence of
$-5/3$ velocity power spectrum over pc-scales in HI.
Solar-wind observations provide {\it in-situ} measurements of the
power spectrum of magnetic fluctuations and 
Leamon et al.~\cite{Lea98} also obtained a slope of $\approx -5/3$.
Is this a coincidence?  
What properties is the magnetized compressible ISM
expected to have? We will deal with these questions, and some
related issues, below.

Our approach here is complementary to that in
Vazquez-Semadeni (this volume) and Mac Low (this volume). 
These reviews discuss attempts to simulate the turbulent ISM in all 
its complexity by including many physical processes
(e.g. heating, cooling, self-gravity) simultaneously. This provides
a possibility of comparing observations and simulations (see review
by Ostriker, this volume). 
The disadvantage is
that such simulations cannot distinguish between the 
consequences of different processes. Note, that in
studies of turbulence the adaptive mesh does not help as
the fine structures emerge through the entire computational volume.

Here we discuss a focused approach which aims at 
obtaining a clear understanding on the fundamental level, and
considering physically relevant complications later. The creative
synthesis of both approaches is the way, we think, that 
studies of
astrophysical turbulence should proceed\footnote{Potentially our approach leads
to an understanding of the relationship between motions at a given time at
small scales
(subgrid scales) and the state of the flow at a previous time at some larger,
resolved, scale. This could lead to a parametrization of the subgrid
scales and to large eddy simulations of MHD.}.  Certainly an understanding
of MHD turbulence in the most ideal terms is a necessary precursor to
understanding the complications posed by more realistic physics and
numerical effects.
For review of general properties of MHD, see a recent book by 
Biskamp \cite{Bis93}.

In what follows, we first consider observational data that motivate 
our study (\S2),
then discuss theoretical approaches to 
incompressible MHD turbulence (\S3). In \S4 we discuss testing and extending
of the Goldreich-Sridhar theory of turbulence,  then in \S5 we deal with
viscous damping of incompressible turbulence and describe a new regime
of MHD turbulence that is present below the viscous cut-off scale. We
move to the effects of compressibility in \S6 and 
discuss implications of our new understanding of MHD turbulence
for the problems of dust motion, cosmic ray dynamics, support of molecular 
clouds,
heating of ISM etc in \S7. We propose observational testing of our results
in \S8 and present the summary in \S9.

\section{Observational Data}

Kolmogorov turbulence is the simplest possible model of turbulence.
Since it is incompressible and not magnetized, it is 
completely specified by its velocity spectrum. If a passive scalar
field, like ``dye particles'' or temperature inhomogeneities, is subjected
to Kolmogorov turbulence, the resulting spectrum of the passive scalar
density is also Kolmogorov (see \cite{Les90,War00}). 
In compressible and magnetized
turbulence this is no longer true, and a complete characterization
of the turbulence requires not only a study of 
the velocity statistics but also the statistics of density and 
magnetic fluctuations.
 
Direct studies of turbulence\footnote{Indirect studies include
the line-velocity relationships \cite{Lar81} where the integrated
velocity profiles are interpreted as the consequence of turbulence.
Such studies do not provide the statistics of turbulence and their
interpretation is very model dependent.} 
have been done mostly for interstellar medium
and for the Solar wind. While for the Solar wind {\it in-situ} measurements
are possible, studies of interstellar turbulence require inverse techniques to
interpret the observational data. 

Attempts to study interstellar turbulence with statistical tools
date as far back as the 1950s \cite{Hor51,Kam55,Mun58,Wil59}
and various directions
of research achieved various degree of success (see reviews by
\cite{Kap70,Dic85,Arm95,Laz99a,Laz99b,Lpe02}).

\subsection{Solar wind} 

Solar wind (see review \cite{Gold95}) 
studies allow pointwise statistics to be measured directly
using spacecrafts. These studies are the closest counterpart of 
laboratory measurements.

The solar wind flows nearly radially away from the Sun, at up to about 
700 km/s. This is much faster than both spacecraft motions and 
the Alfv\'en speed. Therefore, the turbulence is ``frozen'' and
the fluctuations at frequency $f$ are directly related to fluctuations
at the scale $k$ in the direction of the wind, as $k=2\pi f/v$, where $v$ is
the solar wind velocity \cite{Hor99}.

Usually two types of solar wind are distinguished, one being the fast
wind which originates in coronal holes, and the slower bursty wind.
Both of them show, however, $f^{-5/3}$ scaling on small scales.
The turbulence is strongly anisotropic (see \cite{Kle93})
with the ratio of power in motions perpendicular to the magnetic field to 
those parallel to the magnetic field being around 30. The intermittency of 
the solar wind 
turbulence is very similar to the intermittency observed in
hydrodynamic flows \cite{Hor97}.

\subsection{Electron density statistics}

Studies of turbulence statistics of ionized media
(see \cite{Spa90}) have provided information on
the statistics of plasma density at scales $10^{8}$-$10^{15}$~cm. 
This was based on
a clear understanding of processes of scintillations and scattering
achieved by theorists\footnote{In fact, the theory of scintillations was
developed first for the atmospheric applications.}
 (see \cite{Nar89,Goo85}).
A peculiar feature of the 
measured spectrum (see \cite{Arm95}) is the absence of
the slope change at the scale at which the viscosity by neutrals
becomes important. 

Scintillation measurements are the most reliable data in the
``big power law'' plot in Armstrong et al.~\cite{Arm95}. However
there are intrinsic limitations to the scintillations technique
due to the limited number of sampling directions, its relevance only to
ionized gas at extremely small scales, 
and the impossibility
of getting velocity (the most important!) statistics directly. Therefore
with the data one faces the problem of distinguishing actual turbulence
from static density structures. Moreover, the scintillation
data does not provide the index of turbulence directly, but only
shows that the data are consistent with Kolmogorov turbulence.
Whether the (3D) index can be -4 instead of -11/3 is still
a subject of intense debate \cite{Hig84,Nar89}. In physical terms the former
corresponds to the superposition of random shocks rather than
eddies.

Additional information on the electron density is contained in
the Faraday rotation measures of extragalactic radio 
sources (see \cite{Sim88,Sim92}).
However,
there is so far 
no reliable way to disentangle contributions of the magnetic field
and the density to the signal. We feel that those measurements
may give us the magnetic field statistics when we know the
statistics of electron density better.

\subsection{Velocity and density statistics from spectral lines}

Spectral line
data cubes are unique sources of information on interstellar turbulence. 
Doppler shifts due to supersonic motions contain information on the
turbulent velocity field which is otherwise difficult to obtain. Moreover,
the statistical samples are extremely
rich and not limited to discrete directions. In addition, line emission
allows us to study turbulence at large scales, comparable
to the scales of star formation and energy injection.

However, the problem of separating velocity and density fluctuations 
within HI data cubes is far from trivial 
\cite{Laz95,Laz99b,LP00,Lpe02}.
The analytical
description of the emissivity statistics of channel maps (velocity slices)
in Lazarian \& Pogosyan \cite{LP00} (see also \cite{Laz99b,Lpe02} for
reviews)
shows that the relative contribution of the
density and velocity fluctuations 
depends on the thickness of the velocity slice.
In particular, the power-law asymptote of the emissivity fluctuations 
changes
when the dispersion of the velocity at the scale under study 
becomes of the order of the velocity slice thickness (the integrated
width of the channel map).   
These results are the foundation of the Velocity-Channel Analysis (VCA) 
technique which provides velocity and density statistics
using spectral line data cubes.
The VCA has been successfully tested using data
cubes obtained via compressible magnetohydrodynamic simulations and 
has been applied
to Galactic and Small Magellanic Cloud atomic hydrogen (HI) data 
\cite{Lpvp01,LP00,Sta01,Des00d}. 
{}Furthermore,
the inclusion of absorption effects \cite{LP02} has increased 
the power of this technique. 
{}Finally, the VCA can be applied to different species (CO, H$_{\alpha}$ etc.)
which should further increase its utility in the future.

Within the present discussion a number of results obtained with the VCA
are important. First of all, the Small Magellanic Cloud (SMC) HI data
exhibit a Kolmogorov-type spectrum for velocity and HI density from
the
smallest resolvable scale of 40~pc to the scale of the SMC itself, i.e.
4~kpc. Similar conclusions can be inferred from the Galactic data
\cite{Gre93} for scales of dozens of parsecs, 
although the analysis has not been done systematically. Deshpande et al. 
\cite{Des00d} studied absorption of HI on small scales toward Cas A and 
Cygnus A.
Within the VCA their results can be interpreted 
as implying
that on scales less than 1~pc the HI velocity is suppressed by
ambipolar drag and the spectrum of density fluctuations is shallow 
$P(k)\sim k^{-2.8}$. Such a spectrum \cite{Des00}
can account for the small scale structure of HI observed in absorption.     

\subsection{Magnetic field statistics}

Magnetic field statistics are the most poorly constrained aspect
of ISM turbulence. The polarization
of starlight and of the Far-Infrared Radiation (FIR) from aligned dust 
grains
is affected by the ambient magnetic fields. Assuming that dust grains are
always aligned with their longer axes perpendicular to magnetic field
(see the review \cite{Laz00a}), one gets the 2D distribution of the 
magnetic field directions in the sky. Note that the alignment is
a highly non-linear process in terms of the magnetic field and
therefore the magnetic field strength is not available\footnote{The
exception to this may be the alignment of small grains which can
be revealed by microwave and UV polarimetry \cite{Laz00a}.}.

The statistics of starlight polarization (see \cite{Fos02})
is rather rich for the Galactic plane and it allows to establish
the spectrum\footnote{Earlier papers
dealt with much poorer samples (see \cite{Kap70})
and they did not reveal power-law spectra.} $E(K)\sim K^{-1.5}$, where $K$ is a two dimensional
wave vector describing the fluctuations over sky patch.\footnote{
%%%
This spectrum is obtained by \cite{Fos02} in terms of the expansion over the 
spherical harmonic basis $Y_{lm}$. 
For sufficiently small areas of the sky analyzed
the multipole analysis results coincide with the Fourier analysis.}
%%%

For uniformly sampled turbulence it follows from Lazarian \& Shutenkov
\cite{Laz90}
that $E(K)\sim K^{\alpha}$ for $K<K_0$ and $K^{-1}$ for $K>K_0$, where
$K_0^{-1}$ is the critical angular size of fluctuations which is
proportional to the ratio of the injection energy scale to the size
of the turbulent system along the line of sight. For Kolmogorov
turbulence $\alpha=-11/3$.

However, the real observations do not uniformly sample turbulence.
Many more close stars are present compared to the distant ones. 
Thus the intermediate slops are expected. Indeed,
Cho \& Lazarian \cite{CL02b} showed through direct simulations that the 
slope obtained in \cite{Fos02} is compatible with the underlying
Kolmogorov turbulence.
At the moment FIR polarimetry
does not provide maps that are really suitable to study turbulence
statistics. This should change soon 
when polarimetry becomes possible using the
airborne SOFIA observatory.  A better understanding of grain
alignment (see \cite{Laz00a}) is required to interpret
the molecular cloud magnetic data where some of the dust is
known not to be aligned (see \cite{LGM97} and references therein).  

Another way to get magnetic field statistics is to use synchrotron
emission. Both polarization and intensity data can be used. The angular
correlation of
polarization data \cite{BBF01} shows the power-law spectrum 
$K^{-1.8}$ and we believe that the interpretation of it is similar to that
of starlight polarization.
Indeed, Faraday depolarization
limits the depth of the sampled region. 
The
intensity fluctuations were studied in \cite{Laz90}
with rather poor initial data and the results were inconclusive.
Cho \& Lazarian \cite{CL02b} 
interpreted the fluctuations of synchrotron emissivity
 \cite{Gia01,Gia02} in terms of turbulence with Kolmogorov spectrum.

\section{Theoretical Approaches to MHD Turbulence}
   Here we consider mainly Kolmogorov-type theories.
   Other theories not discussed in this section include
   the eddy-damped
   quasinormal Markovian (EDQNM) approximation \cite{Pou76},
   the renormalization group technique \cite{Fou82,Ver99}, and
   the direct interaction approximation \cite{Kra59}.

\subsection{Iroshnikov-Kraichnan theory} 
Attempts to describe magnetic turbulence statistics were made by
Iroshnikov \cite{Iro63} and Kraichnan \cite{Kra65}. Their model
of turbulence (IK theory) 
is isotropic in spite of the presence of the magnetic field.

We can understand the IK theory as follows.\footnote{ We follow arguments
in \cite{Gol97}.}
{}For simplicity, let us suppose that
a uniform external magnetic field (${\bf B}_0$) is present.
In the incompressible limit, any magnetic perturbation propagates
{\it along} the magnetic field line.
To the first order, the speed of propagation is constant and equal to
the Alfv\'en speed $V_A=B_0/\sqrt{4\pi \rho}$, where $\rho$ 
is the density.
Since wave packets are moving along the magnetic field line,
there are two possible directions for propagation. 
If all the wave packets are moving in one direction,
then they are stable to nonlinear order\cite{Par79}.
Therefore, in order to initiate turbulence, there must be
opposite-traveling wave packets and the energy cascade occurs only when
they collide.
The IK theory starts from this observation.

The IK theory assumes that,
when two opposite-traveling wave packets of size $l$ collide,
they lose the following amount of energy to smaller scales:
\begin{equation}
 \Delta E \sim (dv^2/dt) \Delta t 
 \sim {\bf v}_l \cdot \dot{\bf v}_l\Delta t
 \sim v_l ( v_l^2/l) \Delta t
 \sim  (v_l^3/l) (l/V_A), \label{deltaE}
\end{equation}
where the IK theory assumes that only collisions 
between similar size packets are
important and $\Delta t \sim l/V_A$.
The latter means that
the duration of the collision is the size of the wave packet 
divided by the speed of the wave packets.
These assumptions look reasonable at first.
But, it is important to note that they fail when
eddies are anisotropic.
That is, if eddies are elongated along the magnetic field line,
then $\Delta t$ is not $l/V_A$, but $l_{\|}/V_A$, where
$l_{\|}$ is the parallel size of the wave packet (or `eddy').

Equation (\ref{deltaE}) tells us that 
the energy change per collision is $v_l^2 (v_l/V_A)$, which
is only a tiny fraction of $v_l^2$ when $V_A \gg v_l$.
Therefore, in order for the eddy to transfer all the energy
to small eddies, the eddy must go through
many collisions.
When such collisions are incoherent, we require
a total $(v^2/\Delta E)^2$ collisions to complete the cascade.
This means that 
the energy cascade time $t_{cas}$ is 
\begin{equation}
 t_{cas} \sim \left(\frac{v^2}{\Delta E}\right)^2 \Delta t
         \sim \frac{l}{v_l} \frac{V_A}{v_l},
\end{equation}
which means that this new cascade time is $(V_A/v_l)$ times
longer than the eddy turnover time ($l/v_l$).
As in the Kolmogorov theory, the IK theory assumes
the constancy of energy cascade (eq.~\ref{scale_indep}):
$
  ( v_l^4 )/( l V_A ) 
                            =\mbox{constant},
$
which, in turn, yields
$
 v_l \propto l^{1/4},
$
or,
\begin{equation}
 \mbox{\it Iroshnikov-Kraichnan:~~~~~} E(k) \propto k^{-3/2}.
\end{equation}

\subsection{Anisotropy}
A uniform component to the magnetic field defines a special direction,
which will be reflected in the dynamics of turbulent fluctuations. 
One obvious effect is that 
it is easy to mix
field lines in directions perpendicular to the
local mean magnetic field and much more difficult to bend them.
The IK theory assumes isotropy of the energy cascade
in Fourier space, an assumption which has attracted severe criticism
\cite{Mon81,She83,Mon95,Sri94,Mat98}.
Mathematically, anisotropy manifests itself in the 
resonant conditions for 3-wave 
interactions:
\begin{eqnarray}
  {\bf k}_1 + {\bf k}_2 & = & {\bf k}_3, \label{k123} \\
  \omega_1 +  \omega_2 & = & \omega_3, \label{w123}
\end{eqnarray}
where ${\bf k}$'s are wavevectors and $\omega$'s are wave frequencies.
The first condition is a statement of wave momentum conservation
and the second is a statement of energy conservation.
Alfv\'{e}n waves satisfy the dispersion relation: $\omega = V_A |k_{\|}|$,
where $k_{\|}$ is the component of wavevector parallel to the background
magnetic
field.
Since only opposite-traveling wave packets interact, ${\bf k}_1$ and
${\bf k}_2$ must have opposite signs.
Then from equations (\ref{k123}) and (\ref{w123}),
either $k_{\|,1}$ or $k_{\|,2}$ must be equal to 0 and
$k_{\|,3}$ must be equal to the nonzero initial parallel
wavenumber.  That is, zero frequency modes are essential
for energy transfer \cite{She83}.
Therefore, in the wavevector space, 
3-wave interactions produce an energy cascade
which is strictly perpendicular to the mean magnetic field.
However, in real turbulence, equation (\ref{w123}) does not
need to be satisfied exactly, but only to within an
an error of order $\delta \omega \sim 1/t_{cas}$ \cite{Gol95}.
This implies that the energy cascade is not strictly 
perpendicular to ${\bf B}_0$, although clearly very anisotropic.

It is noteworthy that 
there has been a claim that ``by increasing the magnitude
of the mean field in 3-D simulations one finds that
the transition from the isotropic 3-D scaling properties toward
those observed in 2-D'' \cite{Bis02}. (see also 
\cite{Bis01,VDE02,Bis02} for recent development in 2-D MHD turbulence).
This claim has yet to
be substantiated, however. We feel that the
available numerical simulations \cite{Mul00,Cv00a,Mar01} 
are reasonably consistent with the Goldreich \& Sridhar \cite{Gol95} model
that we review in the next section.
It is also worth noting that
the idea of an anisotropic (perpendicular) cascade has been incorporated 
into the framework of the reduced MHD approximation
  \cite{Str76,Ros76,Mon82,Zan92,Bha98}.

\subsection{Goldreich-Sridhar theory}
%\vspace{0.3cm}
%\noindent
%{\bf Goldreich-Sridhar theory.}
We assume throughout this discussion that 
the rms turbulent velocity at the energy
injection scale is comparable to the Alfv\'en speed of the mean field
and consider only scales below the energy injection scale.  Consequently,
we are not concerned with the problem of magnetic
field generation, or the magnetic dynamo
which is considered elsewhere in this volume\hfill\break
(see also \cite{Mof78,Par79,kr80} for reviews, 
         \cite{Men81,NBJ92,BJN96,Cat96} for numerical calculations,
         \cite{Vai92,Gru94,Cat96} for suppression of the $\alpha$ dynamo effect 
          in highly conducting fluids, and
         \cite{Kul92,Bra01,Vis01,Mar01b,Zwe97} for recent
          developments).

An ingenious model very similar in its beauty and simplicity 
to the Kolmogorov model has been proposed by 
Goldreich \& Sridhar \cite{Gol95} (1995; hereinafter GS95) 
for incompressible MHD turbulence. They pointed out that 
motions perpendicular to the magnetic field lines mix them
on a hydrodynamic time scale, i.e. at a rate 
$t_{cas}^{-1} \approx k_{\bot}v_l$,
where $k_{\bot}$ is the wavevector component perpendicular
to the local mean magnetic field and
$l\sim k^{-1}(\approx k_{\perp}^{-1})$.
These mixing motions couple to the wave-like motions parallel to
magnetic field giving a 
{\it critical balance} condition
\begin{equation}
      k_{\|} V_A \sim k_{\bot}v_k,   \label{cr_bal}
\end{equation}    
where $k_{\|}$ is the component of the wavevector parallel
to the local magnetic field.  When the typical $k_{\|}$ 
on a scale $k_{\bot}$ falls below this limit, the magnetic
field tension is too weak to affect the dynamics and the
turbulence evolves hydrodynamically, in the direction of
increasing isotropy in phase space.  This quickly raises the
value of $k_{\|}$.  In the opposite limit, when $k_{\|}$ is
large, the magnetic field tension dominates, the error $\delta\omega$
in the matching conditions is reduced, and the nonlinear cascade
is largely in the $k_{\bot}$ direction, which restores the
critical balance.

If conservation of energy in the turbulent cascade applies
locally in phase space then 
the energy cascade rate  
($ v_l^2/t_{cas}$) is constant (eq.~\ref{scale_indep}):
$
    (v_l^2)/(l/v_l) = \mbox{constant.}
$
Combining this with the critical balance condition
we obtain an\hfill\break
anisotropy that increases with decreasing scale
\begin{equation}
   k_{\|} \propto k_{\perp}^{2/3}, \label{twothirds}
\end{equation}
and a Kolmogorov-like spectrum for perpendicular motions
\begin{equation}
v_l \propto l^{1/3}, \mbox{~~or,~} E(k) \propto k_{\perp}^{-5/3},
\label{GS_K}
\end{equation}
which is not surprising since the magnetic field 
does not influence motions that do not bend it. At the
same time, the scale-dependent anisotropy reflects the fact that it
is more difficult for the weaker, smaller eddies to bend 
the magnetic field.

GS95 shows the duality of motions in MHD turbulence. Those perpendicular to
the mean magnetic field are essentially eddies, 
while those parallel to magnetic
field are waves. The critical balance condition couples these two types of
motions.

\subsection{Weak/Intermediate turbulence.}
Let us reconsider the interaction of two wave packets moving oppositely along
the mean magnetic field line.
As in equation (\ref{deltaE}), the energy loss per collision is
\begin{equation}
  \Delta E \sim \dot{(v^2)} \Delta t 
 \sim  (v_l^3/l_{\perp}) (l_{\|}/V_A),
\end{equation}
where we explicitly distinguish the parallel ($l_{\|}$) and the perpendicular
size ($l_{\perp}$).
The kinetic energy of the eddy is $v^2_l$.
Therefore, the ratio of $\Delta E$ to $E$,
\begin{equation}
  \zeta_l \equiv \frac{\Delta E}{v^2_l} 
                           \sim \frac{ v_l l_{\|} }{ V_A l_{\perp} }
                         = \frac{ v_l k_{\perp} }{ V_A k_{\|} },
\end{equation}
characterizes the strength of the nonlinear interaction \cite{Gol95}.
In the GS95 theory, $\zeta_l\sim 1$.
This means that $V_A \approx V$ is required at the energy injection
scale when energy injection is isotropic ($k_{\perp,L}\sim k_{\|,L}$).
When this condition is satisfied, the turbulence is called
{\it strong} turbulence.

There are some astrophysical situations, 
e.g.~the Jovian magnetosphere \cite{Sau01},
where the parameter $\zeta_L$ is much smaller than the unity
over a broad range of length scales. Although as noted above
the cascade will evolve in the direction of increasing $\zeta_l$
for decreasing $l$, and may reach the strong turbulent regime
on very small scales.
In this regime, the parallel cascade is strongly suppressed so the
turbulence is qualitatively different from the strong turbulence 
discussed
above.
This is the {\it weak} turbulence regime.
We do not discuss this type of turbulence here due to its
limited astrophysical applicability and restricted inertial range.
{}For more information, see
\cite{Oug94,Sri94,Ng96,Gol97,Gal00}.
Note that Galtier et al.~\cite{Gal00} obtained $E(k)\sim k_{\perp}^{-2}$
(see also \cite{Gol97}).

%%%%%444444444444444444444444444444444444444444444444444444444444444

\section{Testing and Extending Incompressible Theory}
   Here we focus on recent direct numerical simulations related to
   the anisotropic structure of MHD turbulence.
   A discussion of earlier pioneering numerical simulations of 
   MHD turbulence can be found in \cite{Pou78,Men81}.

%%%%%444444444444444444444444444444444444444444444444444444444444444

%%%%%%%%%%%%%%%%%%%%%%%%%%%%%%%%%%%%%%%%
\begin{figure}[t]
\includegraphics[width=.5\columnwidth]{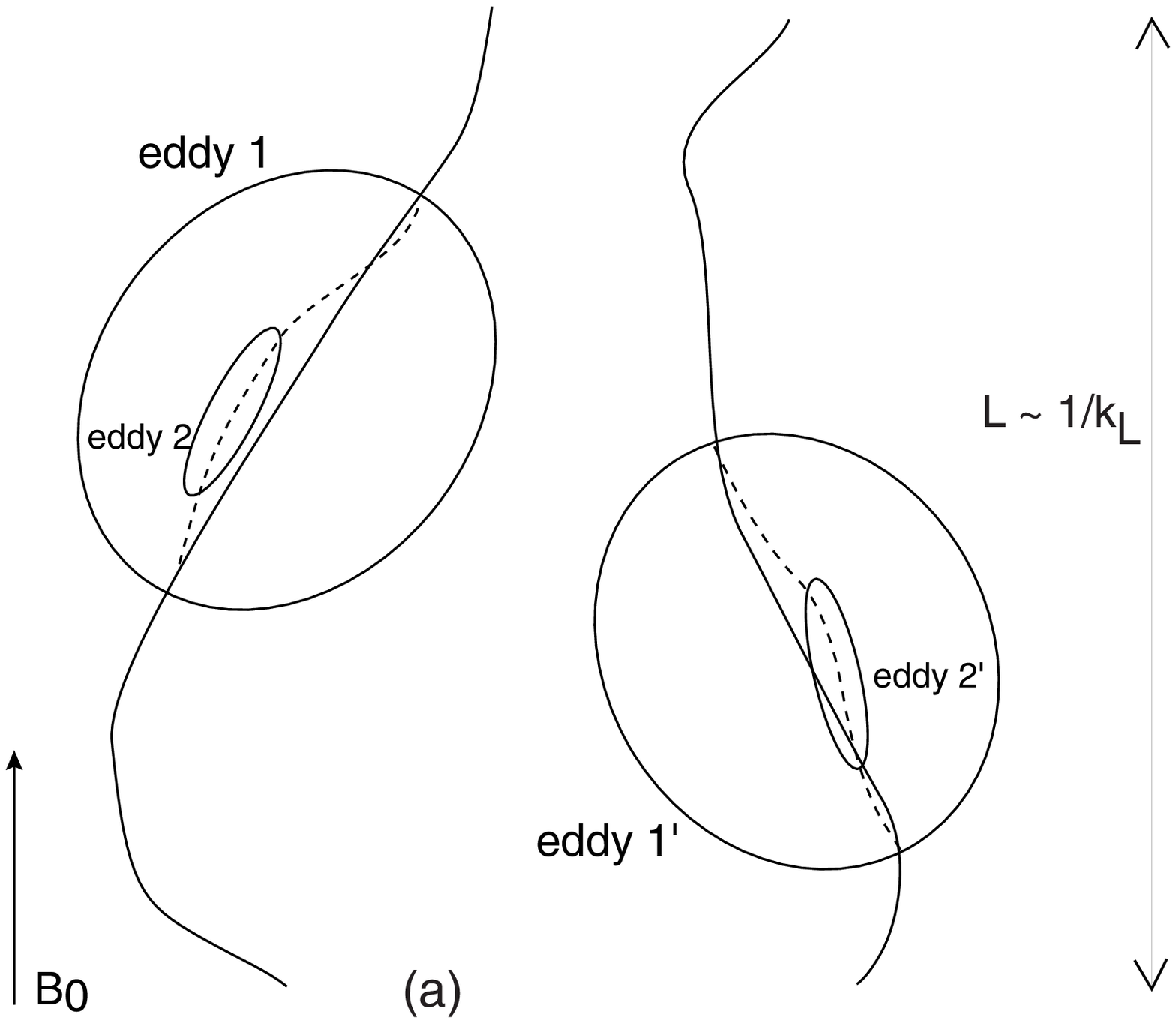}
\hfill
\includegraphics[width=.5\columnwidth]{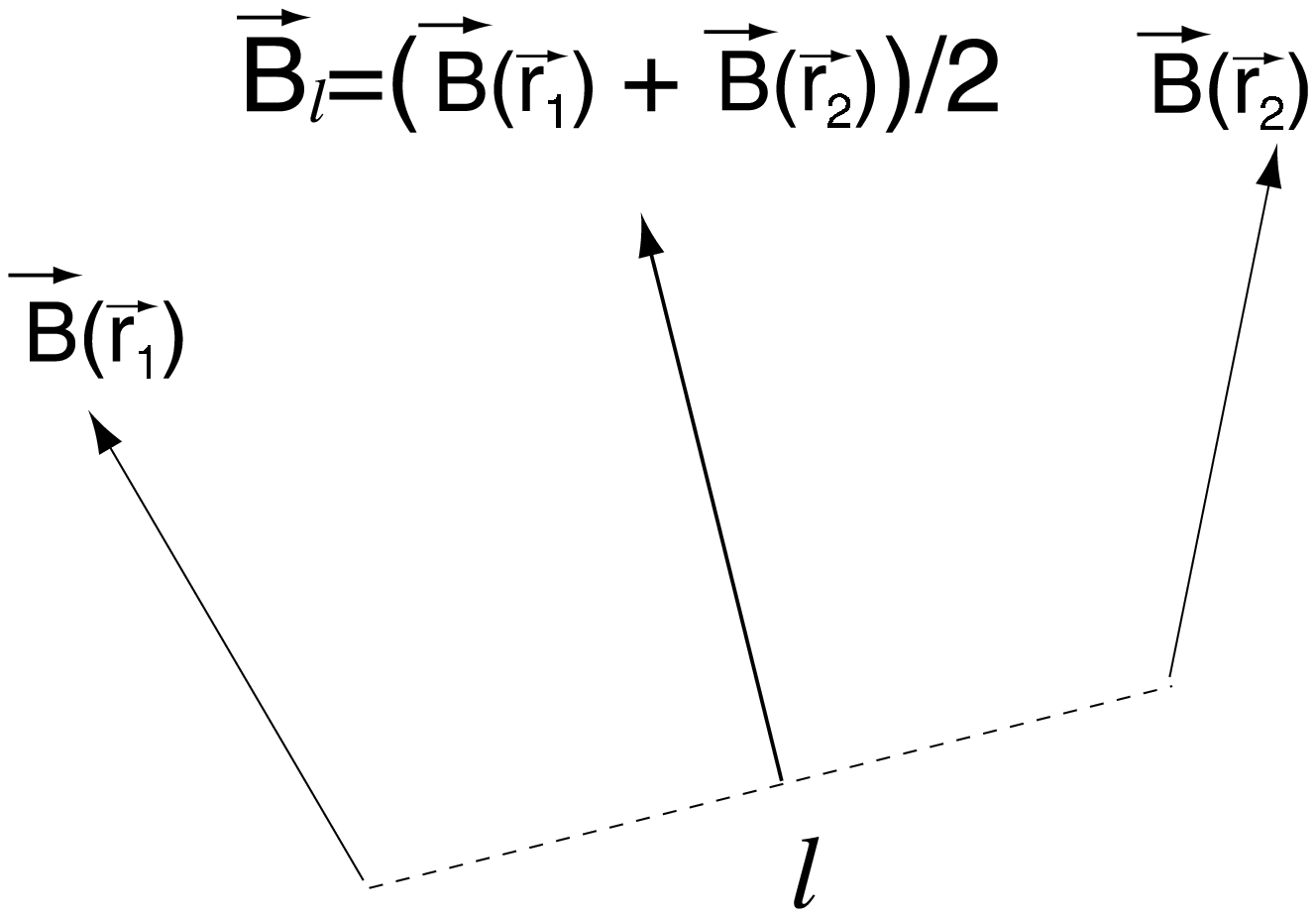}
\caption{
         (a) Eddies and local mean magnetic field.
             Local mean field is the properly averaged magnetic field
             near eddies under consideration.
             Eddies at different locations (e.g.~eddies 1 \& 1$^{\prime}$)
             can have different local mean fields.
             Eddies of different sizes (e.g.~eddies 1 \& 2) can also have
             different local mean fields.
             When we do not take into account the local mean field,
             calculations hardly reveal true eddy structures. 
             {}From \cite{Clv02a}.
         (b) An example of calculating the local mean field.
             The direction of the local mean field ${\bf B}_l$ 
             is obtained by the
             pair-wise average: 
             ${\bf B}_l=({\bf B}({\bf r}_1)+{\bf B}({\bf r}_2))/2$.
       }
\label{local_frame}    
\end{figure}  
%%%%%%%%%%%%%%%%%%%%%%%%%%%%%%%%%%%%%%%
%%%%%%%%%%%%%%%%%%%%%%%%%%%%%%%%%%%%%%% 
\begin{figure}[t]
\begin{center}
\includegraphics{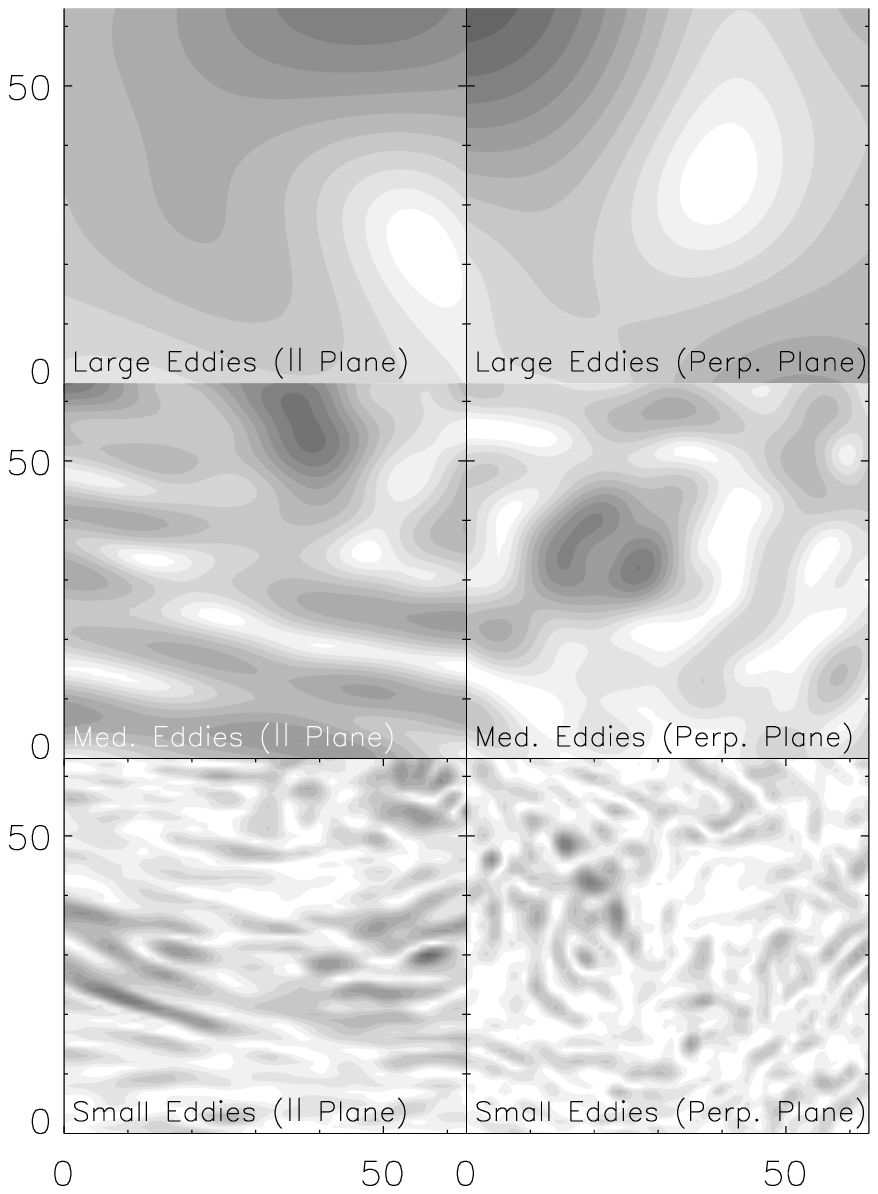}
%\mpicplace{10 cm}{11 cm}
\end{center}
\caption{
      Cross-sections of the data cube.
      ({\it Left panels}) $|{\bf b}_l|$ in a plane ${\|}$ to ${\bf B}_0$.
      ({\it Right panels}) $|{\bf b}_l|$ in a plane ${\perp}$ to ${\bf B}_0$.
      In the left panels, ${\bf B}_0$ is along the horizontal axis.
      {\it Large scale eddies} are obtained from the Fourier components with
      $1\leq k<4$. 
      {\it Medium scale eddies} are obtained from the Fourier components with
      $4\leq k<16$. 
      {\it Small scale eddies} are obtained from the Fourier components with
      $16\leq k<64$. 
      The small scale eddies show a high degree of elongation 
        in the parallel plane.
      However, they do not show a systematic behavior
        in the perpendicular plane.
       }
\label{fig_cross}
\end{figure}  
%%%%%%%%%%%%%%%%%%%%%%%%%%%%%%%%%%%%%%% 

\subsection{Scaling laws}

%\noindent{\bf Scaling of turbulence}\\
Despite its attractiveness, the Goldreich-Sridhar model is a 
conjecture that requires testing.
The first such test was done by Cho \& Vishniac \cite{Cv00a} who
used an incompressible pseudo-spectral MHD code with $256^3$ grid points.
In their simulations, they used $V_A=B_0/\sqrt{4\pi \rho}\sim V$.
Their results for eddy shapes  are shown in Figure \ref{fig_cv00}a, which 
shows a reasonable agreement with 
the predicted scale-dependent anisotropy of the turbulence 
(eq.~(\ref{twothirds})) in the inertial range 
(i.e.~the scales between the energy
injection scale and the dissipation scale).
They also obtained $E(k)\sim k^{-5/3}$ in the inertial range.
Yet although the velocity does show the
expected scaling, the magnetic field scaling is a bit more uncertain.

A subsequent numerical study by Maron \& Goldreich \cite{Mar01} performed with
a different code and in a different physical regime, namely, 
for $V_A\gg V$,
also supported the GS95 model and clarified the role of pseudo-Alfv\'en
and Alfv\'en modes. In particular, they confirmed that the 
pseudo-Alfv\'en modes
are passively carried down the cascade through interactions with the Alfv\'en 
modes.  They also showed that passive scalars adopt the same power
spectrum as the velocity and magnetic field fluctuations.
In addition, they addressed several issues about the imbalanced cascade.
Overall, they obtained results reasonably consistent with 
the GS95 model, but their energy spectra
scale slightly differently: $E(k)\propto k^{-3/2}$.
They attributed this result to intermittency.

 M\"{u}ller \& Biskamp \cite{Mul00} studied MHD turbulence numerically in the
regime of $V \gg V_A$ and
obtained a Kolmogorov spectrum: $E(k)\sim k^{-5/3}$.
They numerically studied the scaling exponents and obtained
 $\zeta_2\approx 0.7$ and $\zeta_3\approx 1$ 
(see the definition of the scaling exponents in the next
subsection). 
 The value of  $\zeta_2$ is consistent with the energy spectrum.

Other related recent numerical simulations include   
     Matthaeus et al.~\cite{Mat98}
    and Milano et al.~\cite{Mil01}.
Matthaeus et al.~\cite{Mat98} showed that the anisotropy of low frequency
MHD turbulence scales linearly with the ratio of
perturbed and total magnetic field strength $b/B$ 
($=b/(b^2+B_0^2)^{1/2}$).\footnote{
In fact we can derive the GS95 scaling using this result.
Although
their analysis was based on comparing the strength of a uniform
background field and the magnetic perturbations on all scales, 
we can reinterpret this result by assuming that the strength
of random magnetic field at a scale $l$ is $b_l$, and that the
background field is the sum of all contributions from larger scales.
Then Matthaeus et al.'s result becomes a prediction that the 
anisotropy ($k_{\parallel}/k_{\perp}$) is proportional to
($b_l/B$).
We can take the total magnetic field strength $B\sim$ constant as long
as the background field is stronger than the perturbations on all scales.
Since $b_l\sim (kE(k))^{1/2} \sim k_{\perp}^{-1/3}$, we obtain
an anisotropy ($k_{\parallel}/k_{\perp}$) proportional to $k_{\perp}^{-1/3}$,
and $k_{\parallel}\propto k_{\perp}^{2/3}$.
In this interpretation, smaller eddies are more elongated because
they have a smaller $b_l/B$ ratio.
}

%%%%%%%%%%%%%%%%%%%%%%%%%%%%%%%%%%
\begin{figure}[t]
%\begin{center}
\includegraphics[width=.5\columnwidth]{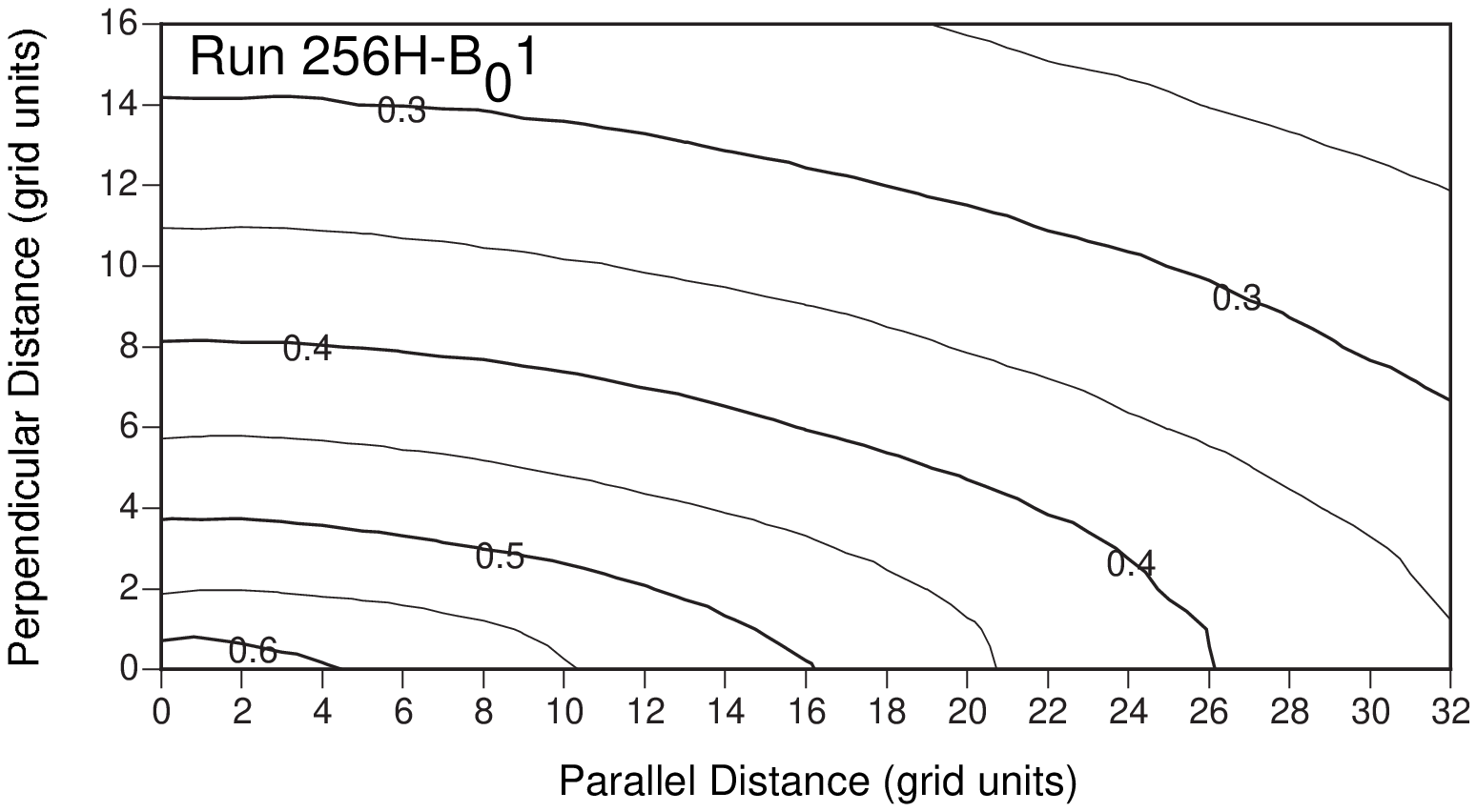}
\hfill
\includegraphics[width=.5\columnwidth]{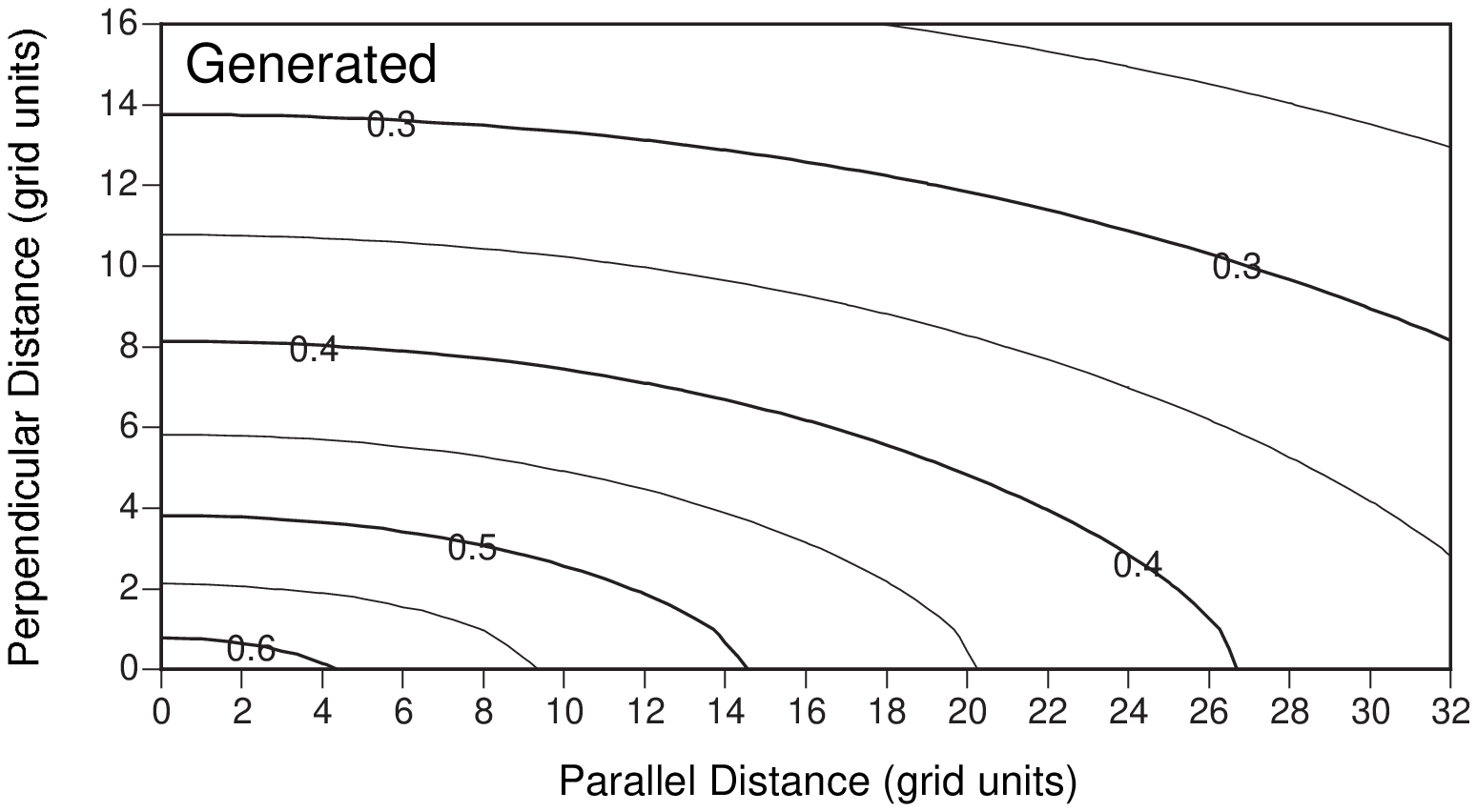}
%\end{center}
%%%\plottwo{cho_laz_yan_1a.eps}{cho_laz_yan_1b.eps}
\caption{
         (a) Velocity correlation function (VCF) from a simulation.
             The contours represent shape of different size eddies.
             The smaller contours (or, eddies) are more elongated.
         (b) VCF generated from equation (\ref{tensor}). 
             {}From \cite{Clv02a}.
       }
\label{fig_conto}   
\end{figure}  
\hfill 

%%%%%%%%%%%%%%%%%%%%%%%%%%%%%%%%%%%
%%%%%%%%%%%%%%%%%%%%%%%%%%%%%%%%%%%
\begin{figure}
\includegraphics[width=.5\columnwidth]{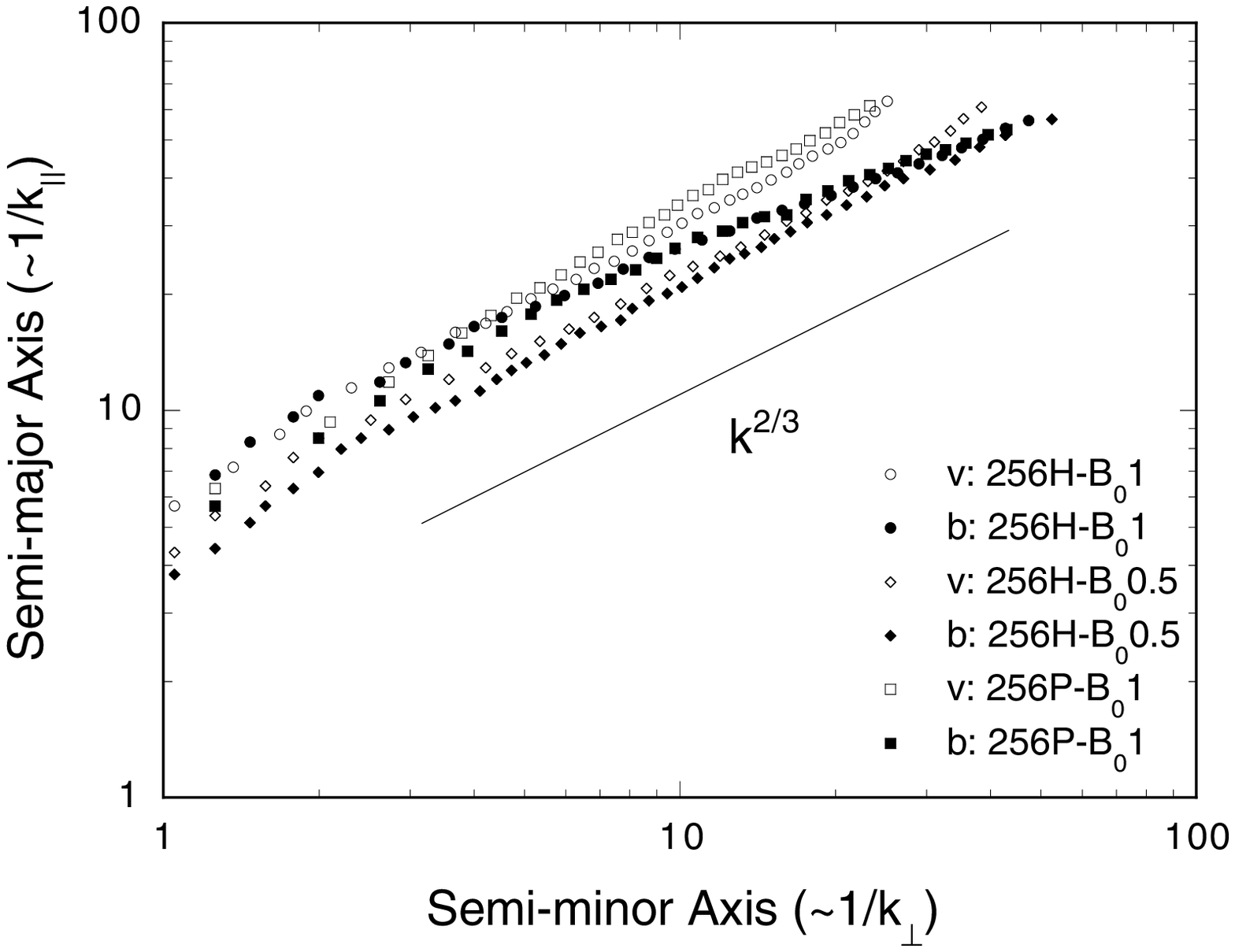}
\hfill
\includegraphics[width=.415\columnwidth]{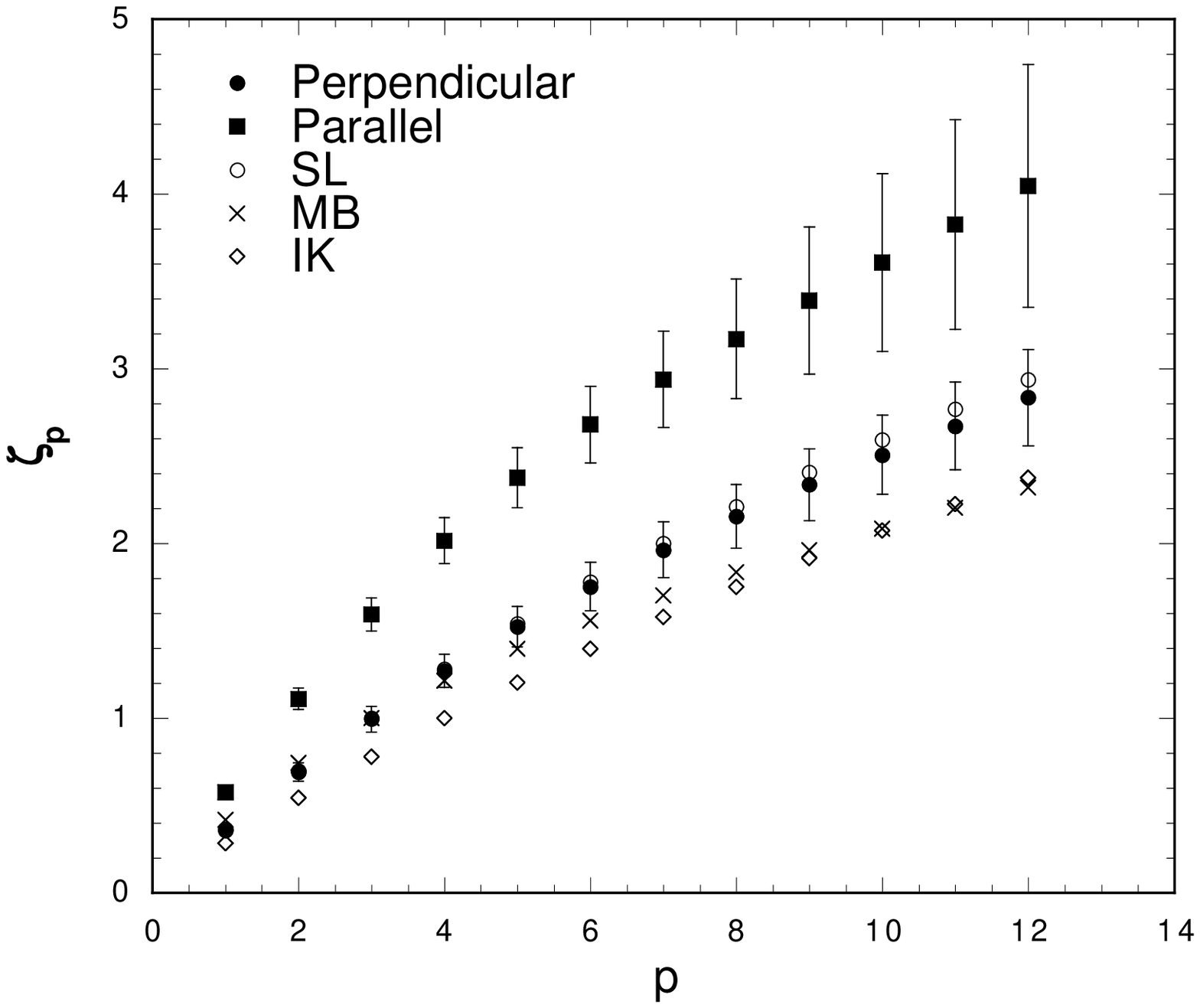}
\caption{
         (a) Semi-major axis and semi-minor axis of contours of
             the VCF shown in Figure \ref{fig_conto}a. 
             The results support $k_{\|}\propto k_{\perp}^{2/3}$.
             {}From \cite{Cv00a}.
         (b) Scaling exponents. From \cite{Clv02a}.
       }
\label{fig_cv00} 
\end{figure}  
%%%%%%%%%%%%%%%%%%%%%%%%%%%%%%%%%%%

All in all, numerical simulations so far have been largely,
but not perfectly, consistent with the GS95
theory, e.g. the Kolmogorov-type scaling
and the scale-dependent anisotropy  
($k_{\|}\propto k_{\perp}^{2/3}$),
and helped to extend it.
An important point, which was not included in some of the earlier
work, is that the 
scale dependent anisotropy can be measured only in a
local coordinate frame which is aligned with the {\it locally averaged}
magnetic field direction \cite{Cv00a}.  
The necessity of using a local frame is due to the fact that
eddies are aligned along the local mean magnetic fields, rather than
the global mean field ${\bf B}_0$.  Since smaller scale eddies
are weaker, and more anisotropic, measurements of eddy shape
based on a global coordinate system are always dominated by the
largest eddies in the simulation.
{}Figure \ref{local_frame} illustrates the concept of
a local frame and one way to identify it.
{}Further research in Cho, Lazarian,
\& Vishniac \cite{Clv02a} (2002a: hereinafter CLV02a) showed that in the 
local system of reference the
mixing motions perpendicular to the magnetic field have statistics
identical to hydrodynamic turbulence (cf. M\"{u}ller \& Biskamp \cite{Mul00}).

{}Fig.~\ref{fig_cross} shows the shapes of eddies of different sizes.
Left panels show an increased anisotropy 
as we move from the top (large eddies) to the bottom (small eddies).
The horizontal axes of the left panels are parallel to ${\bf B_0}$.
Structures in the perpendicular plane do not show a systematic 
elongation. 

{}Fig.~\ref{fig_conto}a  and  Fig.~\ref{fig_cv00}a 
quantify some of these results.  The contours of the correlation
function obtained in
\cite{Cv00a} are shown in
{}Fig.~\ref{fig_conto}a and are 
 consistent with the predictions of the GS95 model.
{}Fig.~\ref{fig_cv00}a shows that the semi-major axis ($1/k_{\|}$)
is proportional to the 2/3 power of the semi-minor axis ($1/k_{\perp}$),
implying that $k_{\|}\propto k_{\perp}^{2/3}$.
While the one dimensional energy spectrum follows Kolmogorov spectrum,
$E(k)\propto k^{-5/3}$,
CLV02a showed that the
3D energy spectrum can be fit by 
\begin{equation}
P(k_{\perp}, k_{\|})=(B_0/L^{1/3}) k_{\perp}^{-10/3}\exp\left(-L^{1/3}
     \frac{ k_{\|} }{ k_{\perp}^{2/3} } \right),
\label{tensor}
\end{equation} 
where $B_0$ is the strength of the mean field and $L$ is the scale of
energy injection.
The velocity correlation from the 3D spectrum provides
an excellent fit to the numerical data (Fig.~\ref{fig_conto}b). 
This allows practical applications illustrated in \S5.

%In conclusion, it appears that the simplest features of the GS95 model,
%i.e. the predicted scale dependent anisotropy and the Kolmogorov-like
%power spectrum, are consistent with the available simulations.  However,
%the increased intermittency and shallower power spectrum reported by
%Maron and Goldreich are not fully understood.  The implications of
%the GS95 model for astrophysics have not yet been fully explored.

\vspace{0.3cm}
\subsection{Intermittency}
%\noindent{\bf Intermittency}\\
Intermittency refers to the non-uniform distribution
of dissipative structures.
Intermittency has an important dynamical consequence:
it affects the energy spectrum.
Highly intermittent turbulent structures
were invoked by Falgarone et al.~\cite{Fal95} and Joulain et al.~\cite{JFD98} 
as the
primary location of endothermic interstellar chemical reactions. 

Maron \& Goldreich \cite{Mar01} studied the intermittency of dissipation 
structures
in MHD turbulence using the fourth order moments of the Elsasser fields and
the gradients of the fields.  Their simulations show strong
intermittent structures.
CLV02a used a different, but complementary,  method to study intermittency, 
based on the higher order longitudinal structure functions.
They found that by this measure the intermittency of velocity field
in MHD turbulence across local
magnetic field lines is as strong as, but not stronger than,
hydrodynamic turbulence.

In fully developed hydrodynamic turbulence, the (longitudinal)
velocity structure functions
$S_p=< ( [ {\bf v}({\bf x}+ {\bf r}) -
      {\bf v}({\bf x})]\cdot \hat{\bf r} )^p>
\equiv < \delta v_L^p({\bf r}) >$ are
expected to scale as $r^{\zeta_p}$.
{}For example, the classical Kolmogorov phenomenology (K41) predicts
$\zeta_p =p/3$.
The (exact) result for p=3 is the well-known 4/5-relation:
$ < \delta v_L^3({\bf r}) > =-(4/5)\epsilon r$, where $\epsilon$
is the energy injection rate (or, energy dissipation rate) 
(see e.g.~\cite{Fri95}).
On the other hand, considering intermittency, 
She \& Leveque (\cite{She94}; hereinafter S-L) 
proposed a different
scaling relation: $\zeta_p^{SL}=p/9+2[1-(2/3)^{p/3}]$.
Note that the She-Leveque model also implies $\zeta_3 =1$.

     So far in MHD turbulence, to the best of our knowledge, 
there is no rigorous intermittency theory which takes into account
     scale-dependent anisotropy. 
Politano \& Pouquet \cite{Pol95} have developed an MHD version of the
She-Leveque model:
\begin{equation}
\zeta_p^{PP}=\frac{p}{g}(1-x)+C \left(1-(1-x/C)^{p/g}\right),
\label{mhdint}
\end{equation}
where $C$ is the co-dimension of the dissipative structure,
$g$ is related to the scaling $v_l \sim l^{1/g}$,
and $x$ can be interpreted as the exponent of the cascade time
$t_{cas}\propto l^{x}$.
(In fact,
$g$ is related to the scaling of Elsasser variable 
z=$v\pm b$: $z_l\sim l^{1/g}$.)
In the framework of the IK theory, where $g=4$, $x=1/2$, and $C=1$ when
the dissipation structures are sheet-like, their model of intermittency becomes
$\zeta_p^{IK}=p/8+1-(1/2)^{p/4}$.  On the other hand,
M\"uller \& Biskamp \cite{Mul00} performed numerical simulations on decaying
isotropic MHD turbulence and obtained
Kolmogorov-like scaling ($E(k)\sim k^{-5/3}$ and $t\sim l^{2/3}$) and
sheet-like dissipation structures, which 
implies $g=3$, $C=1$, and $x=2/3$.
{}From equation (\ref{mhdint}), they proposed that
\begin{equation}
\zeta_p^{MB}=p/9+1-(1/3)^{p/3}.
\end{equation}

The intermittency results from \cite{Clv02a}
are shown in Fig.~\ref{fig_cv00}b.
The filled circles represent 
the scaling exponents of longitudinal
velocity structure functions in directions {\it perpendicular} to
the local mean magnetic field.
It is surprising that the scaling exponents are so close the
original (i.e. hydrodynamic) S-L model.
This raises an interesting  question.
In the simulations of CLV02a,  $t_{cas}\propto l^{2/3}$ and
$E(k) \propto k^{-5/3}$ scaling is observed.
It is observed that MHD turbulence has sheet-like dissipation structures
\cite{Pol95s}.
Therefore, the parameters for CLV02a simulations should be
the same as those of M\"uller \& Biskamp's (i.e. $g=3$, $C=1$, and $x=2/3$)
rather than suggesting $C=2$.
We believe this difference stems from the different
simulation settings (M\"uller \& Biskamp's turbulence is isotropic and 
CLV02a's 
is anisotropic) 
and the way the turbulence is analyzed (global versus local frame).  
In fact, we expect the the small scale behavior of MHD turbulence
should not depend on whether the largest scale fields are uniform or
have the same scale of organization as the largest turbulent eddies.  
Nevertheless,
given the limited dynamical range available in these simulations, it would
not be surprising if the scale of the magnetic field has an impact
on the intermittency statistics.
   It is not clear how scale-dependent anisotropy changes the intermittency
   model
   in equation (\ref{mhdint}) and
   we will not discuss this issue further.
   Instead, we simply stress that a striking 
   similarity exists between ordinary hydrodynamic turbulence and MHD turbulence
   in perpendicular directions, which further supports
   the picture of the GS95 turbulence.\footnote{
       MG01 attributed the deviation of their spectrum
       from the Kolmogorov-type to the intermittency present in their simulation.}

\begin{figure}[t]
\includegraphics[width=.5\columnwidth]{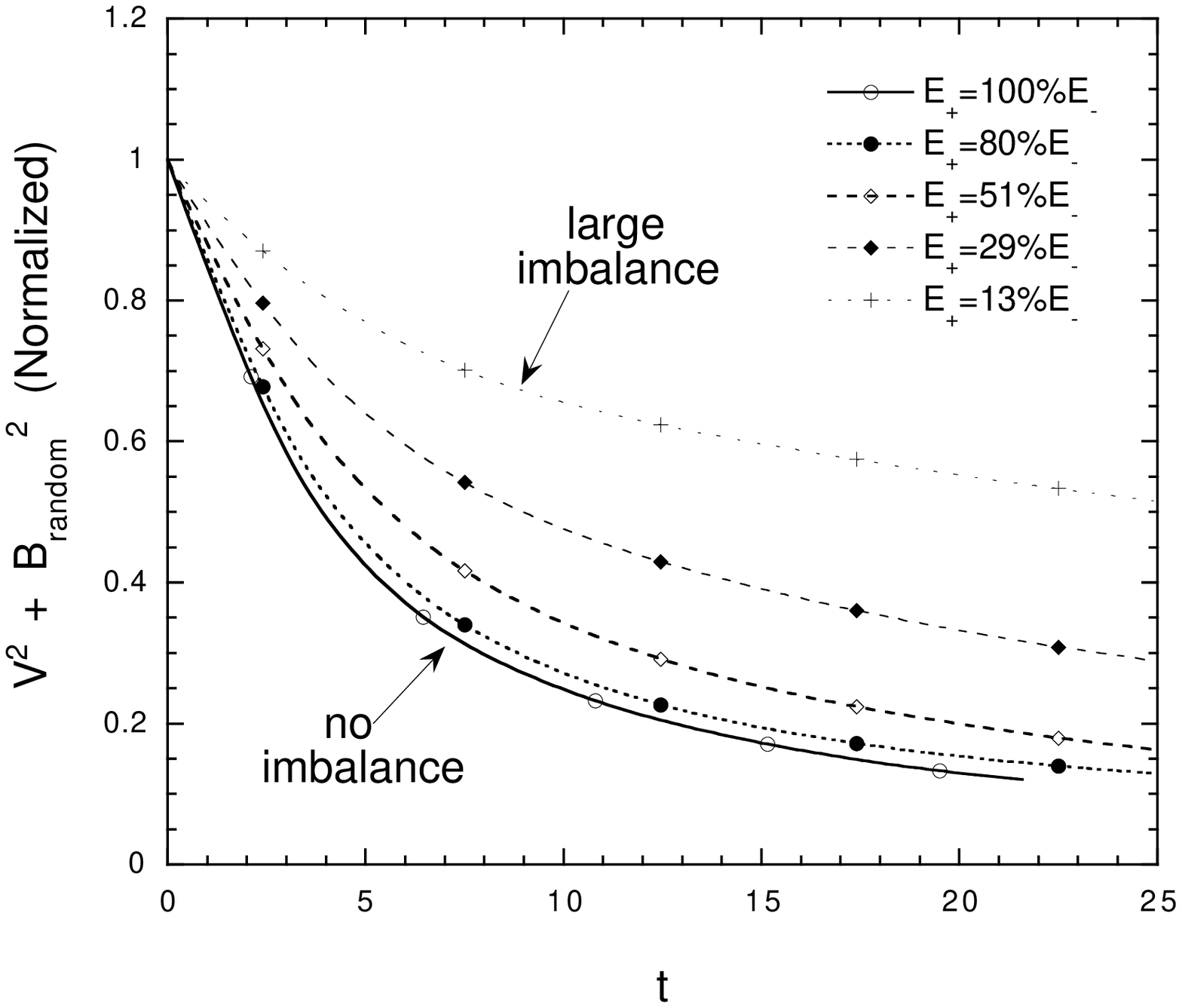}
\hfill
\includegraphics[width=.5\columnwidth]{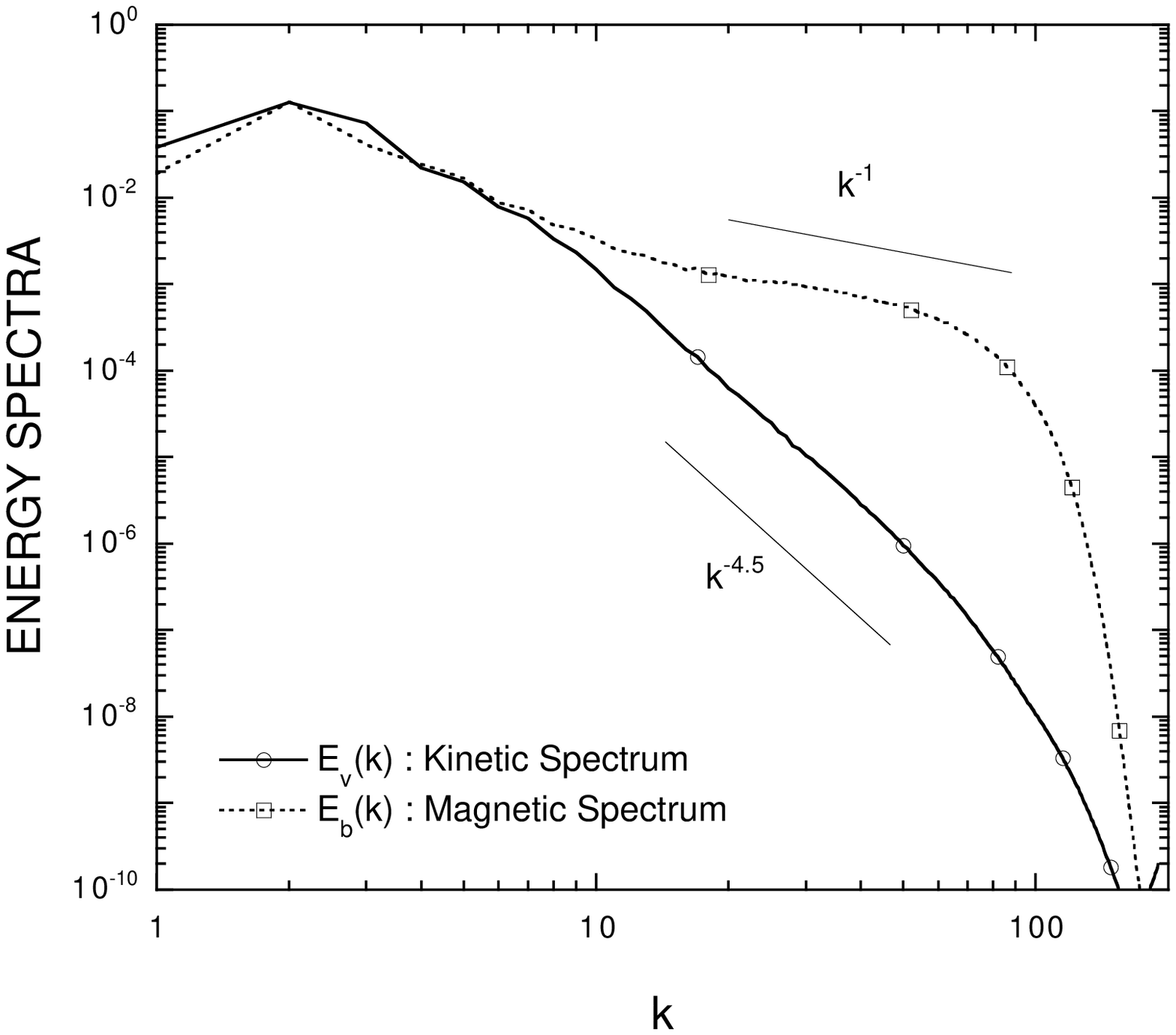}
\caption{
         (a) {\it (Left)} Imbalanced Decay. When imbalance is large,
             turbulence decays slow. From CLV02a.
         (b) {\it (Right)} Viscous damped regime. A new inertial range
             emerges below the viscous cut-off at $k\sim 7$.
             From Cho, Lazarian, \& Vishniac (2002b). 
       }
\label{fig_imb}
\end{figure}

In Figure \ref{fig_imb}a, we also plot the scaling exponents (represented by filled 
squares) of 
longitudinal velocity structure functions {\it along} directions of
the local mean magnetic field.
Although we show only the exponents of longitudinal structure functions,
those of transverse structure functions follow a similar scaling law.
Evidently intermittency along the local mean magnetic field directions
is completely different from the scaling predicted by previous 
(isotropic) models.
Roughly speaking, the scaling exponents along the directions of local 
magnetic field are 1.5 times larger than those of perpendicular
directions. This has an obvious similarity to the scaling of 
eddy shapes.

The second order exponent $\zeta_2$ is related to the the 1-D energy spectra:
$E(k_{\perp})\propto k_{\perp}^{-(1+\zeta_2)}$.
      Previous 2-D driven MHD calculations for $B_0=0$ by 
      Politano, Pouquet, \& 
     Carbone \cite{Pol95c} also found  $\zeta_2\sim 0.7$.
     However,  Biskamp, \& Schwarz \cite{Bis01} obtained $\zeta_2\sim 0.5$
     from decaying 2-D MHD calculations with $B_0=0$.
The result of CLV02a suggests 
that $\zeta_2$ is closer to $2/3$, rather than to $1/2$.
(It is not clear whether or not the scaling exponents follow the original S-L
model exactly. At the same time, our calculation shows that the original
S-L model can be a good approximation for our scaling exponents.
The S-L model predicts that $\zeta_2 \sim 0.696$.)
This is equivalent to our earlier claim that 
our result supports the scaling law
$E(k_{\perp}) \propto k_{\perp}^{-5/3}$ at least for velocity.
For the parallel directions, the results support 
$E(k_{\|}) \propto k_{\|}^{-2}$ although the uncertainty is large.

%%%%%%%%55555555555555555555555555555555555555555555555555555555555555

\section{Damping of Turbulence}

%%%%%%%%55555555555555555555555555555555555555555555555555555555555555

%%%%%%%%%%%%%%%%%%%%%%%%%%%%%%%%%%%%%
%%%%%%%%%%%%%%%%%%%%%%%%%%%%%%%%%%%%%
\subsection{Imbalanced cascade}
Turbulence plays a critical role in molecular cloud support and star 
formation and the issue of the time scale of turbulent decay is vital for
understanding these processes.
If MHD turbulence decays quickly, then serious
problems face the researchers attempting to explain important observational 
facts, e.g.~turbulent  motions seen within molecular clouds without
star formation \cite{Mye99} and rates of star formation \cite{Mck99}.
Earlier studies attributed the rapid decay of turbulence to compressibility
effects \cite{Mac99}. GS95 predicts and numerical simulations, 
e.g.~CLV02a,  
confirm that MHD turbulence decays rapidly even
in the incompressible limit. This can be understood if 
mixing motions perpendicular to magnetic field lines are considered. 
As we discussed earlier, such eddies, as in
hydrodynamic turbulence, decay in one eddy turnover time.

Below we consider the effect of 
imbalance \cite{MGM83,TMM86,GMM88,HGPM95,Mar01}
on the turbulence decay time scale. Duality of the MHD turbulence
means that the
turbulence can be described by colliding wave packets.
`Imbalance' means that the flux of wave packets traveling in one
direction is
significantly larger than those traveling in the other
direction.
In the ISM, many energy sources are localized both in space and time.
For example, in terms of energy injection, stellar outflows are essentially 
point energy sources.
With these localized energy sources, it is natural 
that interstellar turbulence be typically  imbalanced.

Here we show results of the
CLV02a study that demonstrate that imbalance does extend
the lifetime of MHD turbulence (Fig.~\ref{fig_imb}a).
We used a run on a grid of $144^3$ to investigate the decay time scale.
For initial conditions we took a data cube from a driven turbulence simulation.
The initial data cube contains both upward (denoted as $+$) and downward moving
waves (denoted as $-$).
To adjust the degree of initial imbalance, we either increased or decreased 
the energy of the upward moving components
components and, by turning off the forcing terms, let the turbulence decay.
Note that the initial energy is normalized to 1.
The y-axis is the normalized total (=up $+$ down) energy.

The dependence of the turbulence decay time on the degree of imbalance
is an important finding. To what degree the results persist in 
the presence of compressibility is the subject of our current research.
It is obvious that results of CLV02a are applicable to incompressible, namely,
Alfv\'en motions.\footnote{
In the long run, the imbalance will be defeated by the 
parametric instability, which develops through formation of
density inhomogeneities within the beam of Alfv\'{e}n waves 
\cite{Der78,Gol78,JH93,Del01}.
However, this instability takes many wave periods to be established.
A similar argument can be applied when we consider completely
imbalanced cascade. That is, even in the completely imbalanced cascade
decay of energy can occur due to non-linear steepening if waves.
But, this will be very slow for Alfv\'{e}n waves.
}
 We show in \S\ref{sec_coupling} that the Alfv\'en motions
are essentially decoupled from compressible
modes.
As the result we expect that the turbulence decay time may
be substantially longer than one eddy turnover time
provided that the turbulence is imbalanced.

\subsection{Ion-neutral damping: a new regime of turbulence} \label{sec_vis}

In hydrodynamic turbulence viscosity sets a minimal scale for
motion, with an exponential suppression of motion on smaller
scales.  Below the viscous cutoff the kinetic energy contained in a 
wavenumber band is 
dissipated at that scale, instead of being transferred to smaller scales.
This means the end of the hydrodynamic cascade, but in MHD turbulence
this is not the end of magnetic structure evolution.  For 
viscosity much larger than resistivity,
$\nu\gg\eta$, there will be a broad range of
scales where viscosity is important but resistivity is not.  
On these
scales magnetic field structures will be created 
by the shear from non-damped turbulent motions, which
amounts essentially to the shear from the smallest undamped scales.
The created magnetic structures would evolve through
generating small scale motions.
As a result, we expect
a power-law tail in the energy distribution, rather than an exponential
cutoff.  To our best knowledge, this is a completely new regime
for MHD turbulence.

In partially ionized gas
neutrals produce viscous damping of turbulent motions. 
In the Cold Neutral Medium (see Draine \& Lazarian \cite{Dra99} for a list of
the idealized phases) this produces damping on the scale of a fraction of
a parsec. The magnetic diffusion in those circumstances is
still negligible and exerts an influence only 
at the much smaller scales, $\sim 100km$. 
Therefore, there is a large
range of scales where the physics of the turbulent cascade 
is very different from 
the GS95 picture.

Cho, Lazarian, \& Vishniac \cite{Clv02b} have
explored this regime numerically.
Here we used a grid of $384^3$ and a 
physical viscosity for velocity damping. The kinetic Reynolds number is
around 100. 
With this Reynolds number, viscous damping occurs around $k\sim 7$,
or about $\sim 1/7$ of the width of the computational box.
We minimized magnetic diffusion through the use of
a hyper-diffusion term of order 3. To test for possible
``bottle neck'' effects we also did simulations with normal 
magnetic diffusion and reproduced similar results but only over a reduced
dynamical range.
The bottleneck effect is a common feature in 
numerical hydrodynamic simulations with hyperviscosity.
See \cite{BSC98} for MHD simulations.

In Fig.~\ref{fig_imb}b, we plot energy spectra.
The spectra consist of several parts.
{}First, the peak of the spectra corresponds to the energy injection scale.
Second, for $2<k<7$, kinetic and magnetic spectra follow a similar slope.
This part is more or less a severely truncated inertial range for undamped MHD
turbulence.  
Third, the magnetic and kinetic spectra begin to decouple at $k\sim 
7$.
{}Fourth, after $k\sim20$, a new {\it damped-scale inertial 
range} emerges.
In the new inertial range, magnetic energy spectrum follows 
\begin{equation}
  \mbox{\it New Regime:~~~~~} E_b(k)\propto k^{-1}, 
\end{equation}
implying considerable magnetic structures below the 
viscous damping scale.  The velocity power spectrum steepens in this
regime, but does not fall exponentially.

\begin{figure}[t]
%\mpicplace{10 cm}{6 cm}
\includegraphics[width=.5\columnwidth]{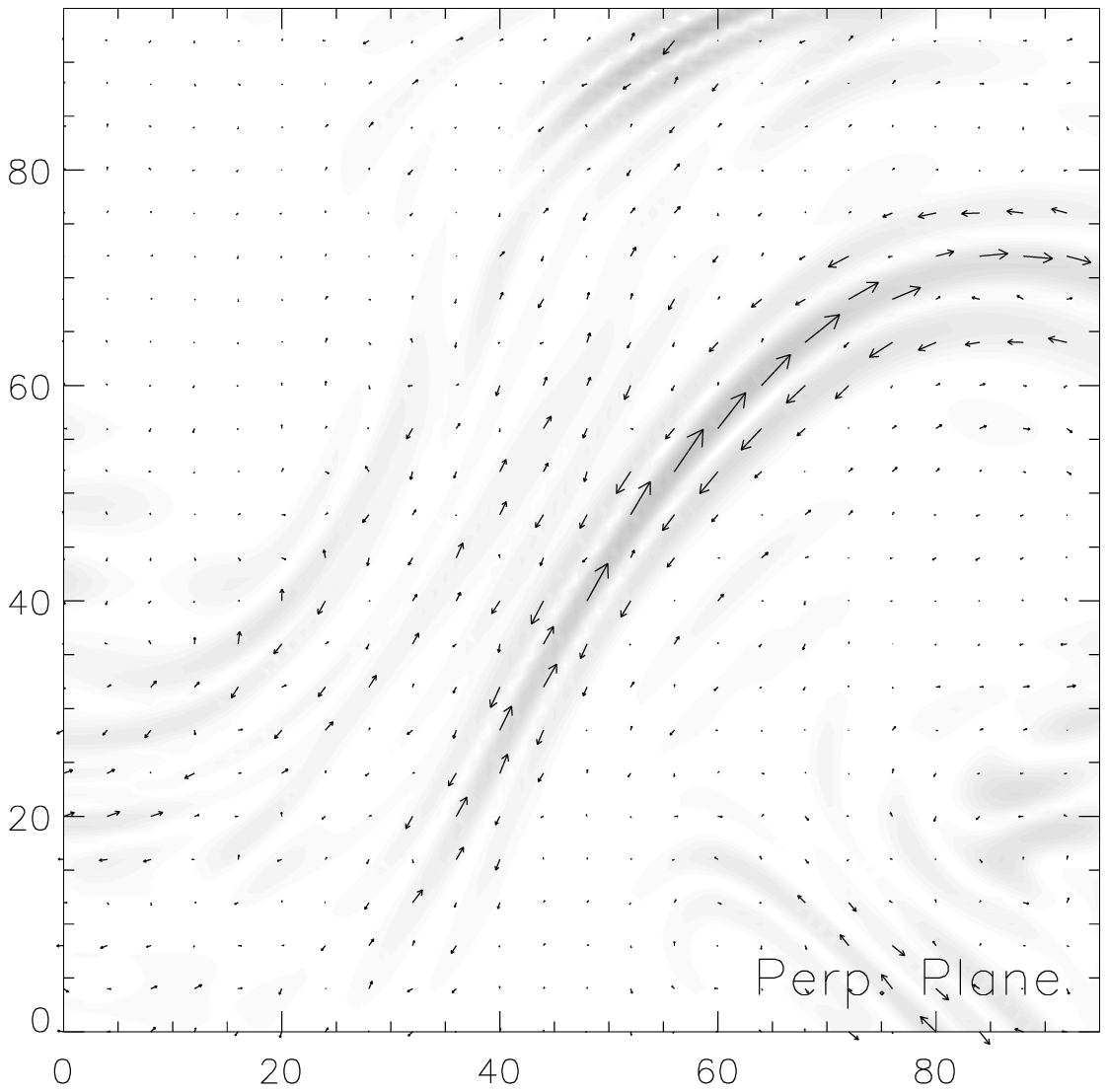}
\hfill
\includegraphics[width=.5\columnwidth]{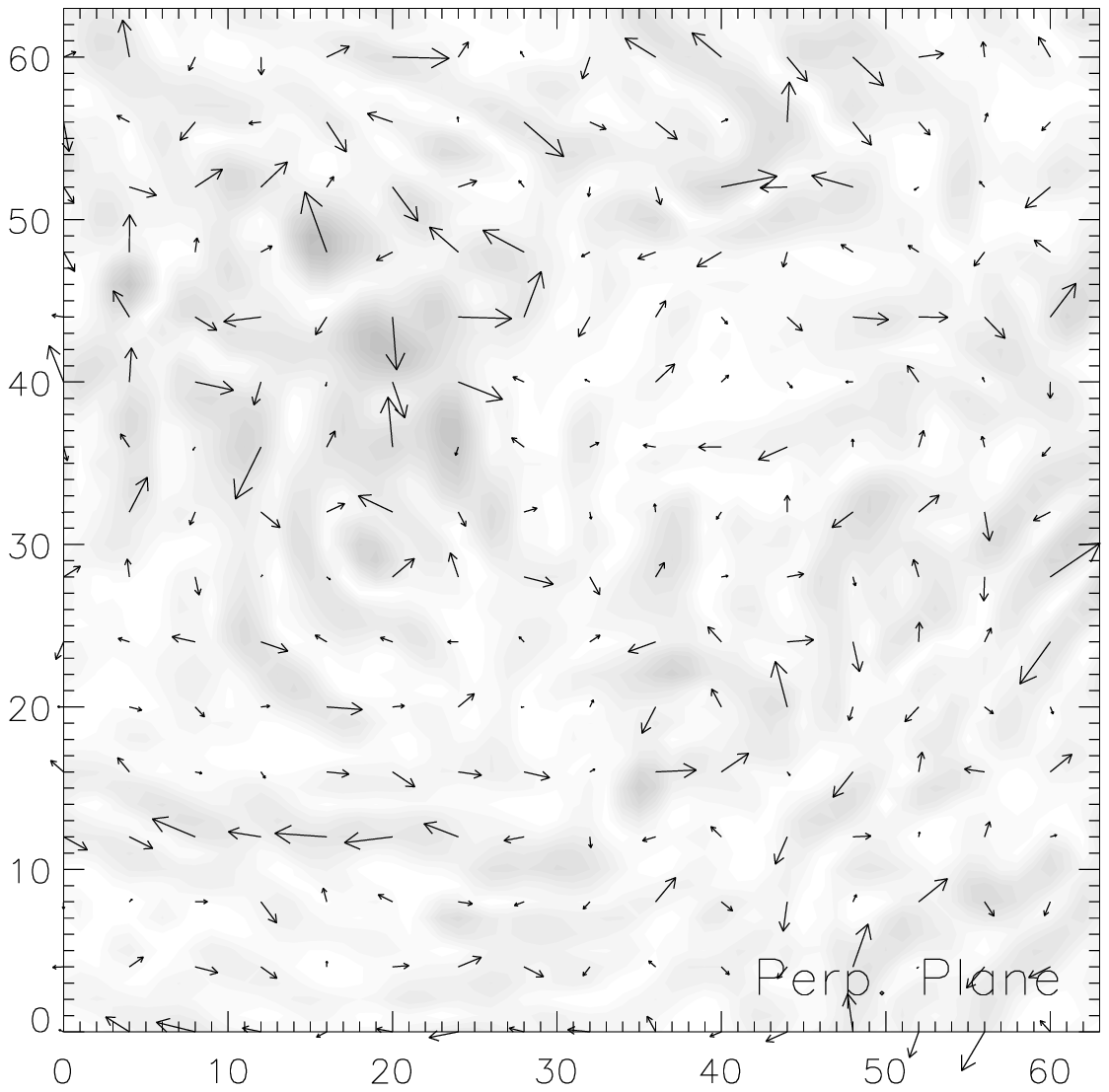}
\caption{
          ({\it Left}) Viscous-damped turbulence.
       Strength of magnetic field in a plane 
       perpendicular to ${\bf B}_0$.
  Arrows are magnetic fields in the plane.
  Only a part of the plane is shown. Note highly intermittent structures.
  {}From \cite{Clv02b}.
      ({\it Right}) 
     Same as {\it Left}, 
     but for ordinary (not viscous-damped) MHD turbulence.
     Structures are less intermittent.
}
\label{fig_vis_ord}
\end{figure}

{}Figure \ref{fig_vis_ord}a shows that the small magnetic structures are highly
intermittent in the viscous-damped regime.  Here we
obtained the small scale magnetic field by eliminating
Fourier modes with $k<20$.
We can see that 
the typical radius of curvature of field lines in the plane
is much larger than the typical perpendicular scale for field reversal.
The typical radius of curvature of field lines corresponds to the
viscous damping scale, indicating that stretched structures are
results of the shearing motions at the viscous damping scale.
There is no preferred direction for these elongated structures.
A similar plot for ordinary ({\it not} viscously damped) MHD turbulence
(Fig. \ref{fig_vis_ord}b)
shows much less intermittent structures.
We remove Fourier modes with $k<20$ also in  Fig.~\ref{fig_vis_ord}b.

%%%%%%%%%%%%%%%%%%%%%%%%%%%%%%%%%%%%%%
%%%%%%%%%%%%%%%%%%%%%%%%%%%%%%%%%%%%%%

A theoretical model for this new regime and its
consequences for stochastic reconnection \cite{LV99} will be found
in an upcoming paper (Lazarian, Vishniac, \& Cho 2002). 
Here we summarize the main points of the model.  We begin
by noting 
that the strong intermittency seen in this regime suggests a new
parameter, $f_l$, which is the volume filling fraction of structures
with scales comparable to $l$.  This in turn implies that we 
need to distinguish between volume averaged means and typical
values of velocity and magnetic field perturbations in fraction of
space where they are concentrated.  We denote the latter with a
`$\hat{~}$', so that
$%%%\begin{equation}
v_l^2=f_l\hat v_l^2
$, %%%\end{equation}
and
$%%%\begin{equation}
b_l^2=f_l\hat b_l^2
$. %%%\end{equation}
This model does not include information about the range of field
strengths or velocities with structure on a scale $l$ in the bulk of the 
volume, aside from assuming that they are sufficiently weak that they
do not contribute to any global averages.

The second new fundamental parameter in this model is the eddy turn over
rate at the damping scale, $k_d$, i.e.
\begin{equation}
\tau_s^{-1}\sim k_d v_d \sim k_d^2\nu_n,
\end{equation}
where $v_d$ is the velocity at $l_d \sim 1/k_d$ and
$\nu_n$ is the viscosity of the plasma due to neutral particles,
which differs from the viscosity of neutral fluid by
the ratio of atomic to total densities.

Since motions on smaller scales are strongly damped, the cascade of magnetic
energy to smaller scales is due to motions on the damping scale.  This
implies that
\begin{equation}
%%%{b_l^2\over\tau_s}\sim \hbox{\ constant},
{b_l^2/\tau_s}\sim \hbox{\ constant},
\label{e1}
\end{equation}
so that $b_l$ is maintained approximately
at the level of magnetic field at the damping scale, $b_d$.

This folding and refolding, perpendicular to the mean field direction,  
has a weak effect on the field line curvature.  (This result is also
seen in the simulations.)  The decrease in the structure length $l$ is
due to an increase in the magnetic field gradient perpendicular to the
mean field direction.  The resulting magnetic pressure gradients are
balanced by plasma pressure gradients.  
Thus on the scales below the viscous cutoff,
the tension forces are balanced
by viscous drag, i.e.
\begin{equation}
{\rho \nu_n\over l^2} \hat v_l\sim \max[{k_d \hat b_l,k_{\|}B_0}] \hat b_l.
\label{e2}
\end{equation}

Finally, this dynamic equilibrium can be maintained only if the small scale
motions are strong enough to counteract the shear, $\tau_s^{-1}$.  In
other words,
$%%%\begin{equation}
%%%{\hat v_l\over l}\sim \tau_s^{-1}.
{\hat v_l/l} \sim \tau_s^{-1}
%%%\label{e3}
$. %%%\end{equation}

Combining these results we see that
$%%%\begin{equation}
f_l\sim k_dl
$, %%%\end{equation}
$%%%\begin{equation}
\hat b_l \sim b_d (k_dl)^{-1/2}
$, %%%\end{equation}
and
$%%%\begin{equation}
v_l\propto l^{3/2}
$. %%%\end{equation}
These scaling laws are at least qualitatively consistent with the
simulation results, although the velocity power spectrum may be 
slightly steeper than the model prediction.

Unfortunately, a realistic treatment of the ISM requires an explicit
recognition of the two fluid nature of the partially ionized plasma,
rather than simply representing neutral drag with an effective viscosity.
Here we are beyond the reach of available simulations, and need to
rely on an extension of the scaling arguments given above.  

First, at sufficiently small values of $l$, the ambipolar diffusion
time 
~($\sim k^2 t_{in} (b^2_l/\rho_{tot}) (\rho_n/\rho_i), where
~t_{in}=$
ion-neutral collision time)
will become less than $\tau_s$, and the magnetic pressure
gradients will be supported entirely by the ionized particle pressure.
This means that only a fraction, $\sim \rho_i/\rho_{tot}$, of the 
energy will continue the cascade to smaller scales.  

Next, for some ion-neutral collision time, $t_{in}$, there will be
some decoupling scale, $l_c$, where
\begin{equation}
\rho \nu_n/ l_c^2 \sim \rho_i/ t_{in}.
\label{e4}
\end{equation}
At smaller scales the ions will be dragged through a more or less uniform
neutral background.  The argument above needs to be modified by replacing
the viscous drag coefficient in equation (\ref{e2}) with the right hand
side of equation (\ref{e4}).  This gives 
\begin{equation}
f_l\propto l^{-1},
\end{equation}
\begin{equation}
\hat b_l \propto l^{1/2}, \mbox{~~~and~~~} v_l\propto l^{1/2}.
\end{equation}

Finally, at some sufficiently small scale the filling factor will rise 
to unity,
and the gradients in the magnetic field will become strong enough that
neutral drag can be ignored.  These conditions are satisfied simultaneously
when
\begin{equation}
k_d l\sim t_{in}/ \tau_s.
\end{equation}
Below this scale we expect to see a resumption of the turbulent cascade,
now involving only the ionized component of the plasma, down to scales
where plasma resistivity and viscosity finally dissipate the remaining energy.
Since the longest cascade time for this regime is $t_{in}\ll \tau_s$, we
expect this small scale turbulence to be intermittent, with a duty cycle
$\sim t_{in}/\tau_s$.

All the consequences of the new regime of the
MHD turbulence have not yet been appreciated, but we expect that it
will have a substantial impact on our understanding of the interstellar
physics.  Moreover, the treatment given above actually applies only
when $\rho_i/\rho$ is not very small.  Otherwise the decoupling scale
can be larger than the viscous damping scale.

%%%%%%%%66666666666666666666666666666666666666666666666666

\section{Compressible Turbulence}

%%%%%%%%66666666666666666666666666666666666666666666666666
%%%%%%%%%TTTTTTTTTTTTTTTTTTTTTTTTTTTTTT
\begin{table}[b!]
\caption{Notations for compressible turbulence}
\begin{center}
\setlength\tabcolsep{5pt}
\begin{tabular}{ll}
\hline\noalign{\smallskip}
 Notation   &  Meaning  \\
\noalign{\smallskip}
\hline
\noalign{\smallskip}
  a, $c_s$, $c_f$, $V_A$    &    sound, slow, fast, and Alfv\'en speed \\
  $\delta V$, $(\delta V)_s$, $(\delta V)_f$, $(\delta V)_A$
                            &  random (rms) velocity  \\
                            & Previously we used $V$ for the rms velocity \\
  $v_l$, $(v_l)_s$, $(v_l)_f$, $(v_l)_A$  &  velocity at scale $l$    \\
  ${\bf v}_{\bf k}$, $({\bf v}_{\bf k})_s$, $({\bf v}_{\bf k})_f$, 
  $({\bf v}_{\bf k})_A$       
                     &               velocity vector at wavevector ${\bf k}$ \\
  $\hat{\bf B}_0$ (=$\hat{\bf k}_{\|}$), $\hat{\bf k}_{\perp}$,
  $\hat{\bf k}$, $\hat{\bf \theta}$, ...      & unit vectors       \\
  ${\bf \xi}_s$, ${\bf \xi}_f$      & displacement vectors       \\
\hline
\end{tabular}
\end{center}
\label{cho_table1} 
\end{table}
%%%%%%%%%TTTTTTTTTTTTTTTTTTTTTTTTTTTTTT
For the rest of the review, we consider MHD turbulence
of a single conducting fluid.
While the GS95 model describes incompressible
MHD turbulence well,
no universally accepted theory exists 
for compressible MHD turbulence despite various
attempts (e.g., \cite{Hig84}).
Earlier numerical simulations of compressible MHD turbulence covered
a broad range of astrophysical problems, such as the decay of turbulence
(e.g.~\cite{Mac98,Sto98}) 
or
turbulent modeling of the ISM 
(see recent review \cite{Vaz02}; 
see also \cite{Pas88,Vaz95,Pas95,Vaz96} for earlier 2D simulations
and \cite{Ost99,Ost01,Pad01,Kle01,Bol02} for recent 3D simulations).
%There are also some attempts to find scaling exponents of
%structure functions (\cite{Bol01}).
In what follows, we concentrate on the fundamental properties of
compressible MHD.

\subsection{Alfv\'en, slow, and fast modes}

%%%%%%%%%%%%%%%%%%%%%%%%%%%%%%%%%%%%%%%%%
\begin{figure}[t]
\begin{center}
%%%\mpicplace{10 cm}{1 cm}
\includegraphics{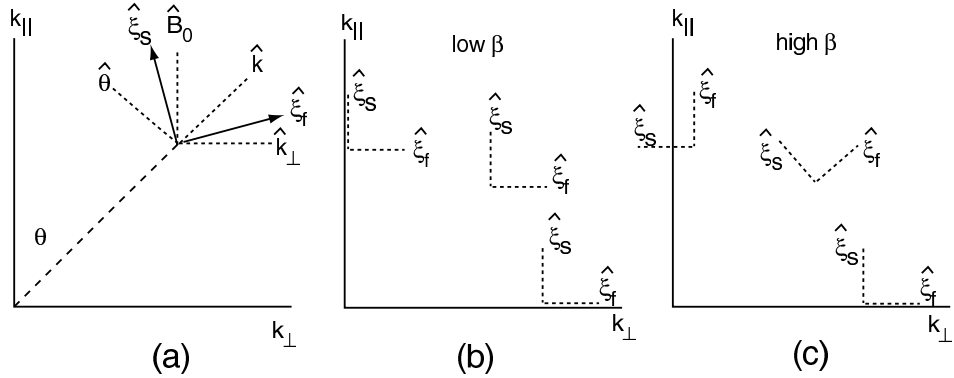} 
%%%\mpicplace{10 cm}{6 cm}
\end{center}
\caption{
         (a) Directions of fast and slow basis vectors.
             $\hat{\xi}_f$ and $\hat{\xi}_s$ represent
             the directions of displacement of fast and slow
             modes, respectively.
             In the fast basis ($\hat{\xi}_f$) is always between
             $\hat{\bf k}$ and $\hat{\bf k}_{\perp}$.
             In the 
             slow basis ($\hat{\xi}_s$) lies between
             $\hat{\theta}$ and $\hat{\bf B}_0$.
             Here, $\hat{\theta}$ is perpendicular to 
             $\hat{\bf k}$ and parallel to the wave front.
             All vectors lie in the same plane formed by
             ${\bf B}_0$ and ${\bf k}$.
             On the other hand, the displacement vector for
             Alfv\'en waves (not shown) is perpendicular to the plane.
         (b) Directions of basis vectors for a very small $\beta$ drawn in the
             same plane as in (a).
             The fast bases are almost parallel to $\hat{\bf k}_{\perp}$.
         (c) Directions of basis vectors for a very high $\beta$.
             The fast basis vectors are almost parallel to ${\bf k}$.
             The slow waves become pseudo-Alfv\'en waves.
       }
\label{fig_modes}
\end{figure}  
%%%%%%%%%%%%%%%%%%%%%%%%%%%%%%%%%%%%%%%%%
%%%%%%%%%%%%%%%%%%%%%%%%%%%%%%%%%%%%%%%%%
\begin{figure}[t]
\begin{center}
%%%\mpicplace{10 cm}{1 cm}
\includegraphics{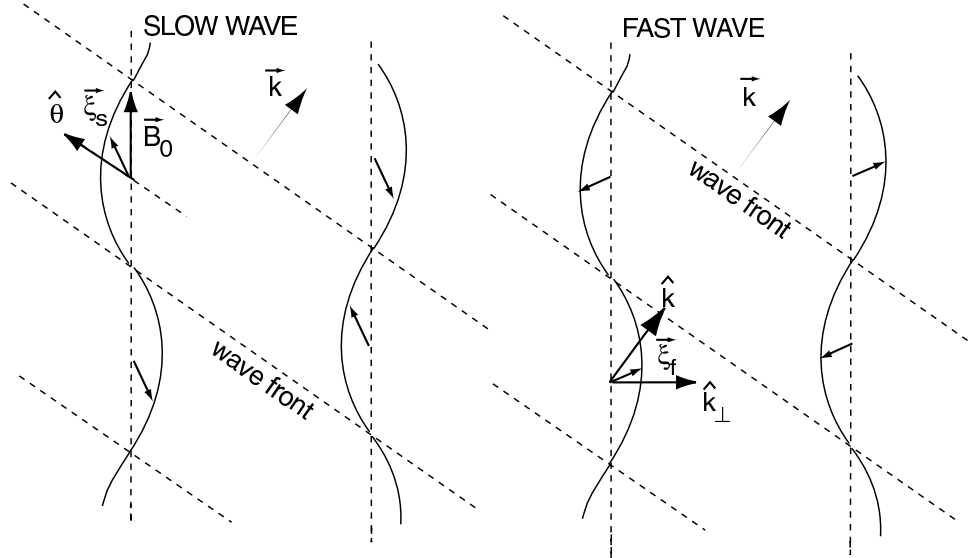}
%%%\mpicplace{10 cm}{6 cm}
\end{center}
\caption{
         Waves in real space.
         We show the directions of displacement vectors for
         a slow wave ({\it left}) and a fast wave ({\it right}).
         Note that $\hat{\xi}_s$ lies between
        $\hat{\theta}$ and $\hat{\bf B}_0 ~(=\hat{\bf k}_{\|})$ and 
                   $\hat{\xi}_f$ between
        $\hat{\bf k}$ and $\hat{\bf k}_{\perp}$.
         Again, $\hat{\theta}$ is perpendicular to 
             $\hat{\bf k}$ and parallel to the wave front.
         Note also that, for the fast wave, for example,
         density (inferred by the
         directions of the displacement vectors)
         becomes higher where field lines are closer, resulting
         in a strong restoring force, which is why fast waves
         are faster than slow waves.
       }
\label{fig_modes-real}
\end{figure}  
%%%%%%%%%%%%%%%%%%%%%%%%%%%%%%%%%%%%%%%%%
  
Let us start by reviewing different MHD waves.
In particular, we describe the Fourier space representation
of these waves.  The
real space representation can be found in   
papers on modern shock-capturing MHD codes  
(e.g.~\cite{Bri88,Ryu95}).
{}For the sake of simplicity, we consider an isothermal plasma.
{}Figure \ref{fig_modes} and Figure \ref{fig_modes-real} give
schematics of slow and fast waves.
{}For slow and fast waves,
${\bf B}_0$, ${\bf v}_{\bf k}$ ($\propto \xi$), and ${\bf k}$ are
in the same plane.
On the other hand, for Alfv\'en waves, the velocity of the
{}fluid element $({\bf v}_{\bf k})_A$ is orthogonal to the 
${\bf B}_0 - {\bf k}$ plane.

As before, the Alfv\'en speed is
$
  V_A=B_0/\sqrt{4\pi \rho_0},
$
where $\rho_0$ is the average density.
{}Fast and slow speeds are
\begin{equation}
 c_{f,s} = \left[   \frac{1}{2} \left\{
         a^2+V_A^2 \pm \sqrt{ (a^2+V_A^2)^2 - 4 a^2 V_A^2 \cos^2{\theta} }
                                \right\} 
           \right]^{1/2},
\end{equation} 
where $\theta$ is the angle between ${\bf B}_0$ and ${\bf k}$.
See Table \ref{cho_table1} for the definition of other variables.
When $\beta$ 
($\beta=P_{g}/P_{B}$=$2a^2/V_A^2$; 
$P_g$= gas pressure, $P_B$= magnetic pressure;
hereinafter $\beta=$ average $\beta \equiv \bar{P}_{g}/\bar{P}_{B}$) 
goes to zero, we have
\begin{eqnarray} 
 c_f &\approx& V_A, \nonumber \\     
 c_s &\approx& a\cos{\theta}.   
\end{eqnarray}

Figure \ref{fig_modes} shows directions of displacement 
(or, directions of velocity) vectors
for these three modes.
We will call them the basis vectors for these modes.
The Alfv\'en basis is perpendicular to both $\hat{\bf k}$ 
and $\hat{\bf B}_0$,
and coincides with the azimuthal vector $\hat{\bf \phi}$ in 
a spherical-polar coordinate system.
Here hatted vectors are unit vectors.
The fast basis $\hat{\bf \xi}_f$ 
lies {\it between} $\hat{\bf k}$ and $\hat{\bf k}_{\perp}$:
\begin{equation}
   \hat{\bf \xi}_f \propto 
     \frac{ 1-\sqrt{D}+{\beta}/2  }{ 1+\sqrt{D}-{\beta}/2  } 
    \left[ \frac{ k_{\perp} }{ k_{\|} } \right]^2
     k_{\|} \hat{\bf k}_{\|}  +
          k_{\perp} \hat{\bf k}_{\perp},
\end{equation}    
where
$
  D=(1+{\beta}/2)^2-2{\beta} \cos^2{\theta}$, and
$  {\beta}$ is the averaged $\beta$ (=$\bar{P}_g/\bar{P}_B$).
The slow basis $\hat{\bf \xi}_s$ lies {\it between} $\hat{\bf \theta}$ 
and $\hat{\bf B}_0$ (=$\hat{\bf k}_{\|}$):
\begin{equation}
   \hat{\bf \xi}_s \propto 
        k_{\|} \hat{\bf k}_{\|}+
     \frac{ 1-\sqrt{D}-{\beta}/2  }{ 1+\sqrt{D}+{\beta}/2  } 
    \left[ \frac{ k_{\|} }{ k_{\perp} }  \right]^2
     k_{\perp} \hat{\bf k}_{\perp}. 
\end{equation}
The two vectors $\hat{\bf \xi}_f$ and $\hat{\bf \xi}_s$ are
mutually orthogonal.
Proper normalizations are required for both bases 
to make them unit-length.

When $\beta$ goes to zero (i.e. the magnetically dominated regime),
$\hat{\bf \xi}_f$ becomes parallel to $\hat{\bf k}_{\perp}$
and $\hat{\bf \xi}_s$ becomes parallel to $\hat{\bf B}_0$ 
(Fig.~\ref{fig_modes}b).
The sine of the angle between $\hat{\bf B}_0$ and $\hat{\bf \xi}_s$ 
is $(\beta/2) \sin\theta \cos{\theta}$.
When $\beta$ goes to infinity (i.e. gas pressure dominated regime)\footnote{
%%%
In this section, we assume that external mean field is strong 
(i.e.~$V_A > (\delta V)$) but finite, so that 
$\beta \rightarrow \infty$ 
means the gas pressure $\bar{P}_g \rightarrow \infty$. 
%%%
},
$\hat{\bf \xi}_f$ becomes parallel to $\hat{\bf k}$
and $\hat{\bf \xi}_s$ becomes parallel to $\hat{\bf \theta}$ 
(Fig.~\ref{fig_modes}c).
This is the incompressible limit.
In this limit, the slow mode is sometimes called the pseudo-Alfv\'en mode 
\cite{Gol95}.

\subsection{Theoretical considerations}
%%%%%%%%%%%%%%%%%%%%%%%%%%%%%%%%%%%%%% 
\begin{figure}[t]
\includegraphics[width=.99\columnwidth]{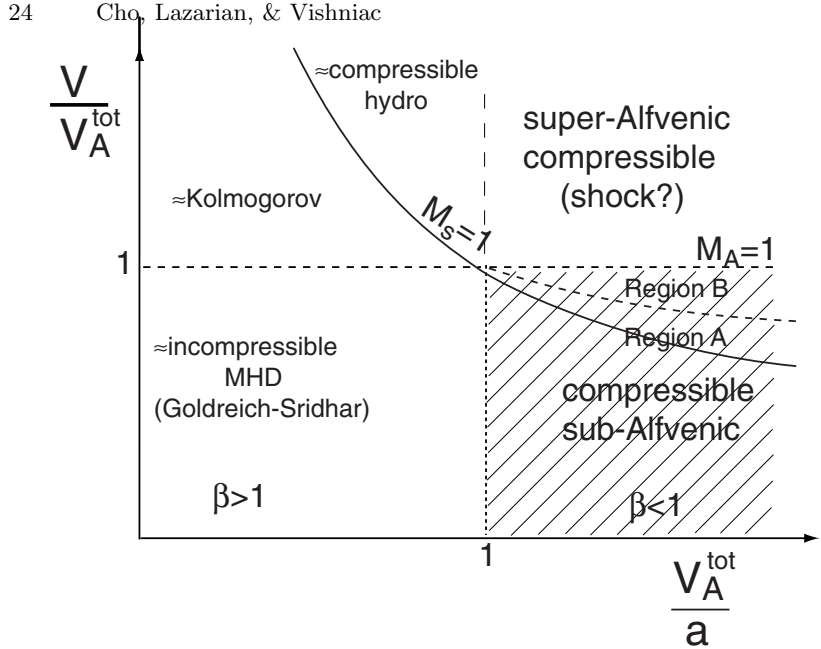}
\caption{
      Different regimes of MHD turbulence.
      We consider the compressible
      sub-Alfv\'{e}nic regime (shaded region).
      In this figure, $V_A^{tot}$ represents
      the total Alfv\'en speed ($=\sqrt{B_0^2+\delta B^2}/\sqrt{4 \pi \rho}$).
      $V\equiv V_{flow} = \delta V$.
      Cho \& Vishniac \cite{Cv00g} argued that,
      even when the external field is weak, small scales can follow
      the GS95-like scaling in incompressible MHD regime.
      Similarly, for compressible case, the small scales of 
      super-Alfv\'{e}nic compressible turbulence
      are expected to fall in the sub-Alfv\'{e}nic compressible regime.
      Moreover, winding of magnetic field by turbulence increases
      the magnetic field energy and the super-Alfv\'{e}nic turbulence
      becomes more and more magnetically dominated with $V/V_A\rightarrow 1$.
      Region B is the region where, within the density fluctuations,
      the velocities can get super-Alfv\'{e}nic.
      In the figure, we used equation (\ref{eq39}) to
      determine the borderline between region A and B.
   }
\label{fig_regimes}
\end{figure}  
%%%%%%%%%%%%%%%%%%%%%%%%%%%%%%%%%%%%%%

Here we address the issue of mode coupling in a low $\beta$
plasma. It is reasonable to suppose that in the limit where $\beta \gg 1$
turbulence for Mach numbers ($M_s=\delta V/a$) less than unity
should be largely similar to the exactly incompressible regime.
Thus, Lithwick \& Goldreich \cite{Lit01} conjectured that the GS95 relations
are applicable to slow and Alfv\'en modes with the fast modes decoupled.
They also mentioned that this relation can carry on
for low $\beta$ plasmas.
{}For $\beta \gg1$ and $M_s>1$, we are in the regime of hydrodynamic 
compressible turbulence for which no theory exists, as far as we know.

In the diffuse interstellar medium $\beta$  is typically less than
unity. 
In addition, it is $\sim 0.1$ or less for molecular clouds.
There are a few simple arguments suggesting that MHD theory can be formulated 
in the regime where the Alfv\'en Mach number ($\equiv \delta V/V_A$) is
less than unity, although this is not a universally accepted
assumption.
Alfv\'en modes describe incompressible motions.
Arguments in GS95 are suggestive that the coupling of Alfv\'{e}n
to fast and slow modes
will be weak.
Consequently, we expect that in this regime the Alfv\'en
cascade should follow the GS95 scaling.
Moreover the slow waves are likely to evolve passively \cite{Lit01}.
{}For $a \ll V_A$ their nonlinear evolution should be governed 
by Alfv\'en modes 
so that we expect the GS95 scaling for them as well.  The phase velocity of 
Alfv\'en waves and slow waves depend on a factor of $\cos{\theta}$ and
this enables modulation of the slow waves by the Alfv\'en ones.
However, fast waves do not have this factor and therefore cannot be
modulated by the changes of the magnetic field direction associated
with Alfv\'en waves. The coupling between the modes is through
the modulation of the local Alfv\'en velocity and therefore
is weak.

{}For Alfv\'en Mach number ($M_A$) larger than unity a
shock-type regime is expected.
However, generation of magnetic field by turbulence \cite{Cv00g} is expected
for such a regime.
It will make the steady state turbulence approach $M_A \sim 1$.\footnote{
We suspect that simulations 
that show super-Alfv\'{e}nic turbulence is widely spread in the ISM 
might not evolve for a long
enough time to reach the steady state.}
 Therefore in 
Cho \& Lazarian \cite{Cl02} we consider turbulence in the limit
$M_s>1$, $M_A<1$, and $\beta <1$ (Fig.~\ref{fig_regimes}).
{}For these simulations, we mostly used $M_s\sim 2.2$, $M_A \sim 0.7$, and
$\beta \sim 0.2$. The Alfv\'en speed of the mean external field
is similar to the rms velocity ($V_A=1,\delta V\sim 0.7, a=\sqrt{0.1}$),
and we used an isothermal equation of state.

Although the scaling relations presented below are applicable to
sub-Alfv\'{e}nic turbulence, we cautiously speculate that
small scales of super-Alfv\'{e}nic turbulence
might follow similar scalings.
Boldyrev, Nordlund, \& Padoan \cite{Bol01} obtained energy spectra
close to $E(k)\sim k^{-1.74}$ in solenoidally driven super-Alfv\'{e}nic
supersonic turbulence simulations.
The spectra are close to the Kolmogorov spectrum ($\sim k^{-5/3}$), rather than
shock-dominated spectrum ($\sim k^{-2}$).
This result might imply that small scales of super-Alfv\'{e}nic MHD turbulence
can be described by our sub-Alfv\'{e}nic model presented below, which predicts
Kolmogorov-type spectra for Alfv\'{e}n and slow modes.

\subsection{Coupling of MHD modes and Scaling of Alfv\'{e}n modes} 
\label{sec_coupling}

Alfv\'en modes are not susceptible to collisionless damping (see
\cite{Spa91,Min97} and references therein) that damps slow and fast modes.
Therefore, we mainly consider the transfer of energy from  
Alfv\'en waves to compressible MHD waves (i.e. to the 
slow and fast modes).

%%%%%%%%%%%%%%%%%%%%%%%%%%%%%%%%%%%%%%
\begin{figure}[t]
\includegraphics[width=.5\columnwidth]{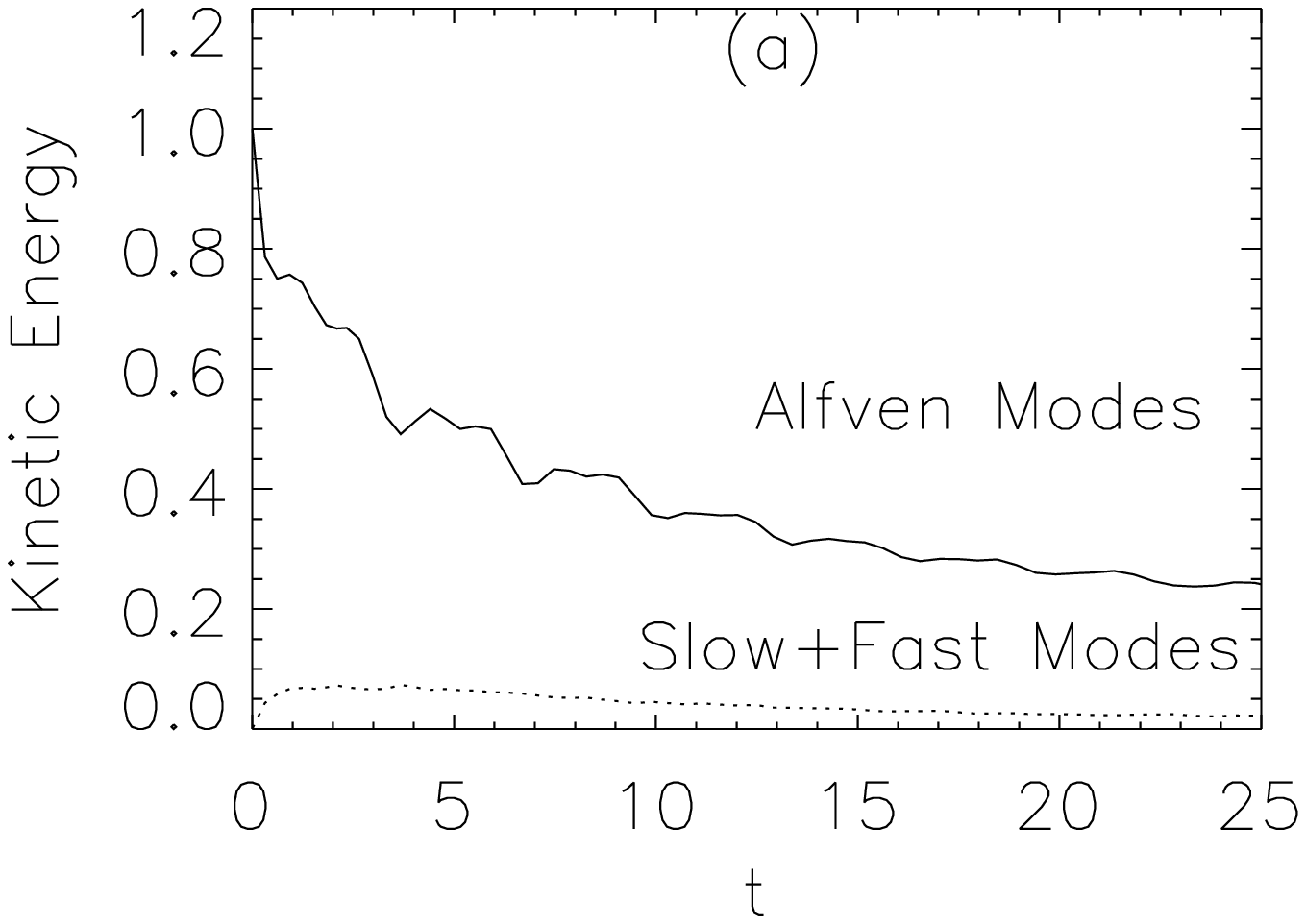}
\hfill
\includegraphics[width=.45\columnwidth]{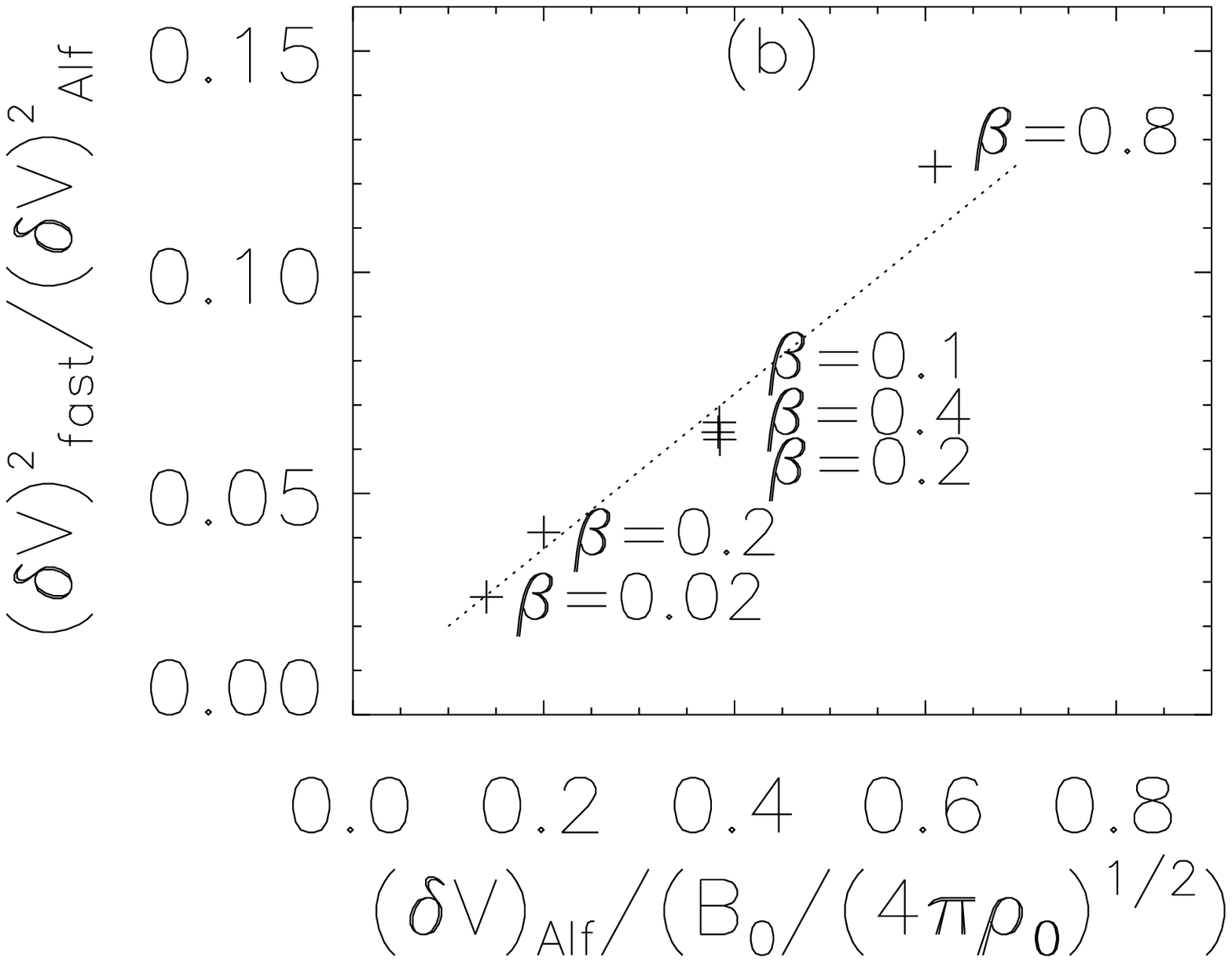}
\caption{
         ({\it Left}) Decay of Alfv\'{e}nic turbulence.
         The generation of fast and slow waves is not efficient.
         $\beta \sim 0.2$, $M_s \sim 3$.
         ({\it Right}) The ratio of $(\delta V)_f^2$ to
          $(\delta V)_A^2$. The stronger the external field ($B_0$) is,
          the more suppressed the coupling is.
          The ratio is not sensitive to $\beta$.
          {}From \cite{Cl02}
       }
\label{fig_coupling}
\end{figure}  
%%%%%%%%%%%%%%%%%%%%%%%%%%%%%%%%%%%%%%

In Cho \& Lazarian \cite{Cl02}, we carry out simulations 
to check the strength of the mode-mode coupling.
We first obtain a data cube from a driven compressible numerical simulation with
$B_0/\sqrt{4 \pi \rho_0}=1$.
Then, after turning off the driving force, 
we let the turbulence decay.
We go through
the following procedures before we let the turbulence decay.
We first remove slow and fast modes in Fourier space and
retain only Alfv\'{e}n modes.
We also change the value of ${\bf B}_0$ preserving its original
direction.
We use the same constant initial density $\rho_0$ for all simulations.
We assign a new constant initial gas pressure $P_g$
 \footnote{
      The changes of both $B_0$ and $P_g$ preserve the 
      Alfv\'{e}n character of perturbations.
      In Fourier space,
      the mean magnetic field (${\bf B}_0$)
      is the amplitude of ${\bf k}={\bf 0}$ component.
      Alfv\'{e}n components in Fourier space are
      for ${\bf k} \neq {\bf 0}$ and their directions are 
      parallel/anti-parallel to
      $\hat{\bf \xi}_A$ (= $\hat{\bf B}_{0} \times \hat{\bf k}_{\perp}$).
      The direction of $\hat{\bf \xi}_A$ does not depend on
      the magnitude of $B_0$ or $P_g$.
}.
After doing all these procedures, we let the turbulence decay.
We repeat the above procedures for different values of $B_0$ and $P_g$.
{}Fig.~\ref{fig_coupling}a shows the evolution of the kinetic 
energy of a simulation.
The solid line represents the kinetic energy of Alfv\'en  modes.
It is clear that Alfv\'en waves are poorly coupled to the compressible
modes, and do not generate them efficiently \footnote{
As correctly pointed out by Zweibel (this volume) there is always 
residual coupling between Alfv\'{e}n and compressible modes due to
steepening of Alfv\'{e}n modes.
However, this steepening happens on time-scales much longer than the
cascading time-scale.}
Therefore, we expect that Alfv\'en modes will follow the same scaling
relation as in the incompressible case.
{}Fig.~\ref{fig_coupling}b shows that the coupling gets weaker as 
$B_0$ increases: 
\begin{equation}
  \frac{ (\delta V)_{f}^2 }{ (\delta V)_{A}^2 }
 \propto \frac{ (\delta V)_{A} }{ B_0 }. \label{vfscale}
\end{equation}
The ratio of $(\delta V)_{s}^2$ to $(\delta V)_{A}^2$ is proportional to
$(\delta V)_{A}^2/ B_0^2$.

%%%%%%%%%%%%%%%%%%%%%%%%%%%%%%%%%%%%%%%%
\begin{figure}[!t]
\includegraphics[width=.5\columnwidth]{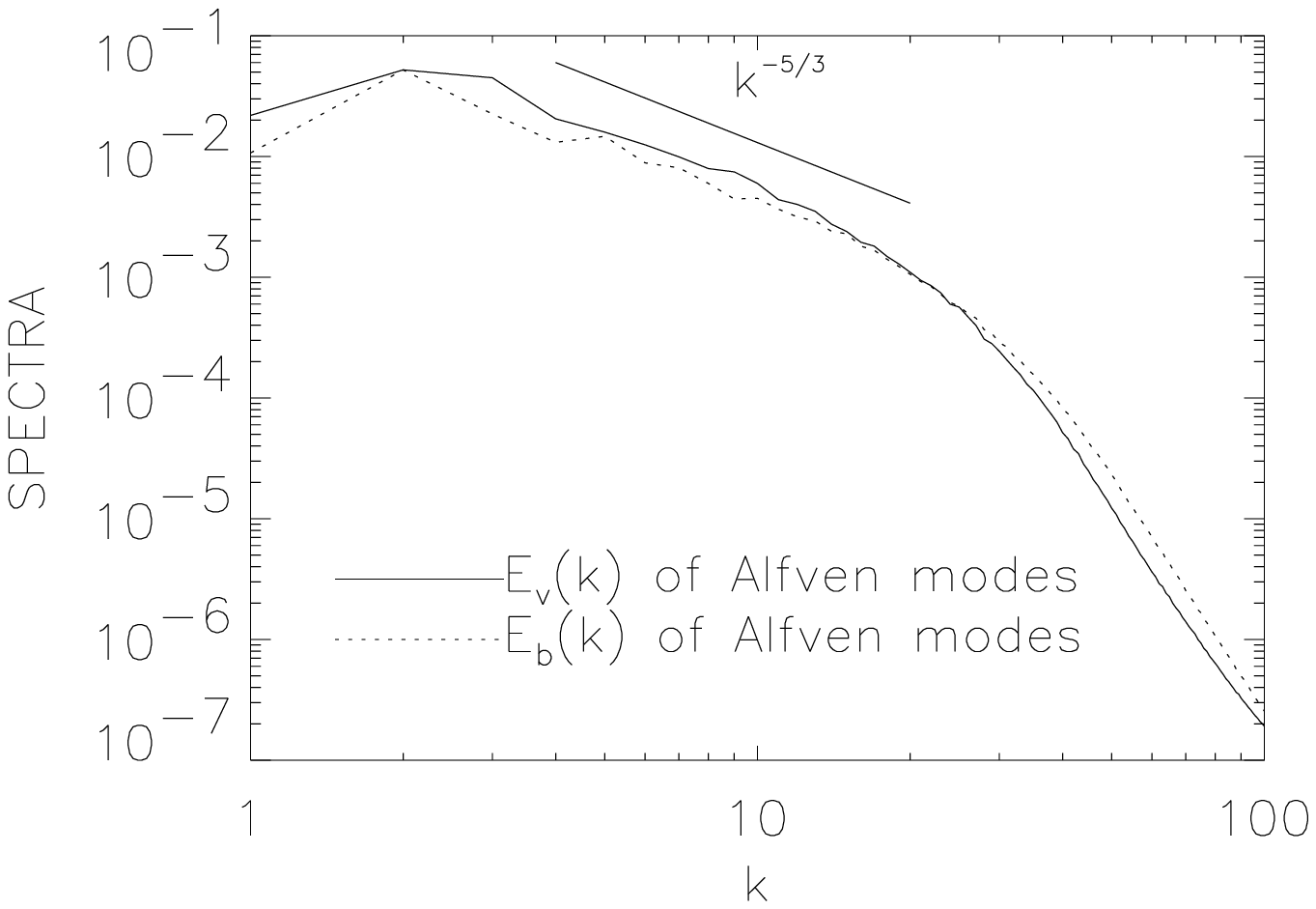}
\hfill
\includegraphics[width=.45\columnwidth]{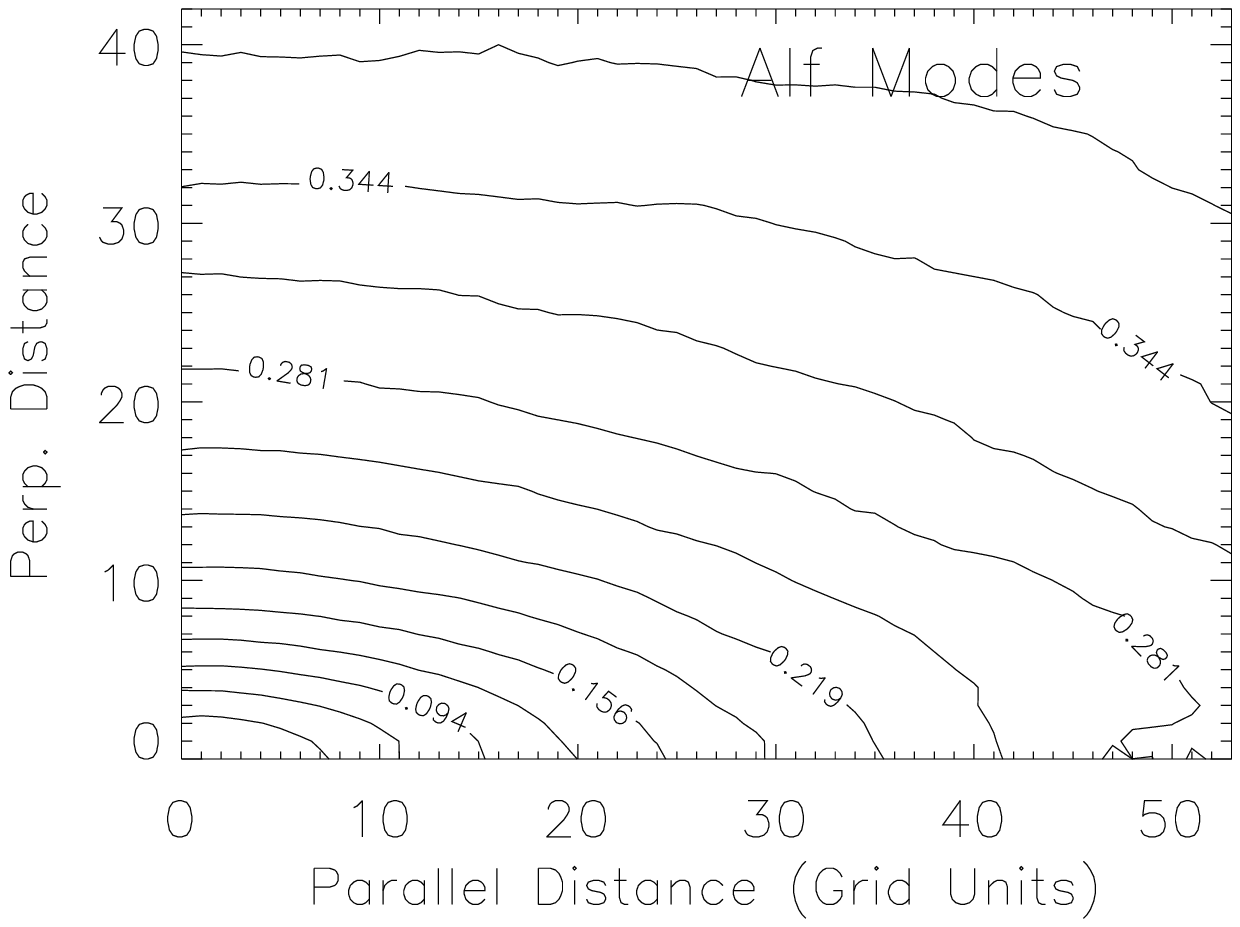}
\caption{
         (a) Alfv\'en spectra follow a Kolmogorov-like power law.
         (b) The second-order structure function 
             ($SF_2=<{\bf v}({\bf x}+{\bf r}) - {\bf v}({\bf x})>$) 
             for Alf\'ven velocity
             shows anisotropy similar to the GS95.
             Conturs represent eddy shapes.
             {}From \cite{Cl02}.
       }
\label{fig_alf} 
\end{figure}  
%%%%%%%%%%%%%%%%%%%%%%%%%%%%%%%%%%%%%%%%

This marginal coupling is in good agreement with a claim in GS95, as well as
earlier numerical studies where the
velocity was decomposed into a compressible component ${\bf v}_C$
and a solenoidal component ${\bf v}_S$. The compressible component 
is curl-free and parallel to the wave vector ${\bf k}$ in Fourier space.
The solenoidal component is divergence-free and perpendicular to ${\bf k}$.
The ratio $\chi = (\delta V)_C/ (\delta V)_S$ is an important parameter 
that determines the
strength of any shock
\cite{Pas88,Pou99}.
Porter, Woodward, \& Pouquet \cite{Por98} performed a 
hydrodynamic simulation of
decaying turbulence with an initial sonic Mach number of unity 
and found that $\chi^2$ evolves toward $\sim 0.11$.
Matthaeus et al.~\cite{Mat96} 
carried out simulations of decaying weakly compressible
MHD turbulence \cite{Zan93} and found that $\chi^2 \sim O(M_s^2)$, where
$M_s$ is the sonic Mach number.
In \cite{Bol01} a weak generation of
compressible components in solenoidally driven super-Alfv\'{e}nic supersonic
turbulence simulations was obtained.

{}Fig.~\ref{fig_alf} shows that the spectrum and 
the anisotropy of Alfv\'en waves in this limit are
compatible with the GS95 model:
\begin{equation}
  \mbox{\it Spectrum of Alfv\'{e}n Modes:~~~~~} E(k)\propto k_{\perp}^{-5/3},
\end{equation}
and scale-dependent anisotropy 
$k_{\|}\propto k_{\perp}^{2/3}$ that is compatible with the GS95 theory.

\subsection{Scaling of the slow modes} \label{sec_slow}
Slow waves are somewhat 
similar to pseudo-Alfv\'en waves (in the incompressible limit).
{}First, the directions of displacement (i.e.~$\xi_s$) of both waves
are similar when anisotropy is present.
The vector ${\bf \xi}_s$ is always between $\hat{\bf \theta}$ and
$\hat{\bf k}_{\|}$.
In Figure \ref{fig_modes}, we can see that the angle between
$\hat{\bf \theta}$ and
$\hat{\bf k}_{\|}$  
gets smaller when $k_{\|} \ll k_{\perp}$.
Therefore, when there is anisotropy (i.e.~$k_{\|} \ll k_{\perp}$), 
$\hat{\bf \xi}_s$ of a low $\beta$ plasma becomes similar to 
that of a high $\beta$ plasma.
Second, the angular dependence in
the dispersion relation $c_s \approx a \cos{\theta}$ is identical to
that of pseudo-Alfv\'en waves (the only difference is that,
in slow waves, the sound speed $a$ is present instead of the Alfv\'en
speed $V_A$).

%%%%%%%%%%%%%%%%%%%%%%%%%%%%%%%%%%%%%%%
\begin{figure}[!t]
\includegraphics[width=.5\columnwidth]{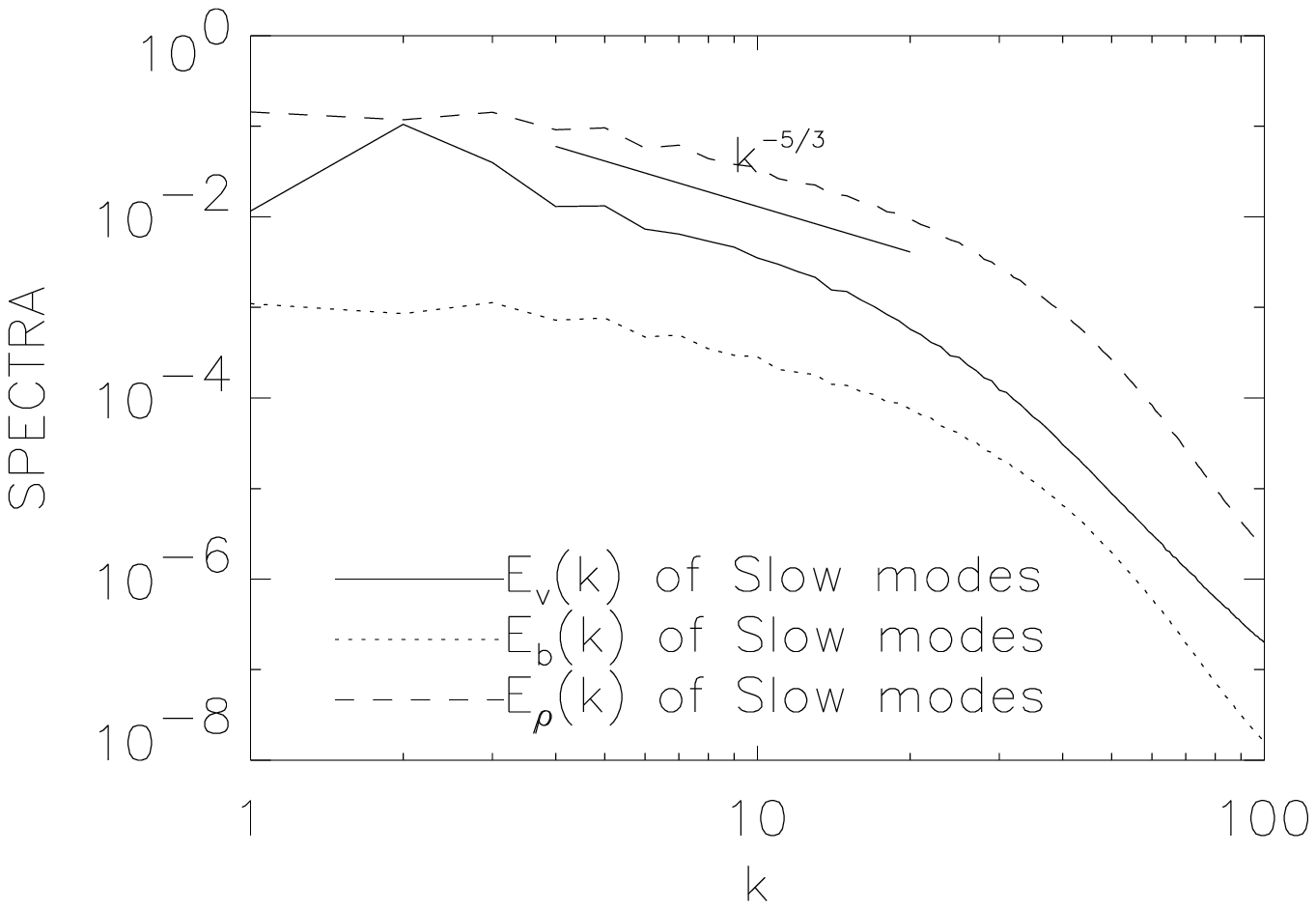}
\hfill
\includegraphics[width=.45\columnwidth]{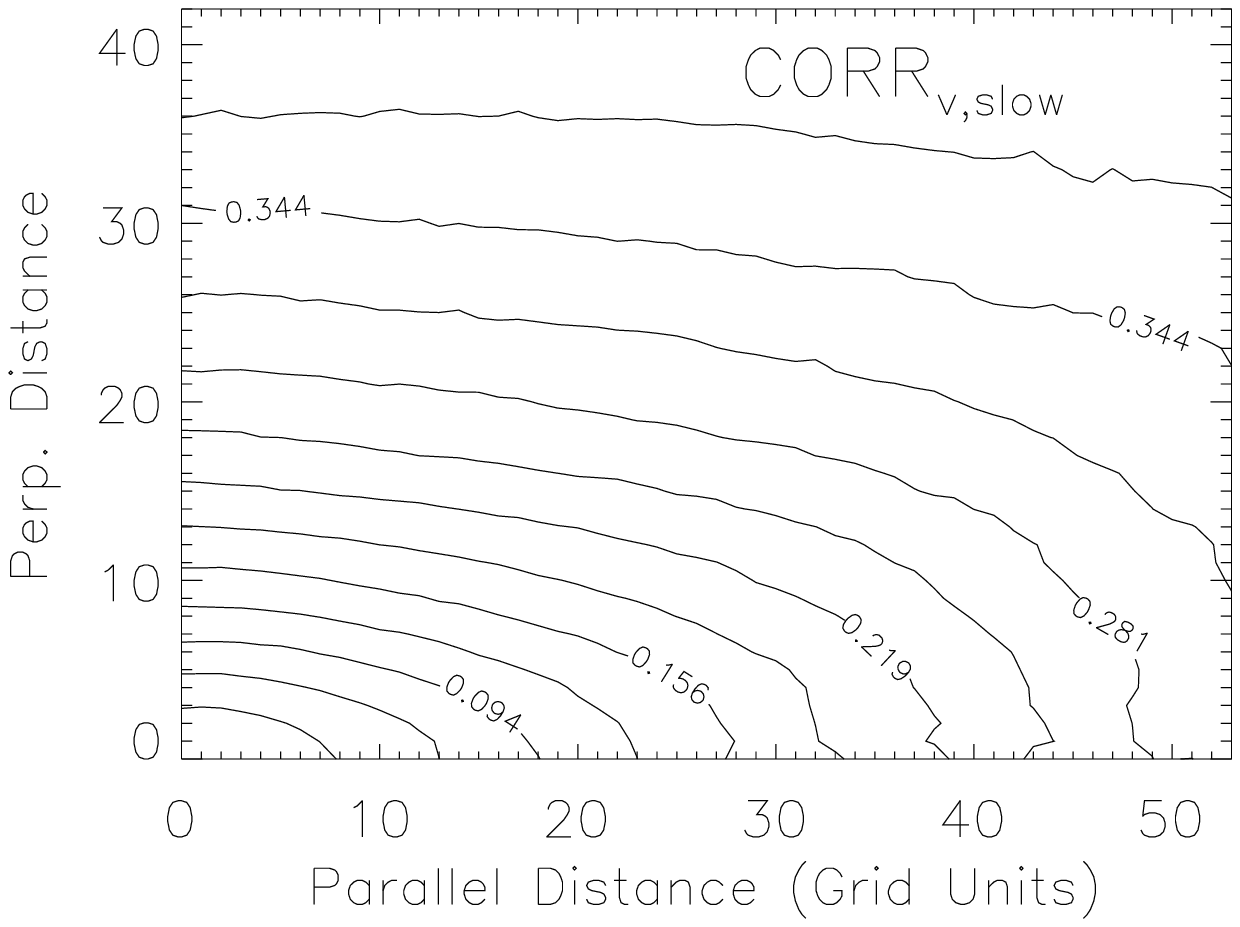}
\caption{   
         (a) Slow spectra also follow a Kolmogorov-like power law.
         (b) Slow modes show anisotropy similar to the GS95 theory.
             {}From \cite{Cl02}.
       }
\label{fig_slow}
\end{figure}  
%%%%%%%%%%%%%%%%%%%%%%%%%%%%%%%%%%%%%%%

Goldreich \& Sridhar \cite{Gol97} argued that
the pseudo-Alfv\'en waves are slaved to the shear-Alfv\'en 
(i.e.~ordinary Alfv\'en)
waves in the presence of a strong ${\bf B}_0$, meaning
that the energy cascade of pseudo-Alfv\'en modes is primarily mediated
by the shear-Alfve\'en waves.
This is because pseudo-Alfv\'en waves do not provide efficient shearing
motions.
Similar arguments are applicable to slow waves
in a low $\beta$ plasma \cite{Cl02} 
(see also \cite{Lit01} for high-$\beta$ plasmas).
As a result, we conjecture that slow modes follow a scaling similar to
the GS95 model \cite{Cl02}:
\begin{equation}
 \mbox{\it Spectrum of Slow Modes:~~~~~}  E^{s}(k) \propto k_{\perp}^{-5/3}.
\end{equation}

{}Fig.~\ref{fig_slow}a shows the spectra of slow modes.
For velocity, the slope is close to $-5/3$.
Fig.~\ref{fig_slow}b shows the contours of equal 
second-order structure function ($SF_2$)
of slow velocity, which are compatible with $k_{\|}\propto k_{\perp}^{2/3}$
scaling.

In low $\beta$ plasmas, density fluctuations are dominated by
slow waves \cite{Cl02}.
{}From the continuity equation $\dot{\rho} = \rho \nabla \cdot {\bf v}$
\begin{equation}
  \omega \rho_k = \rho_0 {\bf k} \cdot  {\bf v}_k,
\end{equation}
we have,
{}for slow modes, 
$
(\rho_k)_s \sim \rho_0 (v_k)_s/a. \label{roks}
$
Hence, this simple argument implies
\begin{equation}
 \left(\frac{ \delta \rho }{ \rho }\right)_s = \frac{ (\delta V)_s }{ a }\sim  
             M_s,   \label{eq39}
\end{equation}
where we assume that $(\delta V)_s \sim (\delta V)_A$ and
$M_s$ is the Mach number.
On the other hand, only a small amount of magnetic field is produced by the
slow waves. Similarly, using the induction equation 
($\omega {\bf b}_k = {\bf k} \times ({\bf B}_0 \times {\bf v}_k)$), we have
\begin{equation}
 \frac{(\delta B)_s}{ (\delta V)_s } \sim \frac{a}{ B_0 } 
                            =O(\sqrt{\beta}),
\end{equation}
which means that equipartition between kinetic and magnetic energy
is not guaranteed in low $\beta$ plasmas.
In fact, in Fig.~\ref{fig_slow}a, the power spectrum 
for density fluctuations has a much larger amplitude than 
the magnetic field power spectrum.
Since density fluctuations are caused mostly by the slow waves and
magnetic fluctuation is caused mostly by Alfv\'{e}n and fast modes,
we {\it do not} expect a strong correlation between density and
magnetic field, which agrees with the ISM simulations
\cite{Pad99,Ost01,Vaz02}.

\subsection{Scaling of the fast modes} 
\begin{figure}[!t]
\includegraphics[width=.5\columnwidth]{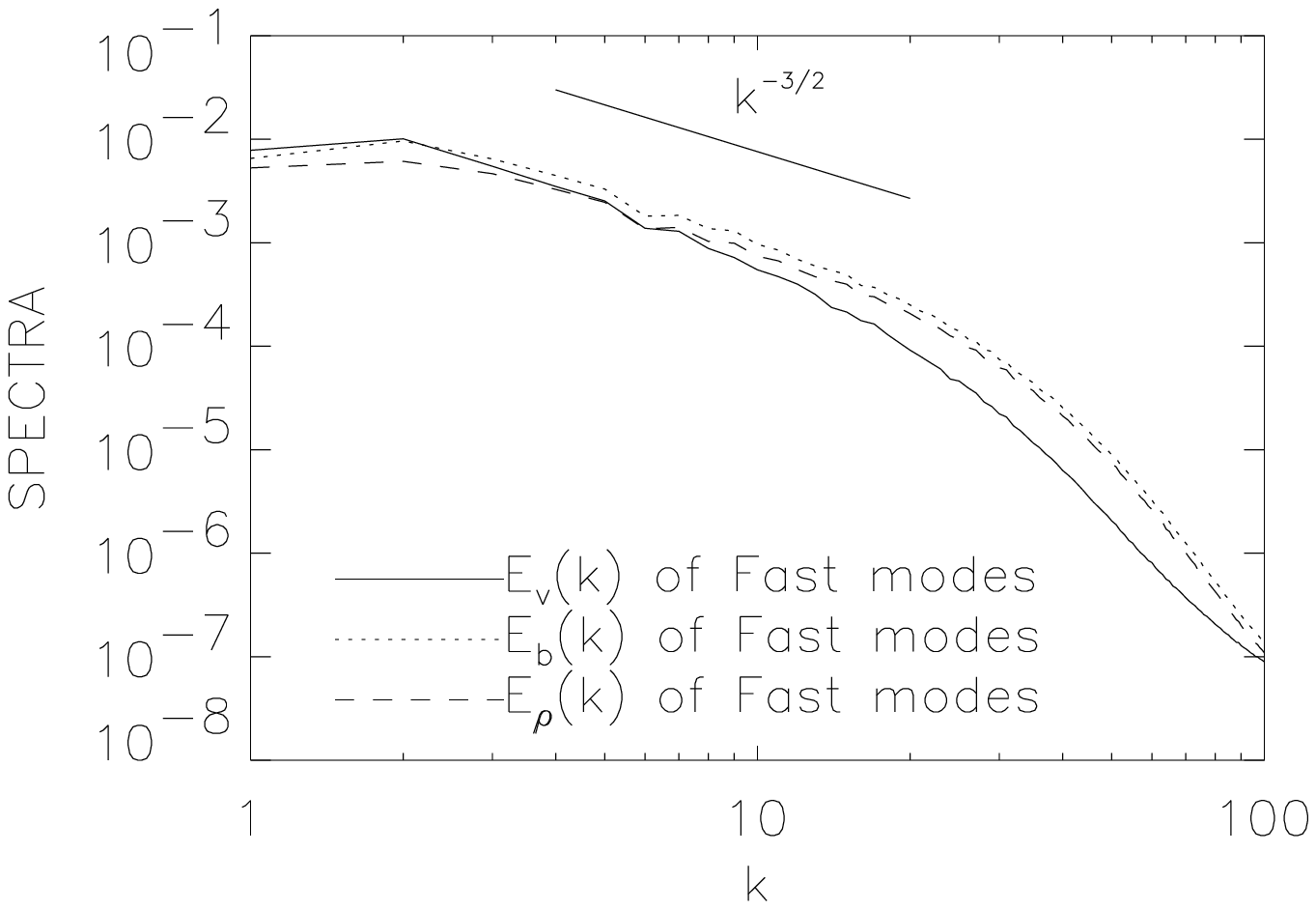}
\hfill
\includegraphics[width=.45\columnwidth]{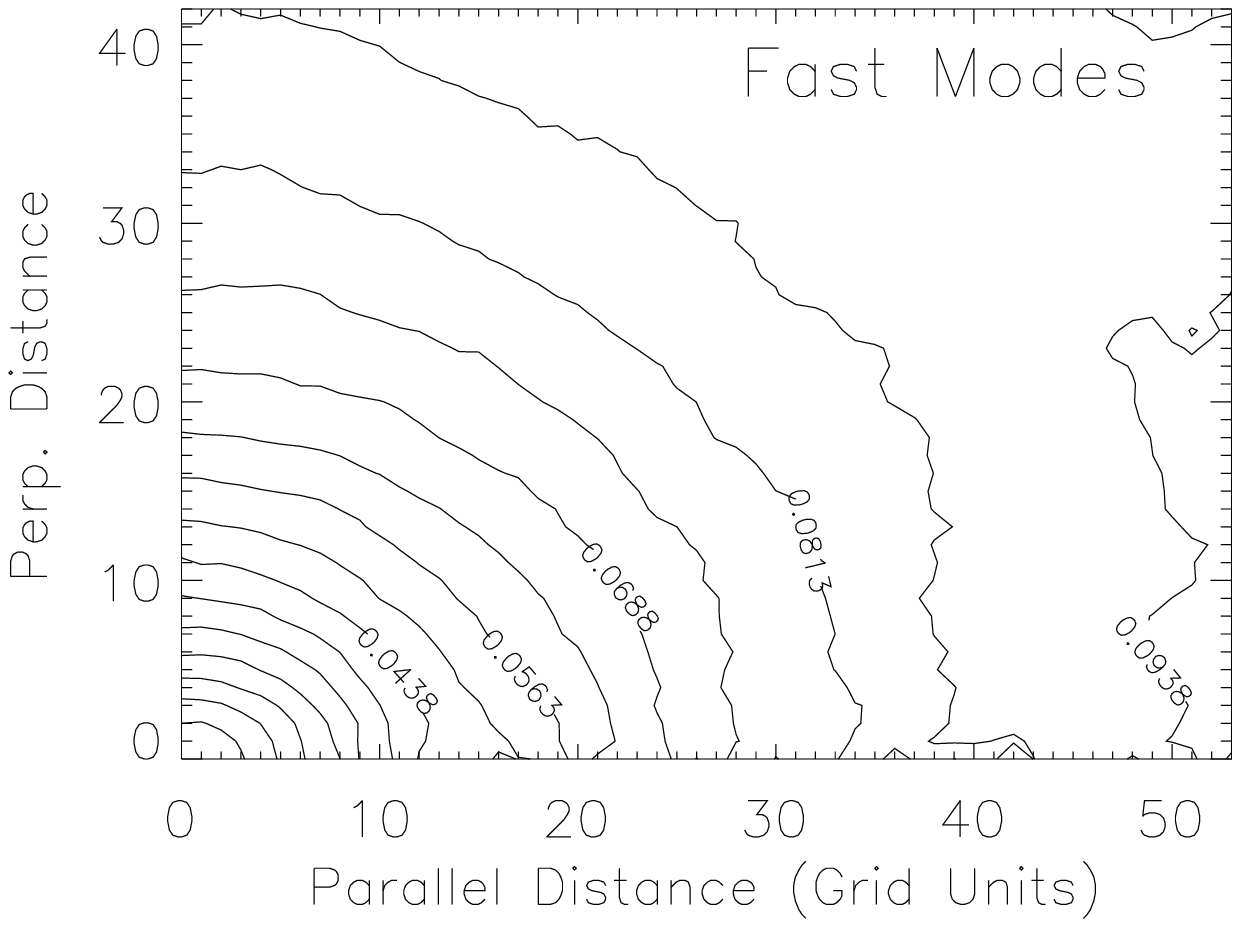}
\caption{
         (a) The power spectrum of fast waves is compatible with
             the IK spectrum.
         (b) The magnetic second-order structure function of 
             fast modes shows isotropy.{}From \cite{Cl02}.
       }
\label{fig_fast}
\end{figure}

{}Figure \ref{fig_fast} shows fast modes are isotropic.
{}The resonance conditions for interacting fast waves are:
\begin{eqnarray}
\omega_1 + \omega_2 &=& \omega_3, \\ 
  {\bf k}_1 + {\bf k}_2 &=& {\bf k}_3.
\end{eqnarray}
Since $ \omega \propto k$ for the fast modes,
the resonance conditions can be met only when
all three ${\bf k}$ vectors are collinear.
This means that the direction of energy cascade is 
{\it radial} in Fourier space, and 
we expect an isotropic distribution of energy in Fourier
space.

%%%%%%%%%%%%%%%%%%%%%%%%%%%%%%%%%%%%%%%
%%%%%%%%%%%%%%%%%%%%%%%%%%%%%%%%%%%%%%%

Using the constancy of energy cascade and uncertainty principle, we
can derive an IK-like energy spectrum for fast waves.
The constancy of cascade rate reads
\begin{equation}
 \frac{ v_l^2 }{ t_{cas} } = \frac{ k^3v_k^2 }{ t_{cas} }=~\mbox{constant.}
\label{cas_rate_fast}
\end{equation}
On the other hand, $t_{cas}$ can be estimated as
\begin{equation}
t_{cas} \sim \frac{ v_k }{ \left( {\bf v}\cdot \nabla {\bf v} \right)_{\bf k} }
        \sim \frac{ v_k }{  \sum_{{\bf p}+{\bf q}={\bf k}}  kv_pv_q }.
\label{eq14_fast}
\end{equation}
If contributions are random,
the denominator can be written by the square root of the 
number of interactions ($\sqrt{\mathcal{N}}$)
times strength of individual interactions
%%%($(k (\Delta k)^2)^{1/2}$) 
($\sim kv_k^2$) \footnote{
To be exact, the strength of individual interactions is 
$\sim kv_k^2 \sin\theta$, where $\theta$ is the angle between ${\bf k}$ and
${\bf B}_0$. Thus marginal anisotropy is expected.
It will be investigated elsewhere.
}. 
Here we assume locality of interactions: $p\sim q\sim k$.
Due to the uncertainly principle,
the number of interactions becomes $\mathcal{N}\sim k(\Delta k)^2$,
where
$\Delta k$ is 
the typical transversal (i.e.~not radial) 
separation between two wave vectors ${\bf p}$ and ${\bf q}$ 
(with ${\bf p}+{\bf q}={\bf k}$).
Therefore, the denominator of equation (\ref{eq14_fast}) is 
$(k (\Delta k)^2)^{1/2} kv_k^2$.
We obtain an independent expression for $t_{cas}$ from 
the uncertainty principle ($t_{cas} \Delta \omega \sim 1$ 
with $\Delta \omega \sim \Delta k (\Delta k/k)$).
{}From this and equation (\ref{eq14_fast}), we get
$%\begin{equation}
t_{cas} \sim  t_{cas}^{1/2}/(k^{2} v_k),
$ %\end{equation}
which yields
\begin{equation}
t_{cas} \sim  1 / k^{4} v_k^2. \label{t_cascade}
\end{equation}
Combining equations (\ref{cas_rate_fast}) and (\ref{t_cascade}), we obtain
$ %\begin{equation}
v_k^2 \sim  k^{-7/2},
$ %\end{equation}
or
$
 E^f(k) \sim k^2 v_k^2 \sim  k^{-3/2}$.
This is very similar to acoustic turbulence, turbulence caused by interacting
sound waves \cite{Zak67,Zak70,Lvo00}. 
Zakharov \& Sagdeev \cite{Zak70} found
$E(k)\propto k^{-3/2}$.
However, there is debate about
the exact scaling of acoustic turbulence.
Here we cautiously claim that our numerical results are compatible
with the Zakharov \& Sagdeev scaling:
\begin{equation}
\mbox{\it Spectrum of Fast Modes:~~~~~} E^f(k) \sim  k^{-3/2}.
\end{equation}

Magnetic field perturbations are mostly affected by
{}fast modes \cite{Cl02} when $\beta$ is small:
\begin{equation} 
\mbox{\it Fast:~~~~~} \frac{(\delta B)_f}{ (\delta V)_A } 
  \sim  \frac{ (\delta V)_f }{ (\delta V)_A },
\end{equation}
if $(\delta V)_A \sim (\delta V)_s$.

The turbulent cascade of fast modes is expected to be slow
and in the absence of collisionless damping
they are expected to propagate in turbulent media
over distances considerably larger than Alfv\'en or
slow modes.
This effect is difficult to observe in
numerical simulations with $\Delta B \sim B_0$.
A modification of the spectrum in the presence of the collisionless
damping is presented in \cite{Yan02a}.

%%%%%777777777777777777777777777777777777777777777777777777777777777777

\section{Astrophysical Implications}

%%%%%777777777777777777777777777777777777777777777777777777777777777777

Many astrophysical problems require some knowledge of the scaling
properties of turbulence.  Therefore we expect a wide range
of applications of the established scaling relations. Here we show how
recent breakthroughs in understanding MHD turbulence affect a few
selected issues. 

%\vspace{0.3cm}
%\noindent
%{\bf Cosmic ray propagation}\\
\subsection{Cosmic ray propagation}

The propagation of cosmic rays is mainly determined by their interactions
with electromagnetic fluctuations in interstellar medium. 
The resonant interaction of cosmic ray particles
with MHD turbulence has been repeatedly suggested as the main
mechanism for scattering and isotropizing cosmic rays. In these
analysis it is usually assumed that the turbulence is
{\it isotropic} with a Kolmogorov spectrum (see  
\cite{Sch98}). How should these calculations be modified?

The essence of the mechanism is rather simple.
Particles moving with velocity $v$ interact with a resonant Alfv\'en wave
of frequency \( \omega =k_{\parallel }v\mu +n\Omega  \) (\( n=\pm 1,2... \)),
where \( \Omega =\Omega _{0}/\gamma  \)
is the gyrofrequency of relativistic particles, \( \mu  \) is the
cosine of the pitch angle. From the resonant
condition above, we know that the most important interaction occurs at 
\( k_{\parallel }\sim \Omega /v\mu \sim (\mu r_{L})^{-1} \),
where \( r_{L} \) is Larmor radius of the high-energy particles. 

\begin{figure} [t!]
\includegraphics[width=.5\columnwidth]{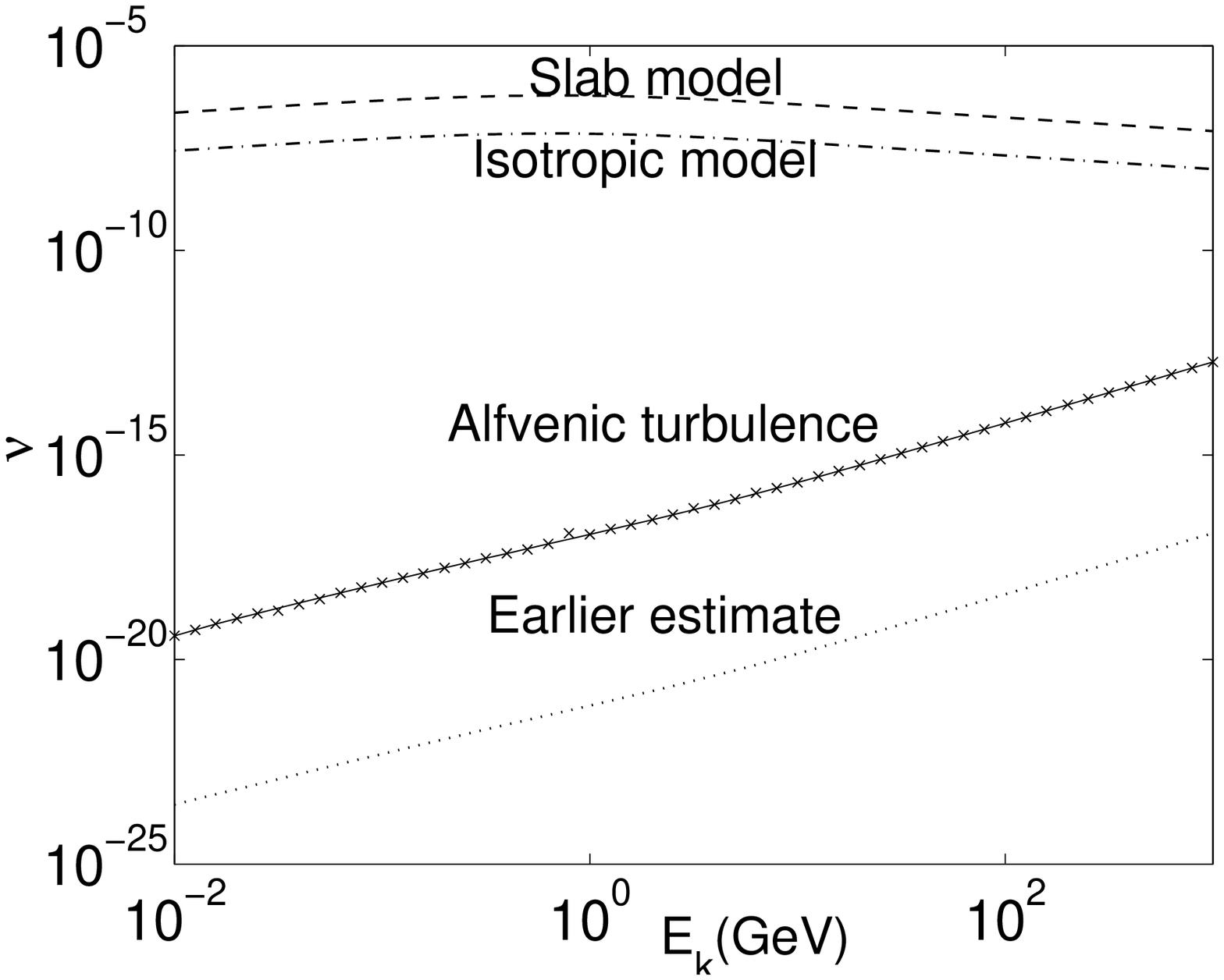}
\hfill
\includegraphics[width=.5\columnwidth]{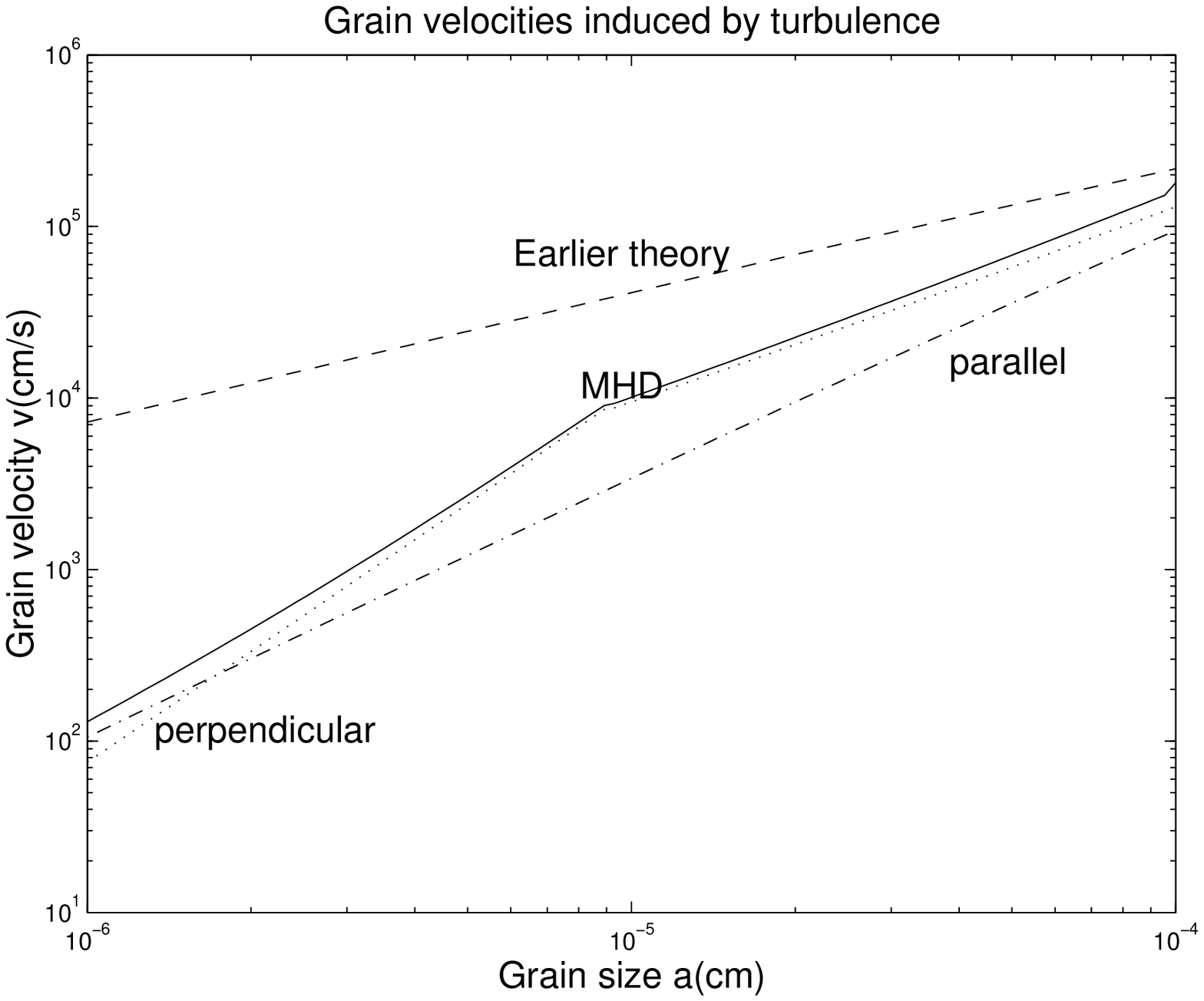}
 \caption{
Applications. (By Lazarian \& Yan).
{\it Left:} Cosmic ray scattering for realistic
MHD turbulence is reduced substantially compared to scattering by
isotropic turbulence, but still larger than estimates in Chandran \cite{Cha01}. 
{\it Right:} The dust acceleration by turbulence is reduced 
compared to the accepted estimates in Draine \cite{Dra85}.}
\label{fig_dust}
\end{figure}

The calculations in \cite{Yan02a} that
made use of tensor (\ref{tensor}) provided 
the scattering efficiency of
anisotropic Alfv\'{e}nic turbulence. The results are compared  
in Fig.~\ref{fig_dust}a with the predictions of the scattering
on isotropic Kolmogorov-type magnetic fluctuations and also
with earlier calculations by Chandran \cite{Cha01}. The latter used rather
{\it ad hoc} form of the tensor to describe magnetic fluctuations
within the Goldreich-Sridhar theory. We see from Fig~\ref{fig_dust}a
that the scattering is substantially suppressed, compared to the 
Kolmogorov turbulence that is usually used for scattering calculations. 
This happens, first of all, 
because most turbulent energy in GS95 turbulence goes to
\( k_{\perp } \) so that there is much
less energy left in the resonance point \( k_{\parallel }=(\mu r_{L})^{-1} \).
Furthermore, \( k_{\perp }\gg k_{\parallel } \) 
means \( k_{\perp }\gg r_{L}^{-1} \)
so that cosmic ray particles cover many eddies during one gyration. 
This random walk decreases the scattering efficiency by a factor of 
\( (\Omega /k_{\perp }v_{\perp })^{1\over 2}=(r_{L}/l_{\perp })^{1\over 2} \),
where \( l_{\perp } \) is the turbulence scale perpendicular to magnetic
field.

Thus the gyroresonance with Alfv\'{e}nic turbulence
is not an effective scattering mechanism for 
cosmic rays if turbulence is injected on the large scales, since the
degree of anisotropy increases on smaller scales.
However, if energy is injected isotropically at small scales, 
the resulting turbulence would be 
more isotropic and scattering will be more efficient.
Scattering by undamped fast modes is more efficient than the Kolmogorov theory
would predict.
Yan \& Lazarian \cite{Yan02a} performed calculations taking into account
the collisionless damping of fast modes and showed that the gyroresonance
scattering by fast modes is the dominant scattering mechanism.

There is another important property of turbulence that was neglected
in earlier work.  
When cosmic rays stream at a velocity much larger than Alfv\'en velocity,
they can excite resonant MHD waves, which in turn scatter cosmic rays.
This is the `streaming instability'. It is usually assumed that this 
instability
can confine cosmic rays with energies less than 100GeV \cite{Ces80}.
However, this is true only in an idealized situation when 
there is no background MHD turbulence. 
As noted earlier, the rates of turbulent
decay are very fast and excited perturbations should
vanish quickly. In \cite{Yan02a} we find that
the streaming instability is only applicable to particles with energies
\( <0.15GeV, \) which is less than the energy of most cosmic ray
particles. This result casts doubt on the self-confinement mechanism discussed
by previous authors. 

All these findings tend to support the alternative
picture of cosmic ray diffusion advocated by Jokipii 
(see \cite{Kot00}). In this picture cosmic rays
follow magnetic field lines, but the magnetic field wanders.
The rate of this wandering can be calculated from the established turbulence
scaling laws.

Knowledge of the scattering rates is essential for understanding both the first
order and the second order Fermi acceleration.
The first order Fermi acceleration, may be important for a wide range
of phenomena from clusters of galaxies and gamma-ray bursts (GRBs) 
to solar flares.
Results obtained in \cite{Yan02a} where the discovered properties of
MHD turbulence are used proved to be very
different from earlier estimates.

%%%%%%%%%%%%%%%%%%%%%%%%%%%%%%%%%%%%%
%%%%%%%%%%%%%%%%%%%%%%%%%%%%%%%%%%%%%

%\vspace{0.3cm}
%\noindent 
%{\bf Grain dynamics}\\
\subsection{Grain dynamics}

Turbulence induces relative dust grain motions and leads to 
grain-grain collisions. These collisions
determine grain size distribution, which affects most dust properties,
including absorption and  H$_2$ formation. Unfortunately, as in the
case of cosmic rays, earlier work appealed
to hydrodynamic turbulence to predict grain relative velocities 
(see \cite{Kus70,Vol80,Dra85,Oss93,Wei94,FP95}).

The differences between the hydrodynamic and MHD calculations stem from
(a) grain charges, which couple grains to the magnetic field, (b) the anisotropy
of MHD cascade, and (c) the direct interaction of charged grains with
magnetic perturbations. Effects (a) and (b) are considered
in Lazarian \& Yan \cite{LY02}, while (c) is considered in Yan \& Lazarian
(2002b; in preparation). As consequence the picture of 
grain dynamics is substantially altered.
 
Consider grain charge first.
If a grain's Larmor period \( \tau _{L}=2\pi m_{gr}c/qB \)
is shorter than the gas drag time \( t_{drag} \), grain perpendicular
motions are constrained by magnetic field. Their velocity dispersion is
determined by the turbulence eddy whose 
turnover period is \( \sim \tau _{L} \) instead of the drag time \cite{Dra85}. 

Accounting for the anisotropy of MHD turbulence it is convenient to consider 
separately
grain motions parallel and perpendicular to magnetic field. The 
perpendicular motion is 
influenced by the Alfv\'en modes, 
which have a Kolmogorov spectrum. The parallel motion is subjected to 
compressible modes which scale as  
\( v_\parallel\propto k_\parallel^{-1/2} \).
In addition we should account for viscous forces. 
When the eddy turnover time 
is of the order of \( t_{damp}\sim \nu _{n}^{-1}k_{\perp }^{-2} \), 
the turbulence is viscously damped. 
Thus grains sample only a part of the eddy before gaining the 
velocity of the ambient gas if \( \tau_{L} \) or \( t_{drag} < t_{damp} \). 
The results are shown in Fig.~\ref{fig_dust}b.

The direct interaction of the charged grains with turbulent magnetic field
results in a stochastic acceleration that can potentially
provide grains with supersonic velocities.

%\vspace{0.3cm}
%\noindent
\subsection{Turbulence in HII regions}

%{\bf Turbulence in HII regions}\\
Lithwick \& Goldreich \cite{Lit01} addressed the issue of the origin of density
fluctuations within HII regions. There the gas
pressure is larger than the magnetic pressure (the `high beta' regime)
and they conjectured that
fast waves, which are essentially sound waves, would be decoupled from
the rest of the cascade.
They found that density fluctuations are due to the slow mode 
and the entropy mode, which are passively mixed by shear Alfv\'en waves and 
follow a Kolmogorov spectrum.
They also found that slow mode density fluctuations are
proportional to $1/\sqrt{\beta}$.
On the other hand, the entropy mode density fluctuations are suppressed when
cooling is faster than the cascade time. 
Lithwick \& Goldreich \cite{Lit01} also gave detailed discussions about
density fluctuations on various scales in the ISM, 
e.g.~proton gyro-radius. These results are important
as radio-wave scintillation observations can constrain
the nature of MHD turbulence in the ISM, especially in the HII regions.
Lithwick \& Goldreich \cite{Lit01} argued that the turbulent 
cascade survives ion-neutral damping only when a high degree of
ionization is present.  However, the study by
Cho, Lazarian \& Vishniac \cite{Clv02b} suggests that the magnetic fluctuations
protrude below the damping scale and the results of \cite{Lit01}
should be revised.

%\vspace{0.3cm}
%\noindent
%{\bf Tiny-Scale Atomic Structures}\\
\subsection{Tiny-Scale Atomic Structures}

The intermittent small scale structures in \S\ref{sec_vis} should have 
important implications for transport processes (heat, cosmic rays, etc.)
in partially ionized plasmas.
We also speculate that they might have some relation to 
the tiny-scale atomic structures (TSAS).
Heiles \cite{Hei97} introduced the term TSAS
for the mysterious
H~I absorbing structures on scales from thousands to tens of
AU, discovered by Dieter, Welch \& Romney \cite{Die76}. Analogs are observed
in NaI and CaII \cite{Mey96,Fai01,And01} 
and in molecular gas 
\cite{Mar93}. 
Recently Deshpande, Dwarakanath 
\& Goss \cite{Des00d} 
analyzed channel maps of opacity fluctuations toward Cas A and Cygnus A.
They found 
that the amplitudes of density fluctuations at scales less than 0.1 pc 
are far larger than expected from extrapolation from larger scales, 
possibly explaining TSAS. This study, however, cannot answer what 
confines those presumably overpressured (but very quiescent!) blobs of 
gas. Deshpande \cite{Des00} related those structures to the shallow spectrum of
interstellar turbulence.
 
{}Figure \ref{fig_imb}b indicates 
that while velocity decreases rapidly, but {\it not}
exponentially, below the viscous damping scale,
the magnetic field fluctuations persist, 
thereby providing nonthermal pressure support. Magnetic structures 
perpendicular to the mean magnetic field are compensated by pressure 
gradients.
Our calculations so far are produced using incompressible
code \cite{Clv02b}. 
In the case of compressible media, we expect the
pressure fluctuations to entail density fluctuations
reminiscent of the Deshpande et al. 
\cite{Des00d} observations. 
 
The calculations in Cho, Lazarian \& Vishniac \cite{Clv02b} 
are applicable on scales from the viscous damping 
scale (determined by equating the energy transfer rate with the 
viscous damping rate; $\sim0.1$ pc in the Warm Neutral Medium with $n$ 
= 0.4 cm$^{-3}$, $T$= 6000 K) to the ion-neutral decoupling scale (the 
scale at which viscous drag on ions becomes comparable to the neutral 
drag; $\ll 0.1$ pc). Below the viscous scale the fluctuations of 
magnetic field obey the damped regime shown in Figure \ref{fig_imb}b and 
produce 
density fluctuations. For typical Cold Neutral Medium gas, the scale of 
neutral-ion decoupling decreases to $\sim70$AU, and is less for denser 
gas. TSAS may be created by strongly nonlinear MHD turbulence! 
 
A simple technique of estimating magnetic field was suggested by 
Chandrasekhar \& Fermi \cite{Cha53} (see also review by Ostriker, this volume).
According to it, the fluctuations of magnetic field
that can be measured from polarization maps are related to
velocity fluctuations measured through Doppler
broadening $\delta b/\sqrt{4\pi \rho} \sim \delta v$.
The existence of the damped regime of MHD turbulence
suggests that this technique is not applicable to very small scales
in partially ionized gas.

%\vspace{0.3cm}
%\noindent
%{\bf Magnetic Reconnection} \\
\subsection{Magnetic Reconnection}

Magnetic reconnection is the fundamental process that 
allows magnetic
fields to change their topology, despite being `frozen' into
highly conducting plasmas.
It is the key process for solar flares, the magnetic dynamo,
the acceleration of energetic particles, etc.
According to the Lazarian \& Vishniac model \cite{LV99} 
(see also review \cite{Lv00} and Vishniac, Lazarian, \& Cho, this volume) of
stochastic reconnection,
this process is controlled by the turbulent wandering of
magnetic field.
The exact properties of the turbulent cascade are especially important for
the viscously damped regime present in partially ionized gas. However,
it is
shown in Lazarian, Vishniac \& Cho \cite{Lvc02} that the reconnection rates
are sufficiently high in this case. The implications of the finding for
the removal of magnetic flux during star formation is to be evaluted yet.

%\vspace{0.3cm} \noindent
%{\bf Support and Compression of Molecular Clouds}\\
\subsection{Support and Compression of Molecular Clouds}

To understand the dynamics of molecular clouds and star formation
it is necessary to understand turbulence. 
In a recent review \cite{Mck99} McKee pointed out that the
fast damping of MHD turbulence observed in numerical simulations is
difficult to reconcile with the fact that ``a GMC such as G216, which
has no visible star formation, can have a level of turbulence that
exceeds that in the Rosette molecular cloud, which has an embedded OB
association''. He pointed out that the conclusions obtained on the
basis of numerics should be treated with caution as they do not
resolve the microscales. 

In typical astrophysical conditions the sources of turbulence are
localized both in space (stars) and time (stellar outflows;
supernovae), and the outgoing waves have much larger amplitudes than
the background waves (we call this situation ``imbalanced
cascade"). Fig.~\ref{fig_imb}a shows that the turbulent damping could be
substantially reduced in this situation. Moreover, even in a balanced
regime, we expect fast modes to be subjected to slow non-linear damping. 

At the same time, it worth mentioning, that turbulence can not only
support, but also compress molecular clouds.  Clouds can
be compressed by external turbulence feeding into them and depositing
energy and momentum. Myers \& Lazarian \cite{ML98} explained
observed infalling motions of molecular gas surrounding dense cores
\cite{Taf98,Lee01} in this way, based on
ion-neutral damping. 
The infall rate is proportional to the rate of turbulence damping.
Therefore, fast non-linear damping associated with the
Alfv\'{e}nic turbulence should enhance the infall.

%\vspace{0.3cm} \noindent
\subsection{Heating of Diffuse Ionized Gas}

The ``Diffuse Ionized Gas" (DIG), or equivalently the ``Reynolds 
layer" within the Milky Way, is detected by rather faint but 
ubiquitous Galactic H$\alpha$ emission \cite{Rey88,RHT99}.
Such emission is found in several other 
spirals as well \cite{Ran90,Ran98,OD99,Ott01}.
In the Galaxy, the Reynolds layer 
contains a substantial portion of the H$^+$ in the ISM.  
Current models generally involve photoionization from the OB 
stars, although how the Lyman continuum 
radiation from OB stars can penetrate the neutral H layer remains 
controversial.

The observations show strong [S\,II] $\lambda$6717 and [N~II]
$\lambda$6583 that increase relative to H$\alpha$ with distance $z$
above the planes of various galaxies, including the Milky Way
\cite{RHT99}.  The only reasonable conclusion is that there
is an additional source of heating in the ISM that dominates over
photoionization heating at low densities. It
has been proposed that carbonaceous molecules provide the excess
heating through the photoelectric effect \cite{WD01},
but this explanation is not unique.
Heating by turbulence, surely present, may dominate. Minter
\& Spangler \cite{Min97} suggested a heating rate that 
is adequate to explain the [S~II]/H$\alpha$ and [N~II]/H$\alpha$
ratios, but did not take nonlinear interactions into account, thereby
underestimating the heating. A new study that would capitalize on
the new understanding of MHD turbulence (damping of Alfv\'en
and fast modes, imbalance etc.) is on our agenda.

\section{Observational Tests}
Comparing numerics with observations is a challenging problem.
Ostriker (this volume) discusses PDF, clump identification, and
linewidth-scale relations as possible diagnostics and outlines
problems with any of those approaches.
A use of spectral line data cubes and application to it of
different techniques (e.g.~spectral correlation function,
principal component analysis, wavelets, etc.) can be found
in the review \cite{Laz99b}. Here we shall concentrate on comparing
spectra from observations with our theoretical 
expectations (see also review \cite{Lpe02}).

\subsection{Is the Big Power Law real?}

We have mentioned above that observations suggest that the Kolmogorov
power law should span from AU to kpc scales. Kolmogorov scaling
is exactly what one would expect from the GS95 picture
when the observations sample magnetic field
in the system of reference aligned with the {\it mean magnetic field}
(see Fig.~\ref{fig_new1} for examples of observable quantities).
Indeed, it is obvious from Fig.~\ref{local_frame} 
in the this system of reference (i.e.
global system of reference) the locally defined scalings of $k_{\|}$
with $k_{\bot}$ are not valid. Indeed, one can easily see that
the fluctuations perpendicular to the {\it local} direction of magnetic 
field dominate both the statistics measured perpendicular to
the {\it mean} magnetic field and parallel to it. As the
result  in the reference system aligned with the {\it mean}
magnetic field $k_{\|}'\propto k_{\perp}'$ (see Fig.~\ref{fig_new1}a,b,c)
and according to equation (\ref{GS_K})
$E(k)\sim k^{-5/3}$ will be measured. 

Ambiguities in measurements reviewed
in \cite{Arm95}, however, make it uncertain whether or
not the Big Power Law should be taken at face value.
Still, we note that agreement between an observed power spectrum
and theoretical expectations is far more significant than
just an approximate fit between observations and numerics. The latter 
is definitely not unique and is {\it a priori} suspect in view of
the huge difference in terms of $Re$ and $Rm$ between any numerical
simulation and the ISM.

To test the Big Power Law properly, it is important to extend the theory 
of Velocity Channel Analysis (VCA) \cite{LP00,Lpe02} 
by including self-absorption in the analysis of turbulent
emission lines. It can then be applied to regions of HI in the
inner Galaxy \cite{Dic01} 
or CO \cite{Stu98,Stu99}\footnote{
Brunt \& Heyer \cite{Bru01} 
applied Principal Component
Analysis (see \cite{Hey97}), to simulated CO data and found
empirical relations between the statistics of velocities, eigenvectors
and eigenimages. However, they note that somehow their relation does
not depend on the absorption coefficient for the limited range of
absorptions they tested. Unfortunately, their analysis does not seem
to be applicable to Galactic HI and the applicability of their technique
to the correlated velocity and density fields is unclear.}. 
Lazarian \& Pogosyan \cite{LP02}
provided results consistent with observations. 
For instance, the study predicts that in the presence of absorption
the emitted power in the line is proportional to $k^{-3}$, exactly 
what is seen in Dickey et al.~\cite{Dic01} for HI in the inner region of the 
Galactic plane. 

An application of the VCA to different emission lines (e.g.  H${\alpha}$, 
[N~II], [S~II] etc.)
would help to answer the question of whether or not ISM turbulence is a large
scale cascade with various phases of the ISM interconnected
through a dynamically important magnetic field.  A contrasting
possibility is that various phases form their own cascades.

HI is rather smoothly distributed
across the sky. Therefore, the effects of image edges does not pose
a problem for the statistical analysis. Results by Stanimirovic
(private communication) show that the Fourier analysis of the SMC
image and a more laborious wavelet analysis\footnote{Wavelet analysis
involves determining the deviation of each pixel from a weighted
average of the pixels at a particular projected distance from it.}
provide identical results. However, when dealing with molecular clouds
we might expect that the cloud edges become important.  Therefore the
incorporation of wavelets (see \cite{Zie99}) within the
VCA is a natural step to make. The only difference would be to apply
the wavelets instead of Fourier transforms to the channel maps.

Synchrotron fluctuations and fluctuations of polarized radiation
arising from aligned dust should be used to study magnetic
field statistics. As we mentioned earlier 
Cho \& Lazarian \cite{CL02b} has shown that those fluctuations
are consistent with Kolmogorov scaling. More studies in this direction
are necessary. The fact that those fluctuctuations interfere with the
CMB studies garantees that in the near future we shall have a lot
of relevant data.

One should remember, however, that the measured power index of fluctuations
may not correspond to the spectral index of the underlying turbulence.
For instance, it is shown in \cite{Sta01} that while the actual turbulence
in SMC is close to being Kolmogorov, depending on the thickness of the
slice, the spectral index of intensity fluctuations within channel maps
span from $\sim -2.8$ to $\sim -3.4$. Similarly, it is shown in
\cite{CL02b} that for Kolmogorov turbulence 
the spectral index of observable fluctuations may vary
from $\sim -1$ to $-3.7$ depending on how observations sample turbulence.

\subsection{Does turbulence reveal magnetic field direction?}

Anisotropy of Alfv\'{e}nic turbulence is a definite prediction of
the GS95 theory. We mentioned earlier that the scale-dependent
anisotropy can only be revealed in the local frame of reference,
which in practical terms require direct measurements, e.g. with
spacecraft. Measurements of the Solar wind magnetic turbulence
have failed so far to reveal the differential scaling of the turbulence
in terms of $k_{\|}$ and $k_{\bot}$, but these measurements are
inconclusive \cite{Hor99}. If measurement are performed in a global system
of reference, as is the case with observations, they should reveal
anisotropy in the direction of the mean magnetic field.

\begin{figure}[t]
\begin{tabbing}
\includegraphics[width=.5\columnwidth]{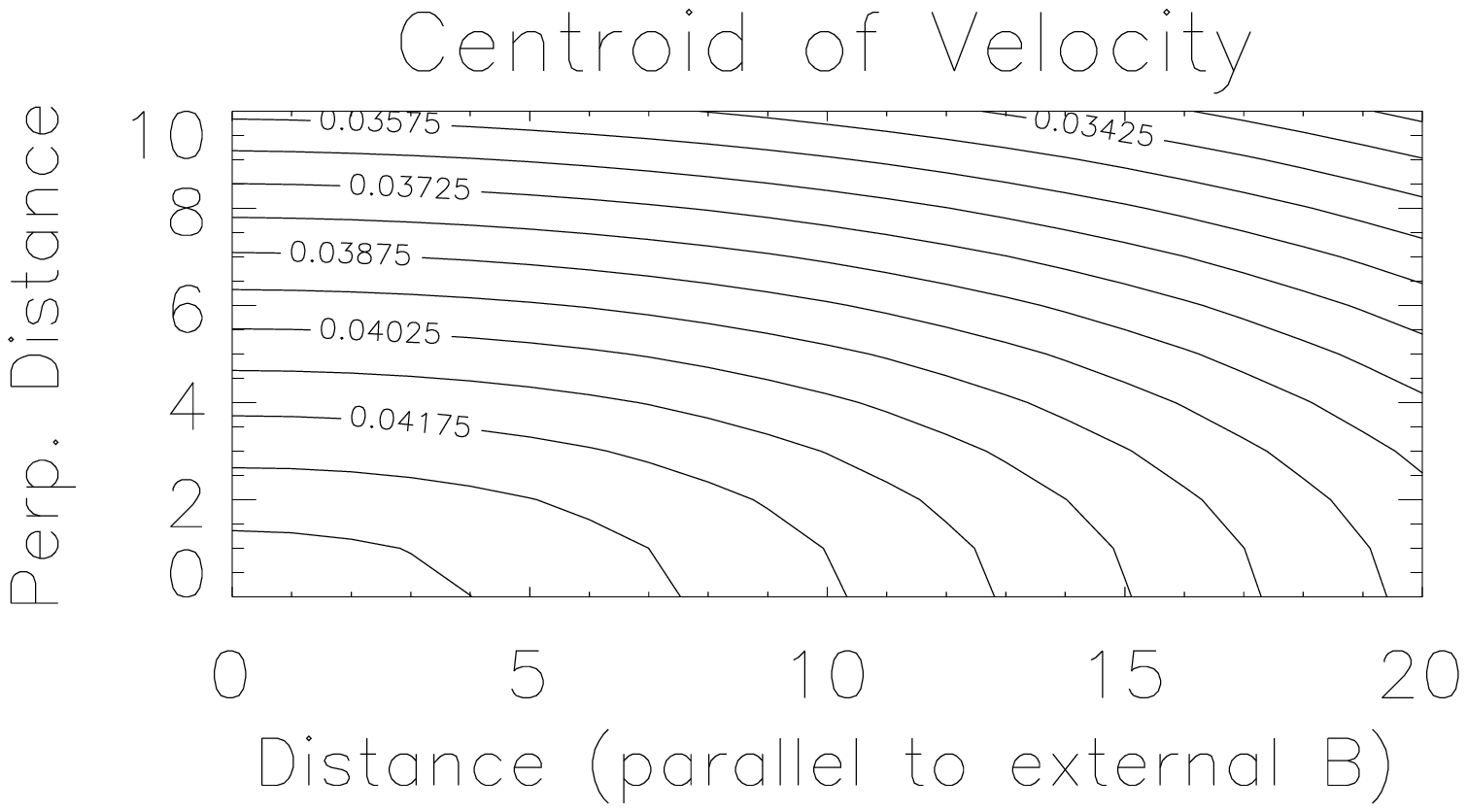} 
\=
\includegraphics[width=.5\columnwidth]{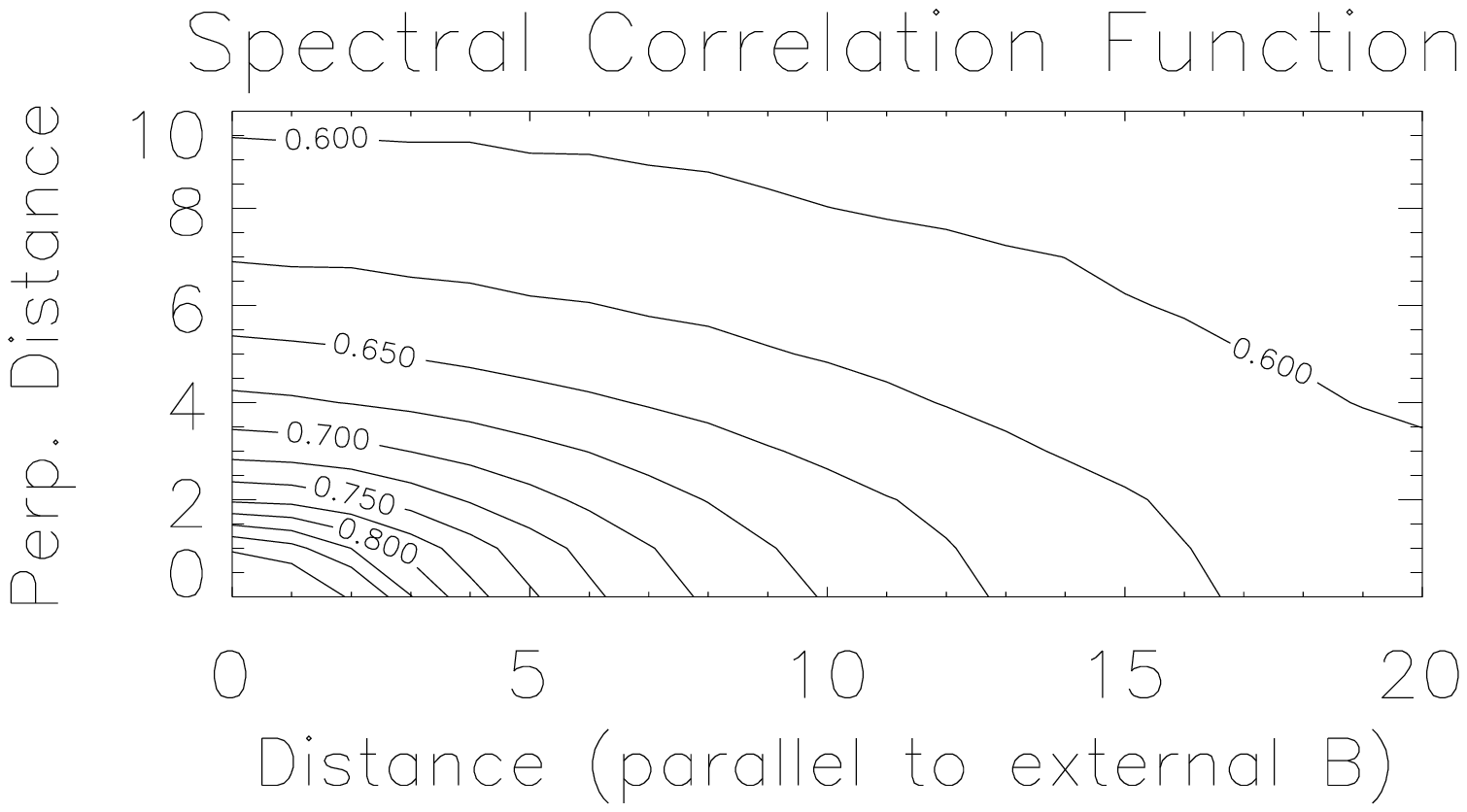}
\\
~~~~~~~~~~~~~~~~~~~~~(a) \> ~~~~~~~~~~~~~~~~~~~~~~~~~~~(b) \\ 
\includegraphics[width=.5\columnwidth]{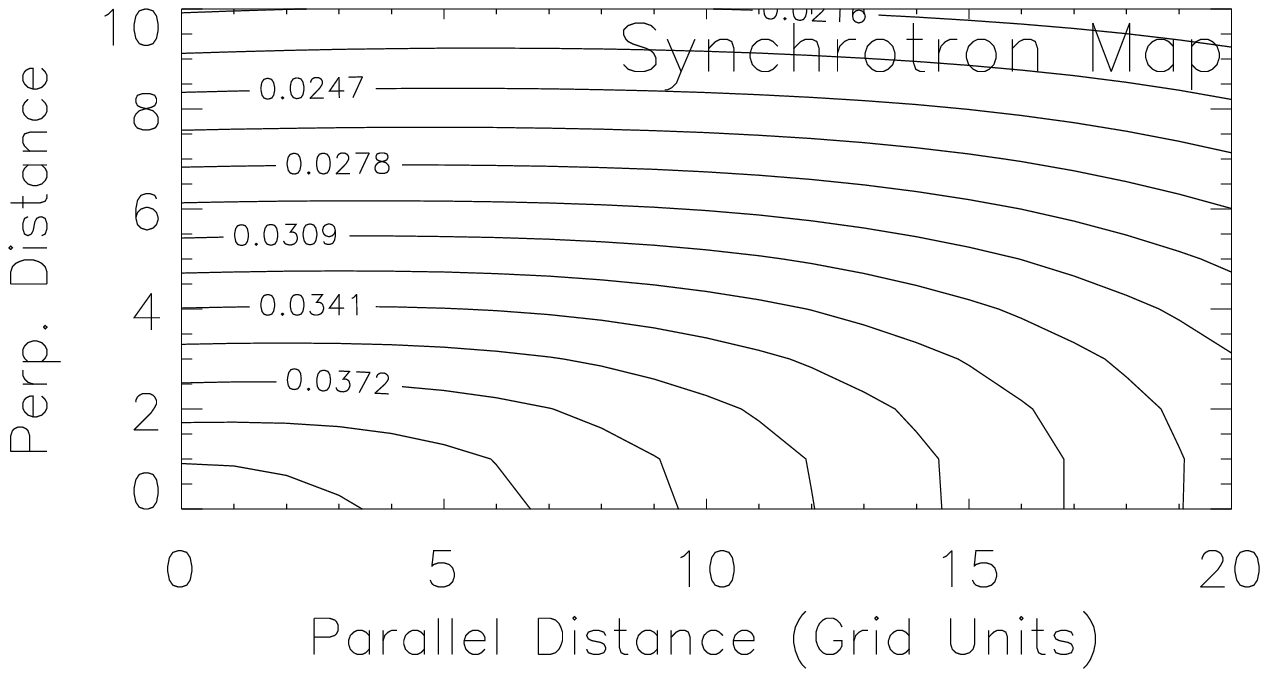}
\>
\includegraphics[width=.5\columnwidth]{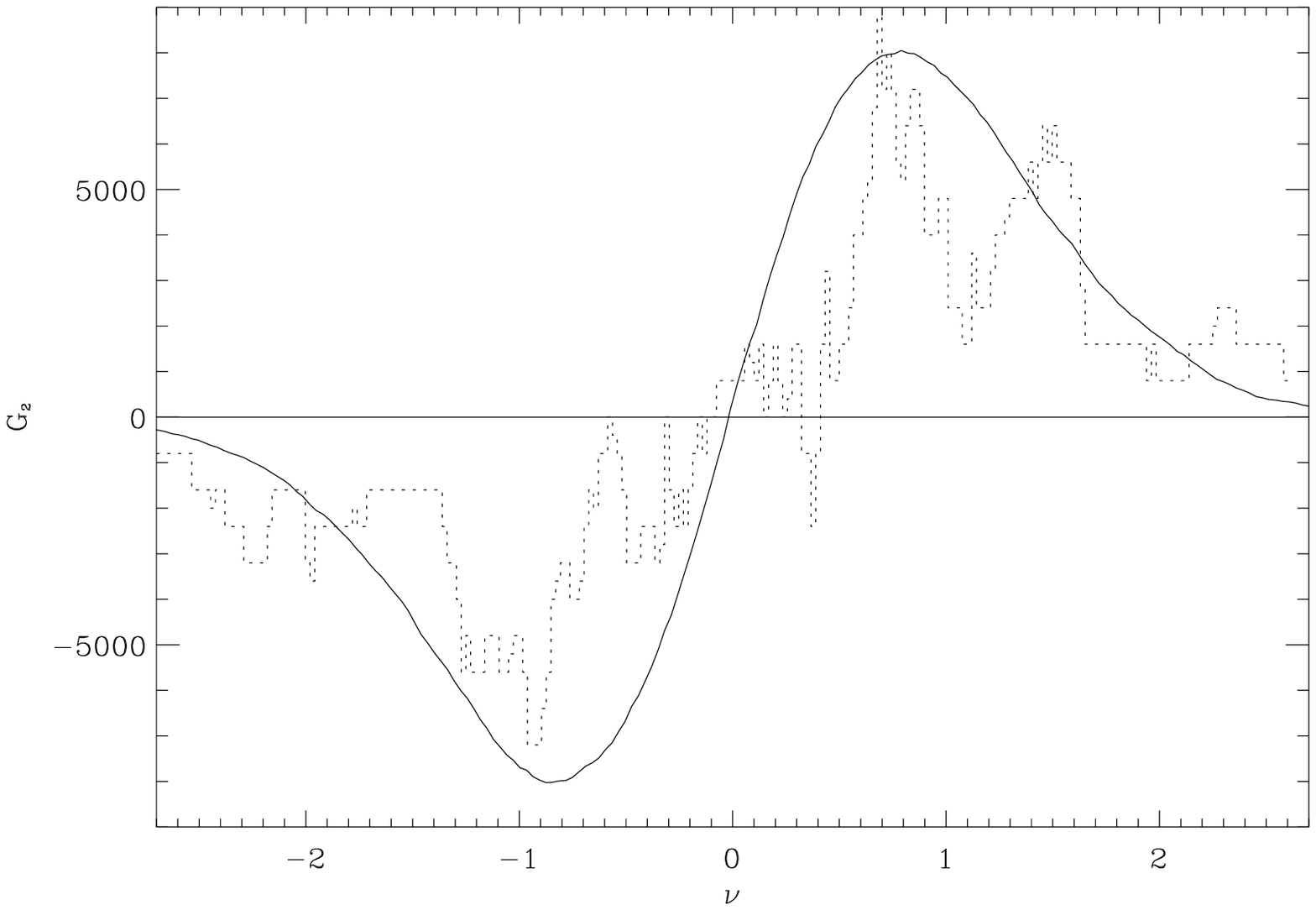}
\\
~~~~~~~~~~~~~~~~~~~~~(c) \> ~~~~~~~~~~~~~~~~~~~~~~~~~~~(d) \\
\end{tabbing}
\caption{{\it 
     Observational test of the synthetic data by Cho, Esquivel} \&
     {\it Lazarian}.
     Contours of equal correlation obtained with
     Centroids of Velocity ({\it (a)})
     and with Spectral Correlation Function (SCF)({\it (b)}). 
     The direction of anisotropy
     reveals the direction of
     projected
     magnetic field. 
     Combined with the anisotropy analysis, the SCF (introduced by 
     Alyssa Goodman) is likely to become even more useful tool.
         {\it (c)} Contours of equal correlation obtained for
                      synthetic synchrotron intensity map. 
         {\it (d)} shows the 2D genus of the Gaussian 
                distribution 
               (smooth analytical curve) 
                against the genus for the isothermal compressible MHD 
                simulations with Mach 
                number $\sim$2.5 (dotted curve).  
} 
\label{fig_new1} 
\end{figure}

In isotropic turbulence, correlations depend only on the distance
between sampling points. Contours of equal correlation are 
circular in this case. The presence of a magnetic field introduces anisotropy
and these contours become elongated with a symmetry axis given by
the magnetic field.  To study turbulence
anisotropy, we can measure contours of equal correlation corresponding
to the data within various
velocity bins. The results obtained with simulated data are shown 
in Fig.~\ref{fig_new1}. 

%%%%%%%%%%%%%%%%%%%%%%%%%%%%%%%%% 
%%%%%%%%%%%%%%%%%%%%%%%%%%%%%%%%%

Since the degree of anisotropy is related
to the strength of the magnetic field, studies of anisotropy can 
provide the means to analyze
magnetic fields. It is important to study different data sets and channel
maps for the anisotropy. Optical and infrared polarimetry
can benchmark the anisotropies in correlation
functions. We hope that 
anisotropies will reveal magnetic field structure  within dark clouds where grain
alignment and therefore polarimetry fails (see \cite{Laz00a} 
for a review of grain alignment).

Not only velocity statistics can be used for such an analysis. Lazarian \& Shutenkov
\cite{Laz90} (see review \cite{Laz92}) showed that the mean magnetic field 
must lead to 
anisotropies in the synchrotron statistics. Lazarian \& Chibisov \cite{Laz91}
pointed out that using HI regions as screens for radiation at the decameter
wavelength it should be possible to study the 3D distribution of the magnetic field.
{}Fig.~\ref{fig_new1}c   
shows the anisotropy of synchrotron statistics available through
simulations. 

%\vspace{0.3cm}
%\noindent{\bf How Else Can We Compare Observations and Simulations?}\\ 

\subsection{How else can we compare observations and simulations?}

 Velocity and density power spectra do not provide a complete description
of turbulence. Intermittency (variations in the strength of the turbulent
cascade) and its topology in the presence of different phases are 
not described by the power spectrum. Use of the higher moments is possible
(see discussion of the 3 point statistics, or bispectrum, in Lazarian
\cite{Laz99b}), but is limited by the noise in the observational data.
 
``Genus analysis" is a good tool for studying the topology of
turbulence (see the review \cite{Lpe02}).  This
tool has already been successfully applied to cosmology
\cite{Got89}. Consider an
area on the sky with contours of projected density. The 2D genus,
$G(\nu)$, is the difference between the number of regions with a
projected density higher than $\nu$ and those with densities lower
than $\nu$. Fig.~\ref{fig_new1}d shows the 2D genus as the function of $\nu$
for a Gaussian distribution of densities (completely symmetric curve),
for MHD isothermal
simulations with Mach number $\sim$2.5.
It is shown in \cite{Lpe02} that the genus of the Small Magellanic Cloud is 
very different from that in Fig.~\ref{fig_new1}d, while the spectra
in both cases are similar.

\section{Summary}

Recently there have been significant advances in the field of
MHD turbulence:
\begin{enumerate}
 
\item
The first self-consistent model (GS95) of incompressible MHD turbulence
that is supported by both numerical simulations and observations is 
now available.
The major predictions of the model are scale-dependent anisotropy 
($k_{\|}\propto k_{\perp}^{2/3}$) and
a Kolmogorov energy spectrum ($E(k)\propto k^{-5/3}$).

\item
There have been substantial advances 
in understanding compressible MHD.
Simulations of compressible MHD turbulence show that
there is a weak coupling between Alfv\'en waves and compressible MHD waves and
that the Alfv\'en modes follow the Goldreich-Sridhar scaling.
{}Fast modes, however, decouple and exhibit isotropy.

\item
Contrary to general belief, in typical interstellar
conditions,
magnetic fields can have rich structures below the scale
at which motions are damped by the viscosity created by neutral
drag (the ambipolar diffusion damping scale).

\item
These advances will have a dramatic impact on our understanding of many
fundamental interstellar processes, like 
cosmic-ray propagation, grain dynamics, turbulent heating
and molecular cloud stability. 

\item
New techniques, e.g.~VCA, allow observational tests of
the theory.
\end{enumerate}

{\bf Acknowledgments:}
We thank Peter Goldreich, John Mathis, Steven Shore, Enrique Vazquez-Semadeni,
and Huirong Yan for
helpful discussions.
AL and JC acknowledge the support of the NSF through grant
AST-0125544. ETV acknowledges NSF grant AST-0098615. 
This work was partially supported by National Computational Science
Alliance under AST000010N and AST010011N
and utilized the NCSA SGI/CRAY Origin2000.
AL thanks the LOC for their financial support. The authors thank
the editors for their patience with the manuscript.


\begin{thebibliography}{8.}
\addcontentsline{toc}{section}{References}


\bibitem{And01} S.~M. Andrews, D.~M. Meyer, J.~T. Lauroesch: 
                Astrophys.~J. \textbf{552}, L73 (2001)

\bibitem{Arm95} J.~W. Armstrong, B.~J. Rickett, S.~R. Spangler:
                Astrophys.~J. \textbf{443}, 209 (1995)

\bibitem{BBF01} C. Baccigalupi, C. Burigana, F. Perrotta, G. De Zotti,
                L. La Porta, D. Maino, M. Maris, R. Paladini:
                Astron.~Astrophys. \textbf{372} 8 (2001)


\bibitem{Bha98} A. Bhattacharjee, C.~S. Ng, S.~R. Spangler: 
                Astrophys.~J. \textbf{494}, 409 (1998)

\bibitem{Bis93} D. Biskamp: {\it Nonlinear Magnetohydrodynamics}
                (Cambridge Uinversity Press, Cambridge, 1993)
\bibitem{Bis02} D. Biskamp: Phys.~Plasmas, \textbf{9}(4), 1486 (2002)
\bibitem{Bis01} D. Biskamp, E. Schwarz:
                Phys. Plasmas, \textbf{8}(7), 3282 (2001)

\bibitem{BSC98} D. Biskamp, E. Schwarz, A. Celani:
                Phys.~Rev.~Lett. \textbf{81} 4855 (1998)

\bibitem{Bol02} S. Boldyrev: Astrophys.~J. \textbf{569}, 841

\bibitem{Bol01} S. Boldyrev, A. Nordlund, P. Padoan: astro-ph/0111345 (2002)

\bibitem{Bra01} A. Brandenburg:
                Astrophys.~J. \textbf{550}, 824 (2001)
\bibitem{BJN96} A. Brandenburg, R.~L.~Jennings, A. Nordlund:
                J.~Fluid Mech. \textbf{306}, 325 (1996)

\bibitem{Bri88} M. Brio, C. Wu: 
                J.~Comput.~Phys. \textbf{75}, 500, (1988) 
 
\bibitem{Bru01} C. Brunt, M. Heyer: 
                Astrophys.~J. in press (2002) (astro-ph/0110155)

\bibitem{Cat96} F. Cattaneo, D.~W. Hughes:
                Phys.~Rev.~E. \textbf{54}, R4532 (1996)

\bibitem{Ces80} C. Cesarsky:
                Annu.~Rev.~Astro.~Astrophys. \textbf{18}, 289 (1980)
\bibitem{Cha01} B. Chandran:
                Phys.~Rev.~Lett. \textbf{85}(22), 4656 (2001)
\bibitem{Cha53} S. Chandrasekhar, E. Fermi:
                Astrophys.~J.~\textbf{118}, 113 (1953)
\bibitem{Cv00g} J. Cho, E.~T, Vishniac:
                Astrophys.~J. \textbf{538}, 217 (2000a)
\bibitem{Cv00a} J. Cho, E.~T. Vishniac: 
                Astrophys.~J. \textbf{539}, 273 (2000b)

\bibitem{Cl02} J. Cho, A. Lazarian: Phys.~Rev.~Lett. accepted (2002)
astro-ph/0205282
\bibitem{CL02b} J. Cho, A. Lazarian: Astrophys.~J. submitted (2002)
astro-ph/0205284
\bibitem{Clv02a} J. Cho, A. Lazarian, E.~T. Vishniac: 
                Astrophys.~J. \textbf{564}, 291 (2002a) (CLV02a)
\bibitem{Clv02b} J. Cho, A. Lazarian, E.~T. Vishniac: 
                Astrophys.~J. \textbf{566}, L49 (2002b)

\bibitem{Del01} L. Del Zanna, M. Velli, P. Londrillo: Astron.~Astrophys. 
                \textbf{367}, 705, (2001)
\bibitem{Der78} N.~F. Derby: Astrophys.~J. \textbf{224}, 1013, (1978)

\bibitem{Des00} A.~A. Deshpande:
                Monthly Not.~Roy.~Astron.~Soc. \textbf{317}, 199 (2000)
\bibitem{Des00d} A.~A. Deshpande, K.~S. Dwarakanath, W.~M. Goss: 
                Astrophys.~J. \textbf{543}, 227 (2000)

\bibitem{Dia90} P.~H. Diamond, G.~G. Craddock: 
                Comments on the Plasma Physics of Controlled Fusion, 
                \textbf{13}, 287 (1990)

\bibitem{Dic01} J.~M. Dickey, N.~M. McClure-Griffiths,
                S. Stanimirovic, B.~M. Gaensler, A.~J. Green:
                Astrophys.~J. \textbf{561}, 264 (2001) 

\bibitem{Dic85} R.~L. Dickman:  
                in  {\em Protostars and Planets II}, 
                ed. by  D.~C. Black, M.~S. Mathews 
                (Tucson: Univ. Arizona Press, 1985) p.150

\bibitem{Die76} N.~H. Dieter, W.~J. Welch, J.~D. Romney:
                Astrophys.~J. \textbf{206}, L113 (1976)

\bibitem{Dra85} B.~T. Draine: 1985, 
                in {\it Protostars and Planets II}, 
                ed. by D.~C. Black, M.~S. Matthews 
               (Tucson: Univ. Arizona Press, 1985) p. 621

\bibitem{Dra99} B.~T. Draine, Lazarian, A.: Astrophys.~J. \textbf{512}, 740
                (1999) 

\bibitem{Elm01} B.~G. Elmegreen:
                in {\it From Darkness to Light},
                ed. by T. Montmerle, P. Andre
                (ASP Conf.~ Series, in press, 2001) (astro-ph:0010582)

\bibitem{Fai01} M.~D. Faison, W.~M. Goss:
                Astron.~J. \textbf{121}, 2706 (2001)
\bibitem{Fal95} E. Falgarone, G. Pineau des Forets, E. Roueff:
                Astron.~Astrophys. \textbf{300}, 870 (1995)
\bibitem{FP95}  E. Falgarone, J.~L.~Puget,
                Astron.~Astrophys. \textbf{293}, 840 (1995)


\bibitem{Fou82} J. Fournier, P. Sulem, A. Pouquet:
                J.~Phys.~A, \textbf{15}, 1393 (1982)
  

\bibitem{Fos02} P. Fosalba, A. Lazarian, S. Prunet, J.~A. Tauber:
                Astron.~J. \textbf{564}, 762 (2002)

\bibitem{Fri95} U. Frisch: {\it Turbulence: the legacy of A.N. Kolmogorov},
                (Cambridge Univ. Press, New York, 1995)
\bibitem{Gal00} S. Galtier, S.~V. Nazarenko, A.~C. Newell, A. Pouquet: 
                J.~Plasma Phys. \textbf{63}, 447 (2000)


\bibitem{Gia01} G. Giardino, A.J. Banday, P. Fosalba, K.M. G\'{o}rski,
             J.L. Jonas, W. O'Mullane, J. Tauber:
             Astron.~Astrophys. \textbf{371} 708 (2001)
\bibitem{Gia02} G. Giardino, A.J. Banday, K.M. G\'{o}rski, K. Bennett,
             J.L. Jonas, J. Tauber:
             Astron.~Astrophys. astro-ph/0202520 (2002)




\bibitem{Gol95} P. Goldreich, H. Sridhar:
                Astrophys.~J. \textbf{438}, 763 (1995) (GS95)
\bibitem{Gol97} P. Goldreich, H. Sridhar:
                Astrophys.~J. \textbf{485}, 680 (1997)

\bibitem{Gol78} M.~L. Goldstein: Astrophys.~J. \textbf{219}, 700 (1978)

\bibitem{Gold95} M.~L. Goldstein, D.~A. Roberts:
                 Annu.~Rev.~Astron.~Astrophys. \textbf{33}, 283, (1995)

\bibitem{Goo85} J. Goodman, R. Narayan:
                Monthly Not.~Roy.~Astron.~Soc. \textbf{214}, 519 (1985)


\bibitem{GMM88} S. Ghosh, W.~M. Matthaeus, D.~C. Montgomery:
                Phys.~Fluids, \textbf{31}, 2171 (1988)

\bibitem{Got89} J.~R. Gott, C. Park, R. Juskiewicz, W. Bies, F. Bouchet,
                A. Stebbins: Astrophys.~J. \textbf{352}, 1 (1990)


\bibitem{Gre93} D.~A. Green: Monthly Not.~Roy.~Astron.~Soc. \textbf{262}, 
                    328 (1993)

\bibitem{Gru94} A.~V. Gruzinov, P.~H. Diamond:
                Phys.~Rev.~Lett. \textbf{72}, 1671 (1994)

\bibitem{Hei97} C. Heiles: Astrophys.~J. \textbf{481}, 193 (1997)
\bibitem{Hey97} M.~H. Heyer, F.~P. Schloerb:
                Astrophys.~J. \textbf{475}, 173 (1997)


\bibitem{Hig84} J.~C.Higdon: Astrophys.~J. \textbf{285}, 109 (1984)

\bibitem{Hor99} T.~S. Horbury:
                in {\it Plasma Turbulence and Energetic particles}, 
                ed. by M. Ostrowski, R. Schlickeiser
                (Cracow, Poland, 1999) p.28

\bibitem{Hor97} T.~S. Horbury, A. Balogh:
                Nonlin.~Proc.~Geophys. \textbf{4}, 185 (1997)


\bibitem{Hor51} S. von Horner: Zs.F.~Ap. \textbf{30}, 17 (1951)

\bibitem{HGPM95} M. Hossain, P.~C. Gray, D.~H. Pontius, W.~H. Matthaeus:
                 Phys.~Fluids, \textbf{7}, 2886 (1995)


\bibitem{Iro63} P. Iroshnikov: Astron.~Zh. \textbf{40}, 742 (1963)
               (English: Sov.~Astron. \textbf{7}, 566 (1964))

\bibitem{JH93} V. Jayanti, J.~V. Hollweg: J.~Geophys.~Res. \textbf{98}, 13247
               (1993)


\bibitem{JFD98} K. Joulain, E. Falgarone, G. Pineau des Forets, D. Flower:
                Astro.~Astrophys. \textbf{340}, 241 (1998)
%%%\bibitem{Jon01} Jones et al. (2001)

\bibitem{Kam55} J. Kamp\'{e} de F\'{e}riet: 
                in: {\em Gas Dynamics of Cosmic Clouds}
                (Amsterdam: North-Holland, 1955) p.134

\bibitem{Kap70} S.~A. Kaplan, S.~B. Pickelner:
               {\it The Interstellar Medium} (Harvard Univ. Press, 1970)


\bibitem{Kle93} L. Klein, R. Bruno, B. Bavassano, H. Rosenbauer:
                J.~Geophys.~Res. \textbf{98}, 17461 (1993)

\bibitem{Kle01} R.~S. Klessen: astro-ph/0106332 (2001)

\bibitem{Kol41} A. Kolmogorov:
                Dokl.~Akad.~Nauk SSSR, \textbf{31}, 538 (1941)

\bibitem{Kot00} J. Kota, J.~R. Jokipii:
                Astrophys.~J. \textbf{531}, 1067 (2000)


\bibitem{Kra59} R. Kraichnan: J.~Fluid Mech. \textbf{5}, 497 (1959)
\bibitem{Kra65} R. Kraichnan:
                Phys.~Fluids \textbf{8}, 1385 (1965)
\bibitem{kr80} F. Krause, \& K.H. Radler: {\it 
               Mean-Field Magnetohydrodynamics 
               and Dynamo Theory} (Oxford: Pergamon Press 1980)
\bibitem{Kul92} R.~M. Kulsrud, S.~W. Anderson: 
                Astrophys.~J. \textbf{396}, 606 (1992)

\bibitem{Kus70} T. Kusaka, T. Nakano, C. Hayashi:
                Prog.~Theor.~Phys.\textbf{44}, 1580 (1970)


\bibitem{Lar81} R.~B. Larson:
                Monthly Not.~Roy.~Astron.~Soc. \textbf{194}, 809 (1981)

\bibitem{Laz92} A. Lazarian:
                Astron.~and Astrophys.~Transactions, \textbf{3}, 33 (1992)

%\bibitem{Laz94} A. Lazarian:
%                Plasma Physics and Controlled Fusion \textbf{36}, 1013 (1994)

\bibitem{Laz95} A. Lazarian:
                Astron.~Astrophys. \textbf{293}, 507 (1995)

\bibitem{Laz99a} A. Lazarian:
                in 
               {\it Interstellar Turbulence}, ed. by J. Franco,  
                A. Carraminana (Cambridge Univ. Press, 1999a) p.95
                (astro-ph/9804024)
\bibitem{Laz99b} A. Lazarian: 
                in {\it Plasma Turbulence and Energetic Particles},
                ed. by M. Ostrowski, R. Schlickeiser  
                (Cracow, 1999b) p.28, (astro-ph/0001001)
\bibitem{Laz00a} A. Lazarian: 2000, 
                in {\it Cosmic Evolution and Galaxy Formation},
                ASP v.215, ed. by J. Franco, 
                E. Terlevich, O. Lopez-Cruz, I. Aretxaga 
               (Astron.~Soc.~Pacific,2000) p.69 
                 (astro-ph/0003414)

\bibitem{Laz91} A. Lazarian, G. Chibisov:
                Sov.~Astron.~Lett. \textbf{17}(3), 208 (1991)

\bibitem{LGM97} A. Lazarian, A. Goodman, P. Myers:
                Astrophys.~J. \textbf{490}, 273 (1997)

\bibitem{Laz90} A. Lazarian, V.~R. Shutenkov: 
                Sov.~Astron.~Lett. \textbf{16}(4), 297 (1990)
\bibitem{LP00}  A. Lazarian, D. Pogosyan:
                Astrophys.~J. \textbf{537}, 720L (2000)
\bibitem{LP02}  A. Lazarian, D. Pogosyan: (2002), in preparation
\bibitem{Lpe02} A. Lazarian, D. Pogosyan, A. Esquivel: 2002, 
                in {\it Seeing Through the Dust}, 
                ed. by R. Taylor, T. Landecker, A. Willis  
                (ASP Conf.~Series, 2002), in press (astro-ph/0112368) (LPE02)

\bibitem{Lpvp01} A. Lazarian, D. Pogosyan, E. Vazquez-Semadeni, B. Pichardo:
                Astrophys.~J. \textbf{555}, 130 (2001)
        
\bibitem{Laz02} A. Lazarian, S. Prunet:
                in {\it Astrophysical Polarized Backgrounds}, 
                ed. by S. Cecchini, S. Cortiglioni, R. Sault, C. Sbarra
                (2002) (astro-ph/0111214)
\bibitem{LV99}  A. Lazarian, E.~T. Vishniac: 
                Astrophys.~J. \textbf{517}, 700 (1999)

\bibitem{Lv00}  A. Lazarian, E.~T. Vishniac: 
                in {Astrophysical Plasmas: Codes, Models, and Observations},
                eds. by J. Arthur, N. Brickhouse, \& J. Franco,
                RevMexAA (Serie de Conferencias), Vol. {\bf 9}, 55

\bibitem{Lvc02} A. Lazarian, E.~T. Vishniac, J. Cho: 2002, in preparation

\bibitem{LY02} A. Lazarian, H. Yan:
                Astrophys.~J. \textbf{566}, 105 (2002)
%\bibitem{LY02} A. Lazarian, H. Yan: Best of Science, submitted (2002)

\bibitem{Lea98} R.~J. Leamon, C.~W. Smith, N.~F. Ness, W.~H. Matthaeus: 
                J.~Geophys.~Res. \textbf{103}, 4775 (1998)

\bibitem{Lee01} C.~W. Lee, P.~C. Myers, M. Tafalla:
                Astrophys.~J.~Supp. \textbf{136}, 703 (2001)

\bibitem{Les90} M. Lesieur:
                {\it Turbulence In Fluids} (Dordrecht: Kluwer, 1990)

\bibitem{Lit01} Y. Lithwick, P. Goldreich: 
               Astrophys.~J. \textbf{562}, 279 (2001)

\bibitem{Lvo00} V.~S.~L'vov, Y.~V.~L'vov, A.~Pomyalov:
                Phys.~Rev.~E, \textbf{61}, 2586 (2000)

\bibitem{Mac98} M. Mac Low: Phys.~Rev.~Lett. \textbf{80}, 2754 (1998)

\bibitem{Mac99} M. Mac Low: Astrophys.~J. \textbf{524}, 169 (1999)

\bibitem{Mar01b} J. Maron, S. Cowley: astro-ph/0111008 (2001)

\bibitem{Mar01} J. Maron, P. Goldreich: Astrophys.~J. \textbf{554}, 1175 (2001)


\bibitem{Mar93} A.~P. Marscher, E.~M. Moore, T.~M. Bania:
                Astrophys.~J. \textbf{419}, L101 (1993)


\bibitem{MGM83}   W.~M. Matthaeus, M.~L. Goldstein, D.~C. Montgomery:
                Phys.~Rev.~Lett. \textbf{51},   1484 (1983)
\bibitem{Mat98} W.~M. Matthaeus, S. Oughton, S. Ghosh, M. Hossain:
             Phy.~Rev.~Lett. \textbf{81}, 2056 (1998)
\bibitem{Mat96} W.~M. Matthaeus, S. Ghosh, S. Oughton, D.~A. Roberts:
                J.~Geophys.~Res. \textbf{101}, 7619 (1996)


\bibitem{Mck99} C.~F. McKee:
                in {\it The Origin of Stars and Planetary 
                    Systems}, ed. by J.~L. Charles, D.~K. Nikolaos 
                (Dordrecht: Kluwer, 1999) p.29 

\bibitem{Men81} M. Meneguzzi, U. Frish, A. Pouquet:
                Phys.~Rev.~Lett. \textbf{47}, 1060 (1981)

\bibitem{Mey96} D.~M. Meyer, J.~C. Blades:
                Astrophys.~J. \textbf{464}, L179 (1996)

\bibitem{Mil01} L.~J. Milano, W.~H. Matthaeus, P. Dmitruk, D.~C. Montgomery:
                Phys.~Plasmas, \textbf{8}(6), 2673 (2001)

\bibitem{Min97} A. Minter, S. Spangler:
                Astrophys.~J. \textbf{485}, 182 (1997)

\bibitem{Mof78} H.~K. Moffatt:
                {\it Magnetic Field Generation in Electrically 
                Conducting Fluids} (Cambridge: Cambridge Univ. Press, 1978)

\bibitem{Mon75} A.~S. Monin, A.~A. Yaglom: 
                {\it Statistical Fluid Mechanics:
                     Mechanics of Turbulence}, 
                Vol. 2 (Cambridge: MIT Press, 1975)

\bibitem{Mon82} D.~C. Montgomery:
                Physica Scripta, \textbf{T2/1}, 83 (1982)
\bibitem{Mon95} D.~C. Montgomery, W.~H. Matthaeus:
                Astrophys.~J. \textbf{447}, 706 (1995)
\bibitem{Mon81} D.~C. Montgomery, L. Turner:
                Phys.~Fluids \textbf{24}(5), 825 (1981)

\bibitem{Mul00} W.-C. M\"uller, D. Biskamp:
                Phys.~Rev.~Lett. \textbf{84}(3) 475 (2000)

\bibitem{Mun58} G. Munch: Rev.~Mod.~Phys. \textbf{30}, 1035 (1958)



\bibitem{Mye83} P.~C. Myers:
                Astrophys.~J. \textbf{270}, 105 (1983)
\bibitem{Mye99} P.~C. Myers:
                in {\it The Origin of Stars and Planetary Systems},
                ed. by J.~L. Charles, D.~K. Nikolaos
                (Dordrecht: Kluwer, 1999) p.67
\bibitem{ML98}  P.~C. Myers, A. Lazarian:
                Astrophys.~J.~Lett. \textbf{507}, 157 (1998) 

\bibitem{Nar89} R. Narayan, J. Goodman:
                Monthly Not.~Roy.~Astron.~Soc. \textbf{238}, 963 (1989)
\bibitem{Ng96} C.~S. Ng, A. Bhattacharjee:
                Astrophys.~J. \textbf{465}, 845 (1996)
\bibitem{NBJ92} A. Nordlund, A. Brandenburg, R. Jennings, M. Rieutord, 
                J. Ruokolainen, R. Stein, I. Tuominen:
                Astrophys.~J. \textbf{392}, 647 (1992)

\bibitem{Oss93} V. Ossenkopf:
                Astron.~Astrophys. \textbf{280}, 617 (1993)


\bibitem{Ost99} E. C. Ostriker, C. F. Gammie, J. M. Stone:
                Astrophys.~J. \textbf{513}, 259 (1999)


\bibitem{Ost01} E. C. Ostriker, J. M. Stone, C. F. Gammie:
                Astrophys.~J. \textbf{546}, 980 (2001)

\bibitem{OD99}  B. Otte, R.-J. Dettmar:
                Astron.~Astrophys. \textbf{343}, 705 (1999)

\bibitem{Ott01} B. Otte, R.~J. Reynolds, J.~S. Gallagher III, A.~M.~N. Ferguson:
                Astrophys.~J. \textbf{560}, 207O (2001)

\bibitem{Oug94} S. Oughton, E.~R. Priest, W.~H. Matthaeus:
                J.~Fluid Mech. \textbf{280}, 95 (1994)

\bibitem{Pad99} P. Padoan, A. Nordlund:
                Astrophys.~J. \textbf{526}, 279 (1999)
\bibitem{Pad01} P. Padoan, A. Goodman, B.~T. Draine, M. Juvela, A. Nordlund,
                O. Rognvaldsson: Astrophys.~J. \textbf{559}, 1005 (2001)


\bibitem{Par79} E.~N. Parker: 
                {\it Cosmical magnetic fields: Their origin and their activity}
                (Oxford University Press, New York, 1979)


\bibitem{Pas88} T. Passot, A. Pouquet, P. Woodward:
                Astron.~Astrophys. \textbf{197}, 228 (1988)
\bibitem{Pas95} T. Passot, E. Vazquez-Semadeni, A. Pouquet:
                Astrophys.~J. \textbf{455}, 536 (1995)



\bibitem{Pol95} H. Politano, A. Pouquet:
                Phys. Rev. E \textbf{52}(1), 636 (1995)

\bibitem{Pol95c} H. Politano, A. Pouquet, V. Carbone: 
                 Europhys.~Lett. 43(5), 516 (1995)

\bibitem{Pol95s} H. Politano, A. Pouquet, P.~L. Sulem:
                Phys.~Plasmas \textbf{2}(8), 2931 (1995)



\bibitem{Por98} D. Porter, P. Woodward, A. Pouquet:
                Phys.~Fluids \textbf{10}, 237 (1998)

\bibitem{Pou76} A. Pouquet, U. Frish, J. L\'{e}orat:
                J.~Fluid Mech. \textbf{77}, 321 (1976)
\bibitem{Pou78} A. Pouquet, G.~S. Patterson:
                J.~Fluid Mech. \textbf{85}, 305 (1978)

\bibitem{Pou99} A. Pouquet: 
                in {\it Interstellar Turbulence}, 
                ed. by J. Franco, A. Carraminana
                (Cambridge Univ. Press, 1999) p.87
\bibitem{Ran90} R.~J. Rand, S.~R. Kulkarni, J.~J. Hester:
                Astrophys.~J. \textbf{352}, 1 (1990)

\bibitem{Ran98} R.~J. Rand:
                Pub.~Astro.~Soc.~Aust. \textbf{15}, 106 (1998)

\bibitem{Rey88} R.~J. Reynolds:
                Astrophys.~J. \textbf{333}, 341 (1988)

\bibitem{RHT99} R.~J. Reynolds, L.~M. Haffner, S.~L. Tufte:
                Astrophys.~J. \textbf{525}, 21 (1999)

\bibitem{Ros76} M.~N. Rosenbluth, D.~A. Monticello, H.~R. Strauss, R.~B. White:
                Phys.~Fluids \textbf{19}, 1987 (1976)

\bibitem{Ryu95} D. Ryu, T.~W. Jones: 
                Astrophys.~J. \textbf{442}, 228 (1995)

\bibitem{Sau01} J. Saur, H. Politano, A. Pouquet, W.~H. Matthaeus:
                American Geophysical Union, Fall Meeting (2001)

\bibitem{Sca84} J.~M. Scalo:
                Astrophys.~J. \textbf{277}, 556 (1984)
\bibitem{Sca87} J.~M. Scalo:
                in {\it Interstellar Processes},
                ed. by D.~J. Hollenbach, H.~A. Thronson 
                (Dordrecht: Reidel, 1987) p.349
\bibitem{Sch98} A. Schlickeiser, J.~A. Miller:
                Astrophys.~J. \textbf{492}, 352 (1998)
\bibitem{She94} Z.-S. She, E. Leveque:
                Phys.~Rev.~Lett.
                \textbf{72}(3), 336 (1994) (S-L)

\bibitem{She83} J.~V. Shebalin, W.~H. Matthaeus, D.~C. Montgomery:
                J.~Plasma Phys. \textbf{29}, 525 (1983)


\bibitem{Sim88} J.~H. Simonetti, J.~M. Cordes: 
                in {\it 
                Radio wave scattering in the interstellar medium; 
                Proceedings of the AIP Conference} 
                (American Institute of Physics, New York, 1988) p.~134
\bibitem{Sim92} J.~H. Simonetti: 
                Astrophys.~J. \textbf{386}, 170 (1992)

\bibitem{Spa91} S.~R. Spangler: Astrophys.~J. \textbf{376}, 540 (1991)
\bibitem{Spa90} S.~R. Spangler, C.~R. Gwinn:
                Astrophys.~J. \textbf{353}, L29 (1990)
\bibitem{Sri94} S. Sridhar, P. Goldreich:
                Astrophys.~J. \textbf{432}, 612 (1994)

\bibitem{Sta01} S. Stanimirovic, A. Lazarian:
                Astrophys.~J. \textbf{551}, L53 (2001)

\bibitem{Sto98} J.~M. Stone, E.~C. Ostriker, C.~F. Gammie: 
                Astrophys.~J. \textbf{508}, L99 (1998)

\bibitem{Str76} H.~R. Strauss:
                Phys.~Fluids, \textbf{19}, 134 (1976)


\bibitem{Stu99} J. Stutzki: 
        in {\it Plasma Turbulence and Energetic particles}, ed. by M.
        Ostrowski, R. Schlickeiser (Cracow, 1999) p.48

\bibitem{Stu98} J. Stutzki, F. Bensch, A. Heithausen, V. Ossenkopf,
                M. Zielinsky: 
                Astron.~Astrophys. \textbf{336}, 697 (1998)

\bibitem{Taf98} M. Tafalla, D. Mardones, P.~C. Myers, P. Caselli, R. Bachiller, B.~J. Benson:
                Astrophys.~J. \textbf{504}, 900 (1998)

\bibitem{TMM86} A. Ting, W.~M. Matthaeus, D.~C. Montgomery:
                Phys.~Fluids, \textbf{29}, 3261 (1986)


\bibitem{Vol80} H.~J. Volk, F.~C. Jones, G.~E. Morfill, 
                S. Roser:
                Astron.~Astrophys. \textbf{85}, 316 (1980)
\bibitem{Wei94} S.~J. Weidenschilling, T.~V. Ruzmaikina:
                Astrophys.~J. \textbf{430}, 713 (1994)
%\bibitem{Vaz00} E. Vazquez-Semadeni, A. Gazol, J. Scalo:
%                Astrophys.~J. \textbf{540}, 271 (2000)

\bibitem{Vaz95} E. Vazquez-Semadeni, T. Passot, A. Pouquet:
                Astrophys.~J. \textbf{441}, 702 (1995)
\bibitem{Vaz96} E. Vazquez-Semadeni, T. Passot, A. Pouquet: 
                Astrophys.~J. \textbf{473}, 881 (1996)

\bibitem{Vaz02} E. Vazquez-Semadeni: 
                in {\it Seeing Through the Dust},
                ed. by R. Taylor, T. Landecker, A. Willis 
                (ASP: San Francisco, 2002) (astro-ph/0201072)


\bibitem{War00} A. Warhaft: 
                Annu.~Rev.~Fluid Mech. \textbf{32}, 203 (2000)
\bibitem{Wil59} O.~C. Wilson, G. Munch, E.~M. Flather, M.~F. Coffeen:
                Astrophys.~J.~Supp. \textbf{4}, 199 (1959)

\bibitem{Vai92} S.~I. Vainstein, F. Cattaneo:
                Astrophys.~J. \textbf{393}, 165 (1992)


\bibitem{Ver99} M.~K.~Verma: Phys.~Plasmas, \textbf{6}(5), 1455, (1999)
\bibitem{VDE02} M.~K.~Verma, G.~Dar, V.~Eswaran: Phys.~Plasmas, \textbf{9}(4),
                1484 (2002)

\bibitem{Vis01} E.~T. Vishniac, J. Cho: 
                Astrophys.~J. \textbf{550}, 752 (2001)

\bibitem{WD01}  J.~C. Weingartner, B.~T. Draine:
                Astrophys.~J.~Supp. \textbf{134}, 263 (2001)

\bibitem{Yan02a} H. Yan, A. Lazarian: Phys.~Rev.~Lett. submitted (2002)
astro-ph/0205285

\bibitem{Yan02b} H. Yan, A. Lazarian: in preparation (2002)


\bibitem{Zak67} V.~E.~Zakharov: Sov.~Phys.~JETP, \textbf{24}, 455 (1967)
\bibitem{Zak70} V.~E.~Zakharov, A.~Sagdeev:
                Sov.~Phys.~Dokl. textbf{15}, 439 (1970)

\bibitem{Zan92} G.~P. Zank, W.~H. Matthaeus:
                J.~Plasma Phys. \textbf{48}, 85 (1992)
\bibitem{Zan93} G.~P. Zank, W.~H. Matthaeus:
                Phys.~Fluids A \textbf{5}(1), 257 (1993)

\bibitem{Zie99} M. Zielinsky, J. Stutzki: Astron.~Astrophys.
                \textbf{347}, 630 (1999)

\bibitem{Zwe97} E.~G. Zweibel, C. Heiles:
                Nature, \textbf{385}, 131 (1997)



\end{thebibliography}
\end{document}